\documentclass[preprint,aps,superscriptaddress,nofootinbib,showkeys,11pt]{revtex4}
\pdfoutput=1
\usepackage{graphicx}
\usepackage{amsmath}
\usepackage{amsfonts}
\usepackage{amssymb}
\usepackage{xcolor, soul}
\usepackage{epstopdf}
\usepackage{float}
\usepackage{subfigure}
\usepackage{hyperref}
\hypersetup{
     colorlinks   = true,
     citecolor    = red,
     linkcolor    = blue,
     urlcolor     = blue,
}
\def\beq{\begin{equation}}
\def\eeq{\end{equation}}
\def\bea{\begin{eqnarray}}
\def\eea{\end{eqnarray}}
\def\be{\begin{equation}}
\def\ee{\end{equation}}

\def\nno{\nonumber}
\def\bse{\begin{subequations}}
\def\ese{\end{subequations}}

\graphicspath{{./figs/}}
\begin{document}

\title{Gravitational dark matter: free streaming and phase space distribution}
\author{Md Riajul Haque}%
\email{riaju176121018@iitg.ac.in}
\author{Debaprasad Maity}
\email{debu@iitg.ac.in}
\affiliation{%
	Department of Physics, Indian Institute of Technology Guwahati,\\
	Guwahati, Assam, India, 781039 
}%

\date{\today}

\begin{abstract}
Gravitational dark matter (DM) is the simplest possible scenario that has recently gained interest in the early universe cosmology. In this scenario, DM is assumed to be produced from the decaying inflaton through the gravitational interaction during reheating. Gravitational production from the radiation bath will be ignored as our analysis shows it to be suppressed for a wide range of reheating temperature $(T_{re})$. Ignoring any other internal parameters except the DM mass $(m_Y)$ and spin, a particular inflation model such as $\alpha$-attractor, with a specific scalar spectral index $(n_s)$ has been shown to uniquely fix the dark matter mass of the present universe. For fermion type dark matter we found the mass $m_f$ should be within $(10^4 - 10^{13})$ GeV, and for boson type DM, the mass $m_{s/X}$ turned out to be within $(10^{-8}-10^{13})$ GeV. Interestingly, if the inflaton equation of state $\omega_{\phi}\rightarrow 1/3$, the DM mass also approaches towards unique value, $m_f \sim 10^{10}$ GeV and $m_{s/X} \sim 10^3\,(\,8\times 10^3\,)$ GeV irrespective of the value of $\omega_\phi$. We further analyzed the phase space distribution $(f_Y)$, and free streaming length $(\lambda_{fs})$ of these gravitationally produced DM. $f_Y$, which is believed to encode important information about DM, is shown to contain a characteristic primary peak at the initial time where the gravitational production is maximum for both fermion/boson. Apart from this fermionic phase-space distribution function contains an additional peak near the inflaton and fermion mass equality ($m_Y=m_\phi$) arising for $\omega_\phi>5/9$.
Furthermore, the height of this additional peak turned out to be increasing with decreasing $T_{re}$, and at some point dominates over the primary one. Since reheating is a causal process and dark matter is produced during this phase, gravitational instability forming small-scale DM structures during this period will encode those phase space information and be observed at present. Crucial condition $\lambda_{fs} <\lambda_{re}$ of forming such small scale DM structure during reheating has been analyzed in detail. We further estimate in detail the range of scales within which the above condition will be satisfied for different dark matter masses. Finally, we end by stating the fact that all our results are observed to be insensitive on the parameter $\alpha$ of the inflaton potential within the allowed range set by the latest Planck and BICEP/Keck results.
\end{abstract}
\maketitle

\section{Introduction}
Cosmological observation over more than half a century made us believe that the observable universe is made of visible and invisible components \cite{Planck:2018jri,Aubourg:2014yra,Baumann:2008bn,Vazquez:2018qdg,Kawasaki:2000en,Steigman:2007xt,Fields:2014uja,Kawasaki:1999na}. Regarding the visible components, we have acquired and inculcated a great deal of knowledge about its very existence and fundamental properties. However, apart from the existential evidences through multiple observations such as galaxy rotation curve, large scale structure, CMB \cite{Planck:2018jri,CMB-S4:2016ple,Sofue:2000jx,Sahni:2004ai,Catena:2009mf,Guo:2009fn}, invisible components are far from our present understanding. Dark matter is one of the invisible components which attracts lot of attention due to its seemingly unavoidable entente with the visible components in quantum field theoretic framework \cite{Arkani-Hamed:2008hhe,Feng:2010gw,Pospelov:2007mp,Tenkanen:2016twd,Heikinheimo:2016hid,Hall:2009bx,Chu:2013jja,Blennow:2013jba,Elahi:2014fsa,Mambrini:2015vna,Nagata:2015dma,Chen:2017kvz,Bernal:2018qlk,Bernal:2018ins,Garcia:2020eof,Garcia:2021gsy,Hochberg:2014dra,Hochberg:2014kqa,Hochberg:2015vrg,Bernal:2015xba,Falkowski:2011xh,LopezHonorez:2010eeh}. Even though very few effective field theory parameters such as mass and cross-section are sufficient to explain the very existence of dark matter (DM), ignorance/non-detection\cite{XENON100:2012itz,XENON:2018voc,PandaX-II:2016vec,LUX:2016ggv,BOSS:2013rlg} of its fundamental characters may seem indicative to suffering of going beyond the present framework of experimental and theoretical approaches \cite{Capozziello:2006dj,Capozziello:2011et,Boehmer:2007kx,Nojiri:2017ncd}. List of conventional particle physics approaches towards DM production being nearly exhaustive, ideas of gravitational mechanism of dark matter \cite{Donoghue:1994dn,Choi:1994ax,Holstein:2006bh,Mambrini:2021zpp,Bernal:2021kaj,Barman:2021ugy,Ema:2015dka,Ema:2016hlw,Ema:2018ucl,Garny:2015sjg,Garny:2017kha,Tang:2017hvq} seems to suggest that simplest possibilities going beyond the convention, still have a lot of unexplored provisions. Gravity so far plays an extremely passive role in understanding the physical properties of standard model particles. However, difficulties in incorporating gravity in the quantum field theory framework are the fundamental reason behind this. Nevertheless, based on our present understanding, the physical laws depend on the energy scale of interest. At low energy ($\lesssim 1$ TeV), the SM particles may have effectively isolated themselves from gravity as long as their fundamental properties are concerned. At high energy, however, this must not be true; rather, particles and gravity may not have independent identities on their own. String theory is an elegant example that subscribes to such an idea. The gravitational production of dark matter may fall along this line of thought. At the classical level, Einstein's equivalence principle suggests that gravity universally couples with particles irrespective of their intrinsic properties except for mass. However, if we tend to apply this at the quantum level, where two different particle sectors are coupled through gravitons, the production cross-section does depend on the intrinsic properties such as spin, charge, hence violating the equivalence principle. In this paper, we study one of such scenarios where DM is produced through inflaton annihilating into fermion/bosonic through s-channel graviton exchange. Given an inflationary model, our focus will be on the reheating phase of the universe. Considering the reheating phase with matter domination, such a scenario has already been studied \cite{Mambrini:2021zpp,Bernal:2021kaj}. We generalize such a study for an arbitrary reheating equation of state. We also include the effect of production from the radiation bath for completeness. Hence, the produced dark matter will have thermal and non-thermal components that are generically non-cold in nature. These different production mechanisms of non-cold DM lead us to further study in detail their phase-space distribution and free streaming length depending upon the reheating equation of state. We will see how depending upon the type of dark matter; the distribution function contains distinct features and its dependence on the reheating equation of state. Those properties play significant role in clustering of matter on galactic and sub-galactic scales\cite{Lazar:2020pep,BOSS:2016wmc,Bode:2000gq}. Observing those small scale matter power spectrum by mapping the Lyman-$\alpha$ \cite{Viel:2013fqw, Narayanan:2000tp, Viel:2005qj,Baur:2015jsy,Irsic:2017ixq,Palanque-Delabrouille:2019iyz,Garzilli:2019qki,Ballesteros:2020adh} forest of absorption lines of light from low redshift $(z=2-4)$ quasars can differentiate different non-cold DM production mechanism and its intrinsic properties.

\section{Boltzmann Framework}\label{method}
After the period of exponential expansion, the inflaton field begins to oscillate around its minima with decaying amplitude. In the framework of quantum field theory, the time-dependent inflaton field can naturally decay into various daughter fields such as radiation, dark matter particles, etc. However, the decay process non-trivially depends on the inflaton coupling with those daughter fields. 
In order to have successful reheating, the inflaton is generically assumed to have direct coupling with the radiation field, which will be the dominating component after the end of reheating. However, due to its sub-dominant nature, the probability of solely gravitational production of dark matter can survive in some region of parameter space, which has already been observed in \cite{Mambrini:2021zpp}. In this section, we first describe the framework of such a scenario. For completeness, dark matter is assumed to be produced both from the radiation bath with a thermal-averaged cross-section $\langle\sigma v \rangle$ as a free parameter and from the gravitational decay of the inflaton field. The gravitational production of dark matter has been proved to be dominated by the annihilation of inflaton zero modes through the s-channel graviton exchange process; namely, $\phi\phi \to SS/ff/XX$, where $\phi$ is the inflaton and $S$, $f$, and $X$ indicate scalar, fermionic and vector dark matter, respectively \cite{Ema:2015dka,Ema:2016hlw,Ema:2018ucl}. The interaction Lagrangian for s-channel gravitational production of dark matter can be universally described by the coupling of the dark matter energy-momentum tensor $T^{\mu\nu}$ with the tensor metric perturbation $h_{\mu\nu}$ as \cite{Donoghue:1994dn,Choi:1994ax,Holstein:2006bh}
\bea \label{lagrangian}
\mathcal{L}=\frac{1}{2M_p}\left(h_{\mu\nu} T_\phi^{\mu\nu}+h_{\mu\nu}T_{S/f/X}^{\mu\nu}\right)~~.
\eea
Associated with this action, the corresponding decay widths can be calculated as \cite{Mambrini:2021zpp,Barman:2021ugy} 
\begin{eqnarray}\label{decaywidthscalar}
&&\Gamma_{\phi \phi \to SS}=\frac{\rho_\phi\, m_\phi}{1024\,\pi\, M_p^4}\left(1+\frac{m_s^2}{2\,m_\phi^2}\right)^2\sqrt{1-\frac{m_s^2}{ m_\phi^2}}~~,\\
&&\Gamma_{\phi \phi \to ff}=\frac{\rho_\phi \,m_f^2}{4096\,\pi\, M_p^4 \,m_\phi}\left(1-\frac{m_f^2}{ m_\phi^2}\right)^{\frac{3}{2}}~~,\\ \label{fermion}
&&\Gamma_{\phi \phi \to XX}=\frac{\rho_\phi\, m_\phi}{32768\,\pi\, M_p^4}\,\sqrt{1-\frac{m_X^2}{ m_\phi^2}}\,\left(4+\,4\,\frac{m_X^2}{ m_\phi^2}+\,19\,\frac{m_X^4}{ m_\phi^4}\right)~~.\label{gauge}
\end{eqnarray}
Where $m_{s/f/X}$ is the mass of the scalar, fermionic, and vector dark matter, respectively, and the effective mass of the inflaton is symbolized as $m_\phi$. At this point, we would like to point out that gravitational production of dark matter from radiation is also possible, and production rate per unit time per unit volume is followed by Eq.\ref{SMgrav}. Such production is strongly suppressed compared to the production from inflaton, which we have shown in Sec-\ref{comparison rad-inflaton}
. However, for high reheating temperature $ 10^{15} \gtrsim  T_{re} \gtrapprox 10^{13}$ GeV, fermion type dark matter gravitationally produced from the radiation bath has been observed to satisfy correct abundance in a certain range of fermion mass. We have numerically checked the results, which is shown in Fig.\ref{purely gravitational dark matter}. We will not include this possibility in detail in our subsequent mathematical discussions. However, we will describe the numerical results of such a scenario as we go along. 

We investigate the detailed dynamics of reheating by solving the following Boltzmann equations with three density components for inflaton $\rho_\phi$, radiation $\rho_r$, and the total dark matter number density $n_{Y} = \left( n_{Y}^r+n_{Y}^{\phi}\right)$ as \cite{Chung:1998rq,Giudice:2000ex,Maity:2018dgy}
\begin{eqnarray}\label{boltzmann}
&&\dot{\rho}_\phi+3\,H\,\left(1+\omega_\phi\right)\,\rho_\phi+\left(\,\Gamma_\phi+\Gamma_{\phi\phi\to YY}\,\right)\rho_\phi\left(1+\omega_\phi\right)=0~~,\\
&&\dot{\rho}_r+4\,H\,\rho_r-\Gamma_\phi\,\rho_\phi\left(\,1+\omega_\phi\,\right)-2\,\langle\sigma v\rangle \,\langle E_{Y}\,\rangle^r\left[\,\left(n_{Y}^{r}\right)^2-\left(n_{Y}^{eq}\right)^2\,\right]=0~~,\\
&&\dot{n}_{Y}^r+3\,H\,n_{Y}^r+\langle\sigma v\rangle\,\left[\,\left(n_{Y}^r\right)^2-\left(n_{Y}^{eq}\right)^2\,\right]=0~~,\\
&&\dot{n}_{Y}^\phi+3\,H\,n_{Y}^\phi-\frac{\rho_\phi\,\left(\,1+\omega_\phi\,\right)}{\langle E_{Y}\,\rangle^\phi}\,\Gamma_{\phi \phi\to YY}=0~~,\label{B2}
\label{boltzmann1}
\end{eqnarray}
where $n_{Y}^r$ and $n_{Y}^\phi$ are the DM number density gravitationally produced from the thermal bath and the decay of the inflaton field respectively. $\langle \,E_{Y}\,\rangle^r=\sqrt{m_{Y}^2+\left(3\,T_{rad}\right)^2}$ and $\langle \,E_{Y}\,\rangle^\phi=\sqrt{m_{Y}^2+\,m_\phi^2}$ are the average energy per dark matter particle produced from the thermal bath \cite{Giudice:2000ex} and inflaton decay respectively. The equilibrium number density of the dark matter particles can be expressed by the modified Bessel function of the second kind as,
\bea
n_Y^{eq}=\frac{\tilde{g_Y}\, T_{rad}^3}{2\,\pi^2}\left(\,\frac{m_{Y}}{T_{rad}}\,\right)K_2\left(\,\frac{m_{Y}}{T_{rad}}\,\right)~~,
\eea
Additionally, the energy associated with each gravitationally produced dark matter particle can be calculated from the energy and momentum conservation of the annihilation-like $\phi\phi \to SS/ff/XX$ process as
\bea
0=p_1+p_2~~;~~2\,m_\phi=\sqrt{p_1^2+m_{Y}^2}+\sqrt{p_2^2+m_{Y}^2}=2\,\sqrt{p_1^2+m_{Y}^2}~~.
\eea
Here $p_1$ and $p_2$ are the momenta of two gravitationally produced dark matter particles. The above equations assume the fact that the homogeneous background inflaton is at rest, and hence the energy stored in each gravitationally produced dark matter particle will be of the order of $m_\phi$.
   In order to solve the above set of Boltzmann equation we define the following dimensionless variables corresponding to different energy components,
    \bea
\Phi=\frac{\rho_\phi\,A^{3\,(\,1+\omega_\phi\,)}}{(\,m_\phi^{end}\,)^{4}}~,~R=\frac{\rho_r\,A^4}{(\,m_\phi^{end}\,)^{4}}~,~Y^r=\frac{n_{Y}^r\,A^3}{(\,m_\phi^{end}\,)^{3}}~,~Y^\phi=\frac{n_{Y}^\phi\,A^3}{(\,m_\phi^{end}\,)^{3}}~~.
\eea
Where, $A={a}/{a_{end}}$ and $m_\phi^{end}$ are the normalized scalar factor and the effective mass of the inflaton field at the end of the inflation respectively. $m_\phi^{end}=\partial _\phi^2 V(\phi_{end})$.
This modification factor $m_\phi^{end}$ increases the stability of the numerical solution. 
In terms of new dimensionless variable Eqns.(\ref{boltzmann}-\ref{boltzmann1}) can be written as
   \bea \label{Boltz1}
&&   {\Phi'} = -c_1\,\left(\,\Gamma_\phi +\Gamma_{\phi \phi\to YY}\,\right)\frac{A^{1/2}\,\Phi}{\textbf{H}}~~,\nno \\
&&   {R'} =  c_1\,\Gamma_\phi\,\frac{ A^{\frac{3\left(1-2\omega_\phi\right)}{2}}\,\Phi}{\textbf{H}} + 2\,\sqrt{3}\,\frac{A^{-3/2}\,\langle\, \sigma v\,\rangle \,\langle\, E_{Y}\,\rangle^r\, M_{p} }{\textbf{H}}\,\left[\,(Y^{r})^{2}-Y_{eq}^2\right]~~,\nno\\
&& {(Y^r)'} = - \sqrt{3}\, \frac{A^{-5/2}\,\langle \,\sigma v\,\rangle M_{p}\, m_\phi^{end} }{\textbf{H}}\,\left[\,(Y^{r})^{2}-Y_{eq}^2\right]~~,\nno \\
 &&{(Y^\phi)'}= c_1\,\Gamma_{\phi \phi\to YY}\,\frac{A^{\frac{1}{2}-3\,\omega_\phi}\,\Phi}{\textbf{H}}\,\left(\,\frac{m_\phi^{end}}{\langle \,E_{Y}\,\rangle^\phi}\,\right)~~,
   \eea
where 
\bea
\textbf{H}=\sqrt{\frac{\Phi}{A^{3\omega_\phi}}+ \frac{R}{A}+ \frac{Y^r\, \langle\, E_{Y}\,\rangle^r}{m_\phi^{end}}+\frac{Y^\phi\,\langle \,E_{Y}\,\rangle^\phi}{m_\phi^{end}}}~~;~~c_{1}= \frac{\sqrt{3}\, M_{p} \,\left(\,1+\omega_\phi\,\right)}{(m_{\phi}^{end})^2}~~.
\eea

\subsection{Model of inflation}
We will focus on a class of models called the $\alpha$-attractor model \cite{Kallosh:2013hoa,Kallosh:2013yoa}, which unifies the large class of inflationary models parameterized by $\alpha$ and $n$. Here we would like to mention that the canonical property of this class of models predicts inflationary observables ($n_s,r$) in favor of Planck observation \cite{Planck:2018jri,Planck:2015sxf}. The $\alpha$-attractor E-model has the defining inflaton potential,
\begin{equation} \label{alphapotential}
V(\phi) = \Lambda^4 \left[  1 - e^{ -     \sqrt{\frac{2}{3\alpha} }    \frac{\phi}{M_p}     } \right]^{2n}~~,
\end{equation}
where $\Lambda$ is the mass scale that can be fixed by the CMB power spectrum, which is of the order of $\sim 8\times 10^{15}$ GeV. 
Moreover, for $n=1,\alpha=1$, the $\alpha-$attractor model turns out as Higgs-Starobinsky model \cite{Bezrukov:2007ep,Starobinsky:1980te}. 
To this end we would also like to point out that recent Planck and BICEP/Keck combined result has put constraint on the $\alpha $ to be $\sim(1-12)$ \cite{Ellis:2021kad} within $1\sigma$ of $n_s$ value \cite{BICEP:2021xfz}.
Throughout our study we set $\alpha=1$ and varying $n$. 
Importantly, we have checked that changing $\alpha$ within the aforementioned range does not significantly change our results.  
 Let us first try to establish the relationship between the potential parameters with inflationary parameters. For the potential (\ref{alphapotential}), the inflationary e-folding number $N_k$ and tensor to scalar ratio $r_k$ can be expressed as \cite{Drewes:2017fmn}
\bea \label{inflationaryparameter}
  N_k=\frac{3\alpha}{4n} \left[e^{\sqrt{\frac{2}{3\alpha}}\frac{\phi_k}{M_p}}-e^{\sqrt{\frac{2}{3\alpha}}\frac{\phi_{end}}{M_p}}-\sqrt{\frac{2}{3\alpha}}\frac{(\phi_k-\phi_{end})}{M_p}\right]~,~r_k=\frac{64 n^2}{3\alpha \left(e^{\sqrt{\frac{2}{3\alpha}}\frac{\phi_k}{M_p}}-1\right)^2}~~.
\eea
Here $\phi_k$ and $\phi_{end}$ denote the values of the scalar field $\phi$  at the Hubble crossing of a particular mode $k$ and the end of the inflation, respectively. From the condition of the end of the inflation $\epsilon(\phi_{end})=\frac{1}{2M_p^2}\left({V'(\phi_{end})}/{V(\phi_{end})}\right)^2=1$, the value of the field and the potential at the end of the inflation are,
\bea\label{infpotentialfield}
\phi_{end}=\sqrt{\frac{3\alpha}{2}}M_p~ ln\left(\frac{2n}{\sqrt{3\alpha}}+1\right)~~,~~V_{end}=\Lambda^4 \left(\frac{2n}{2n+\sqrt{3\alpha}}\right)^{2n}~~.
\eea
For a given canonical inflaton potential $V(\phi)$, the inflationary observables can be related to the slow-roll parameters and Hubble parameter at the point when the mode with wave number $k$ crosses the horizon,
\begin{equation}\label{measurephik}
 n_{s}= 1- 6 \epsilon(\phi_k)+ 2 \eta(\phi_k)~,~r_k=16\epsilon(\phi_k)~,~~~H_k =  \frac{\pi M_p\sqrt{r_k A_s}}{\sqrt{2}}~,
\end{equation}
where $A_s$ is the amplitude of the inflaton fluctuation, which is of the order of $\sim 10^9$ measured from CMB observation. The above equation (\ref{measurephik}) can be  inverted to give the field value $\phi_k$ and the mass scale $\Lambda$ as
\bea
\phi_k=\sqrt{\frac{3\alpha}{2}}~M_p~ln\left(1+\frac{4n+\sqrt{16~n^2+24~\alpha ~n~(1+n)~(1-n_s)}}{3~\alpha~(1-n_s)}\right)~,\\
 \Lambda= M_p \left(\frac{3\pi^2 r A_s}{2}\right) \left[\frac{2n(1+2n)+\sqrt{4n^2+6\alpha(1+n)(1-n_s)}}{4n(1+n)}\right]^{\frac{n}{2}}~.
\eea
In order to solve the Boltzmann Eqs.\ref{Boltz1}, one needs to replace the inflaton field variable in terms of its oscillation average, which can be further expressed in terms of the energy density of the inflaton. Such average over oscillation period provides the following relation $V(\phi(t))=\rho_\phi(t)$ \cite{Shtanov:1994ce}. By using this, the effective mass $m_\phi$ of the inflaton can be expressed in terms of inflaton energy density during reheating as:
\bea\label{effectivemass}
m_\phi^2=-\frac{4n \rho_\phi\left(\left\{1-\left(\frac{\rho_\phi}{\Lambda^4}\right)^\frac{1}{2n}\right \}^{-1}-2n\right)}{3\alpha \left(\left\{1-\left(\frac{\rho_\phi}{\Lambda^4}\right)^\frac{1}{2n}\right \}^{-1}-1\right)^2M_p^2}~~,
\eea
For example, for $n=1$ model, the inflaton mass turns out to be $
m_\phi\simeq ({2\Lambda^2})/({\sqrt{3\alpha}M_p})$.
After identifying all of the required parameters during inflation, one can set initial conditions for subsequent reheating dynamics, which in turn provide important relationships among the reheating parameters, namely the temperature $T_{re}$ and e-folding number $N_{re}$ in terms of $(n_s,r)$. Therefore, we can establish the relations between the CMB anisotropy and reheating era via inflation.
\subsection{Reheating parameters and observable constraints}\label{methodology}
In order to solve Boltzmann equations numerically, the initial conditions for the dimensionless comoving densities are set at the end of inflation
 \begin{equation}\label{boundary1}
 \varPhi(A=1)= \frac{3}{2} \,\frac{V_{end}(\phi)}{\left(m_{\phi}^{end}\right)^4}~~,~~R(A=1)=\,Y^r(A=1)=\,Y^\phi(A=1)=0~~.
\end{equation}
Once we solve the Boltzmann equations numerically during reheating, the reheating temperature can be identified at the point where the decay rate is maximized,  
\begin{equation}\label{reheating 1}
 H(A_{re})^2= \left(\frac{\dot A_{re}}{A_{re}}\right)^2= \frac{\rho_\phi\,(\,\Gamma_\phi,\,A_{re},\,n_{s})+ \,\rho_{r}\,(\Gamma_\phi,\,A_{re},\,n_{s})+\,\rho_{Y}\,(\Gamma_\phi,\,A_{re},\,n_{s}))}{3 M_p^2}=\left(\Gamma_{\phi}+\Gamma_{\phi\phi\to YY}\right)^2~,
\end{equation} 
where $A_{re}= {a_{re}}/{a_{end}}$, the normalized scale factor at the end of reheating. To this end let us point out that the above condition may not necessarily always be satisfied. For those situation, $\rho_{\phi} = \rho_R$, can be used. Such situation may arise when dilution of radiation due to expansion is faster than the production, which may happen for $\omega_\phi > 1/3$. We will discuss such scenarios in detail in our subsequent paper. Nonetheless, at the point of reheating end the production rate of the dark matter mediated by gravity from inflaton is sub-dominant compared to the decay rate of the inflaton. Therefore, we can approximate the total decay width at the end point of reheating as $\Gamma_T = \Gamma_\phi+ \Gamma_{\phi\phi\rightarrow YY} \simeq \Gamma_{\phi}$. This will be useful for our analytic computation of the reheating parameters. Furthermore, the reheating temperature can be expressed in terms of radiation temperature as
\bea \label{reheating 2}
T_{re}= T_{rad}^{end}= \left(\frac{30}{\pi^2 g_{re}}\right)^{1/4}\rho_{r}(\Gamma_\phi,A_{re},n_{s})^{1/4}~.
\eea
We can also relate the reheating temperature to the present CMB temperature under the assumption that after reheating entropy is preserved in CMB and neutrino background today, that leads to the following {\bf constraint relation}
\bea \label{eqtre}
T_{re}= \left(\frac{43}{11 g_{s, re}}\right)^{\frac 1 3}\left(\frac{a_0T_0}{k}\right) H_k e^{-N_k} e^{-N_{re}} = {\cal G} e^{-N_{re}},\\\mbox{where}~~ \mathcal{G}=\left(\frac{43}{11\, g_{s, re}}\right)^{\frac{1}{3}}\left(\,\frac{a_0\,T_0}{k}\,\right) H_k\,e^{-N_k}.
\eea
The present CMB temperature, $T_0=2.725\,k$, CMB pivot scale of Planck $k/a_0=0.05$ $Mpc^{-1}$, $a_0$ is the cosmological scale factor at present and $g_{s,re}$ is the degrees of freedom associated with entropy at reheating. Combining  equations (\ref{reheating 1}), (\ref{reheating 2}) and (\ref{eqtre}), we can fixed the decay width $\Gamma_\phi$ in terms of $n_s$ and obtain one to one correspondence between $T_{re}$ and $\Gamma_\phi$. \\
Furthermore, cosmological observation on the dark matter abundance $\Omega_{Y}h^2$ provides second {\bf constraint relation} as  \cite{WMAP:2010sfg,Planck:2018vyg}
\bea\label{darkmatter relic}
\Omega_Y h^2= m_{Y}\, \frac{\left(\,Y^r(A_F)+\,Y^\phi\,(A_F)\,\right)}{R\,(A_F)}\,\frac{T_F\,A_F}{T_{now}\,m_\phi^{end}}\,\Omega_r h^2~ = 0.12~~,
\eea
where the present day radiation abundance $\Omega_R h^2=4.3 \times 10^{-5}$ and $T_F$ is the radiation temperature determined at a very late time $A_F$, when both comoving radiation and dark matter energy density became constant. Solving Boltzmann equations and utilizing the conditions mentioned in Eqs.\ref{eqtre}, \ref{darkmatter relic}, we can constrain the dark matter parameters $(\langle\,\sigma v\,\rangle,\,m_Y)$ in terms of $(T_{re}, n_s)$. The dark matter particles produced from radiation bath populated the early universe with two possible mechanisms: 1) The produced dark matter particles reach thermal equilibrium, and as the temperature falls below the dark matter mass, the number density of dark matter freezes out. This mechanism is referred to as the freeze-out mechanism \cite{Hut:1977zn,Lee:1977ua,Vysotsky:1977pe,Srednicki:1988ce,Gondolo:1990dk,Griest:1990kh,Arcadi:2017kky,Hochberg:2018rjs}. 2) The interaction of the dark matter particles with radiation bath could be too weak to attain thermal equilibrium before it freezes out. This mechanism is referred to as the freeze-in \cite{Bernal:2020qyu} mechanism, and the produced dark matter particles are generally known as feebly interacting dark matter (FIMP) \cite{Tenkanen:2016twd,Heikinheimo:2016hid,Hall:2009bx,Chu:2013jja,Blennow:2013jba,Elahi:2014fsa,Mambrini:2015vna,Nagata:2015dma,Chen:2017kvz,Bernal:2018qlk}.
For gravitationally produced dark matter freeze-in mechanism will be effective, and dark matter produced from the radiation bath will have both possibilities of freeze-in and freeze-out production.  However, we will consider the freeze-in mechanism for both the dark matter sector.

\section{Constraining  the dark sector}
\begin{figure}[t!]
\begin{center}

\includegraphics[height=3.4cm,width=5.40cm]{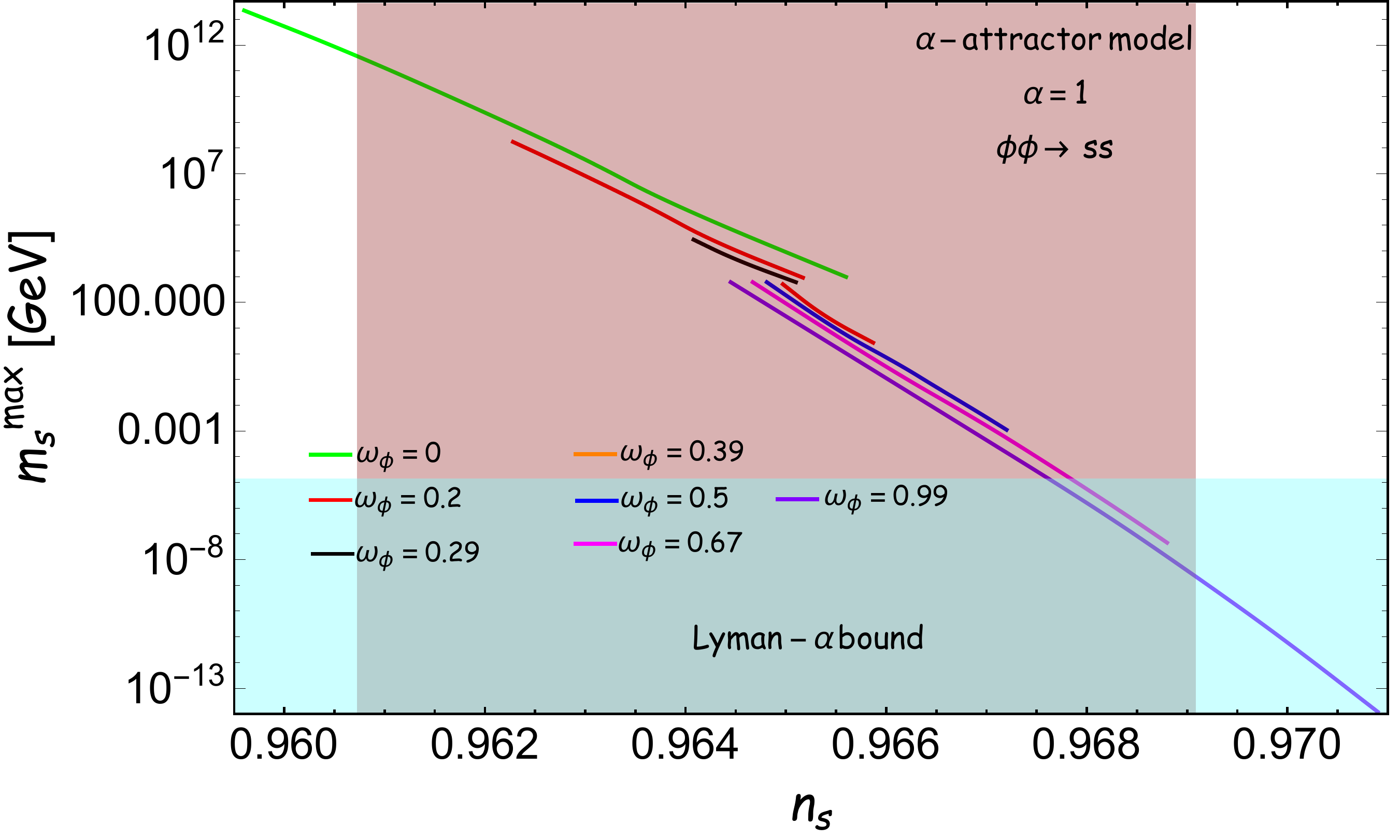}
\includegraphics[height=3.4cm,width=5.40cm]{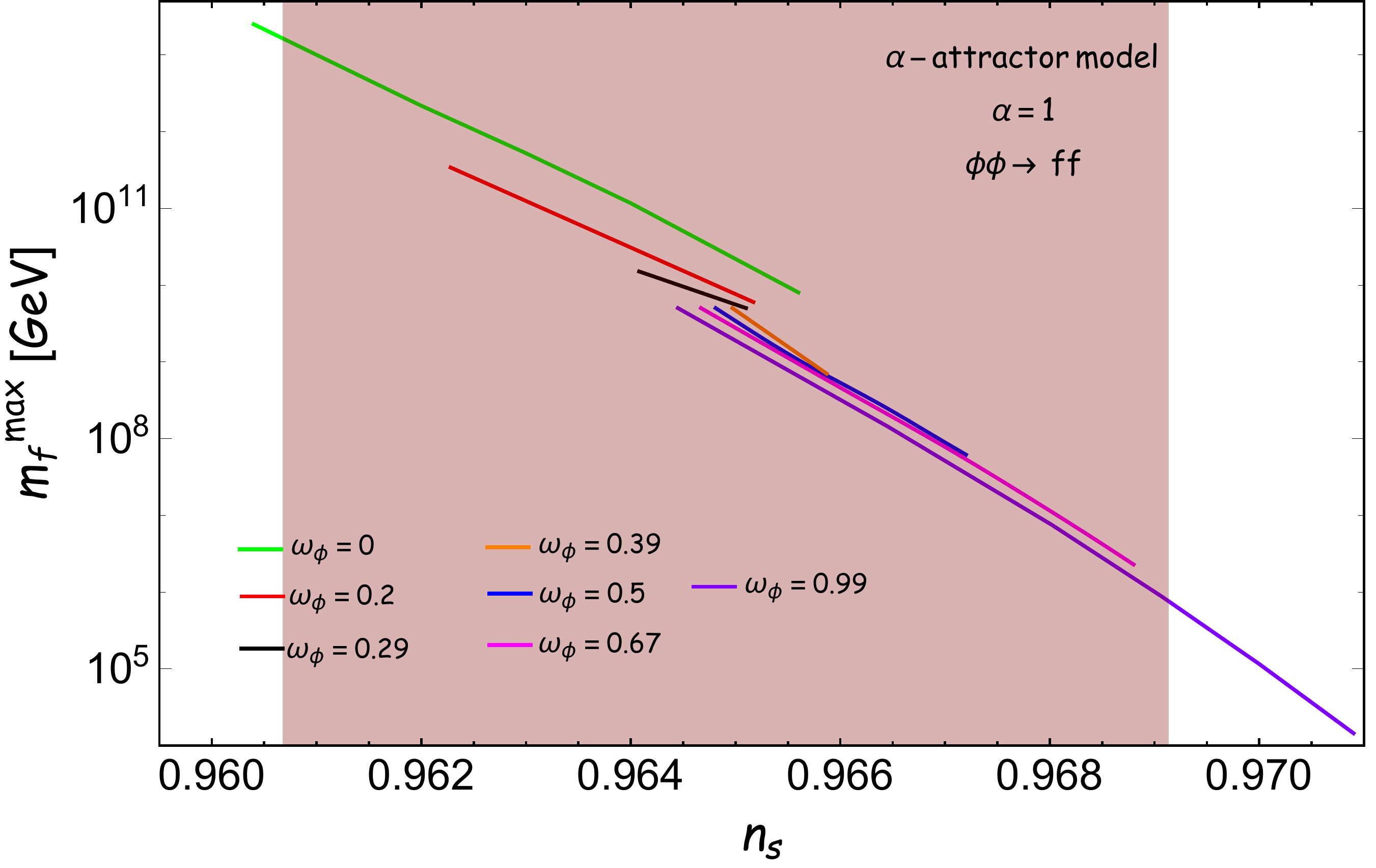}
\includegraphics[height=3.4cm,width=5.40cm]{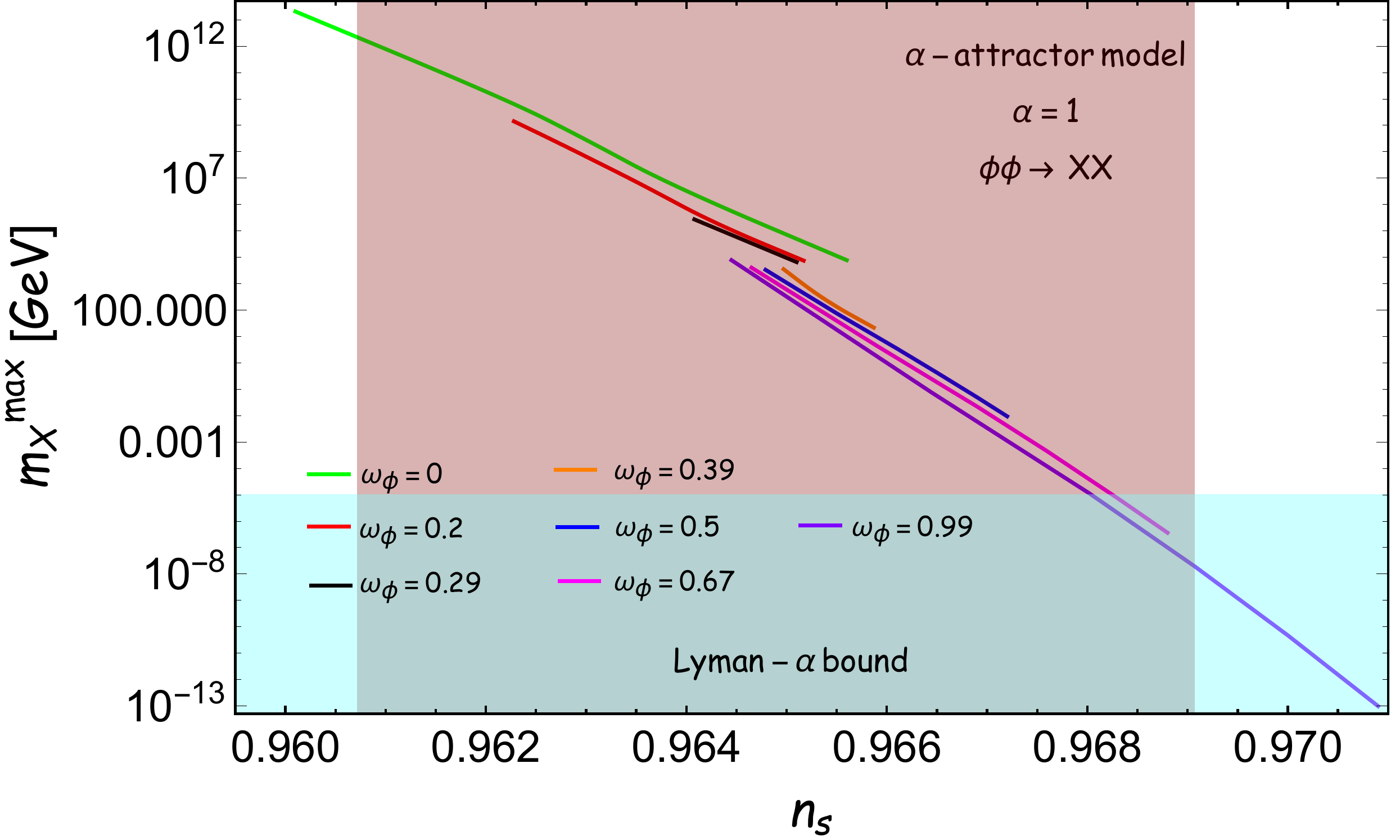}
\includegraphics[height=3.4cm,width=5.40cm]{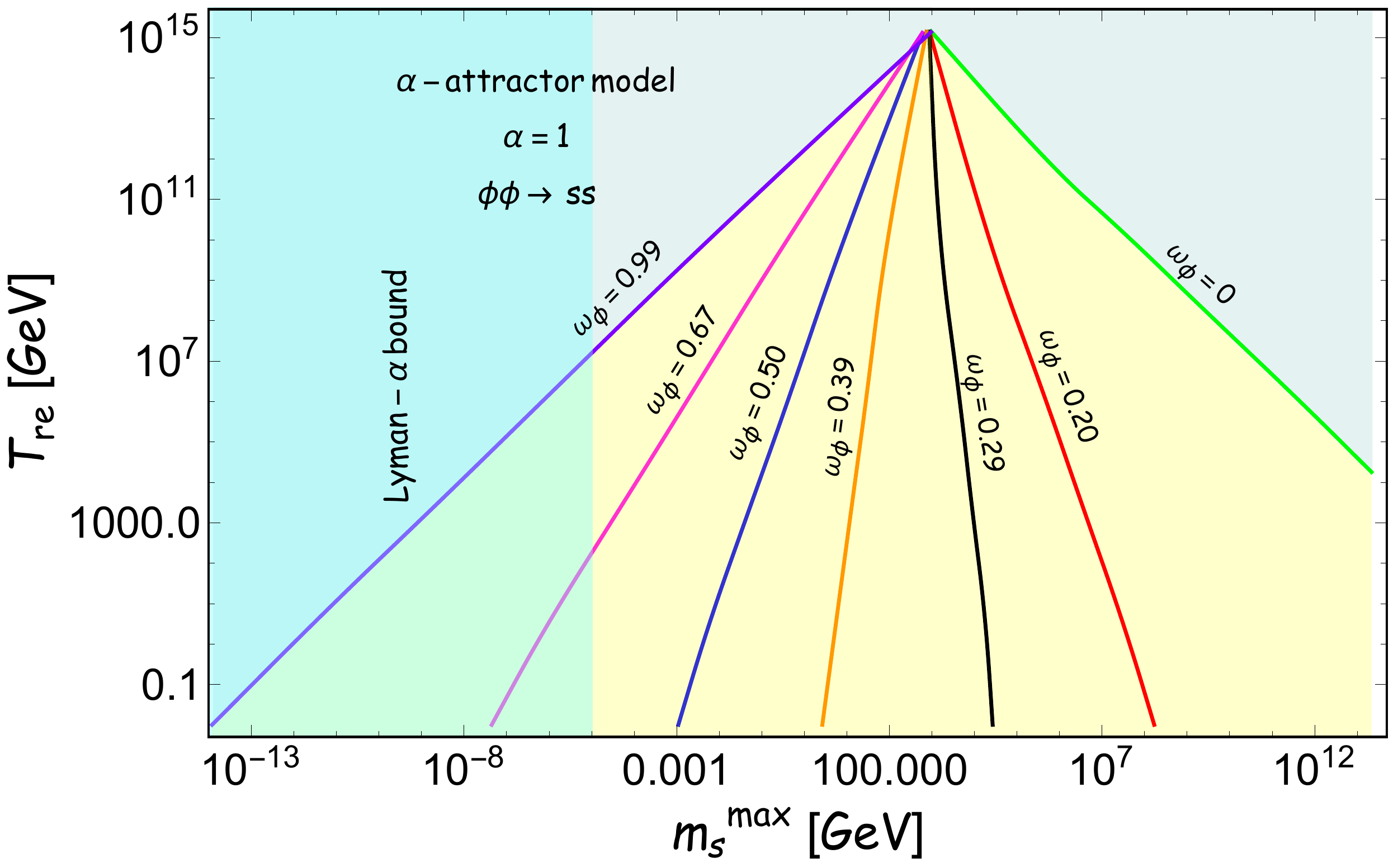}
\includegraphics[height=3.4cm,width=5.40cm]{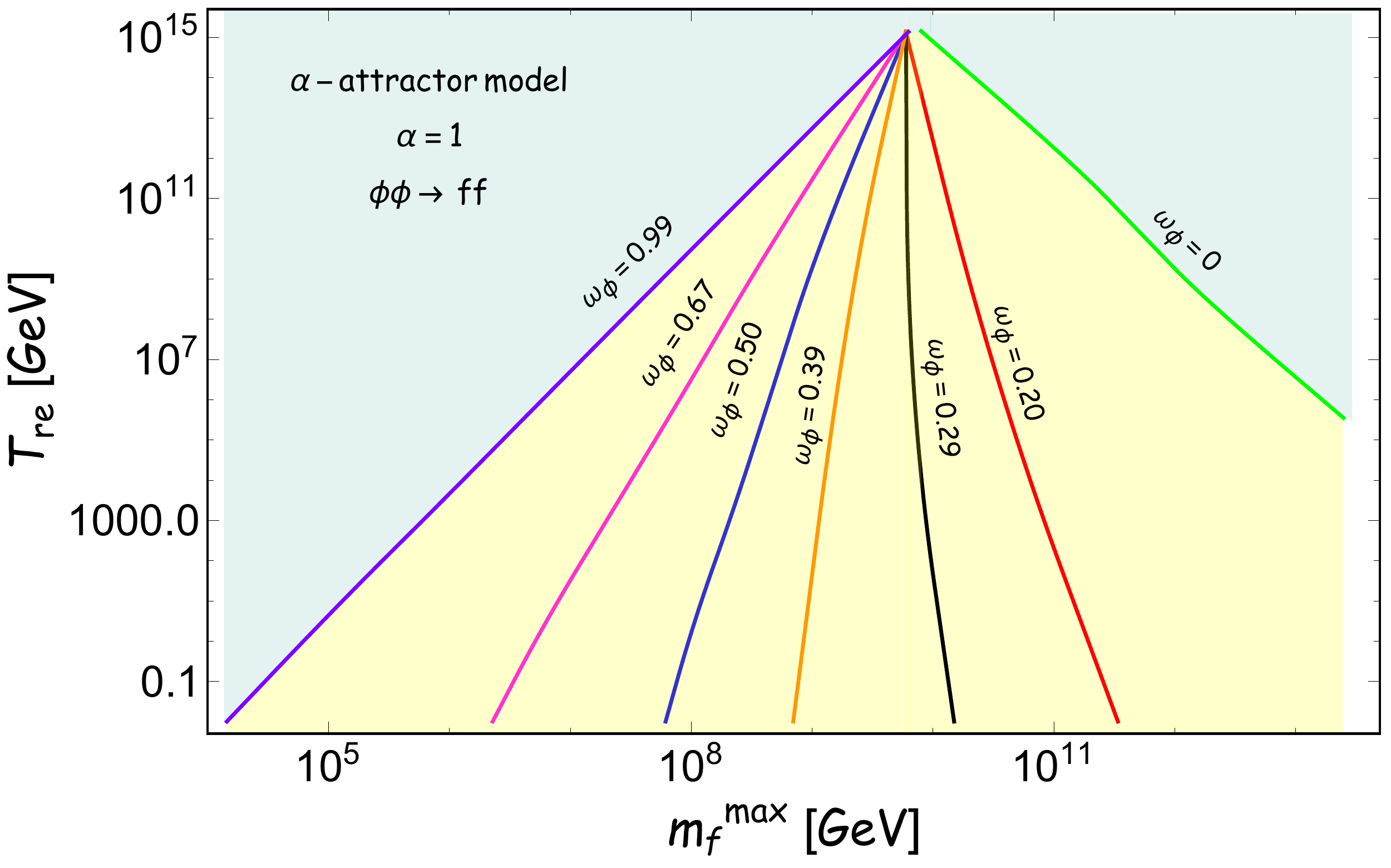}
\includegraphics[height=3.4cm,width=5.40cm]{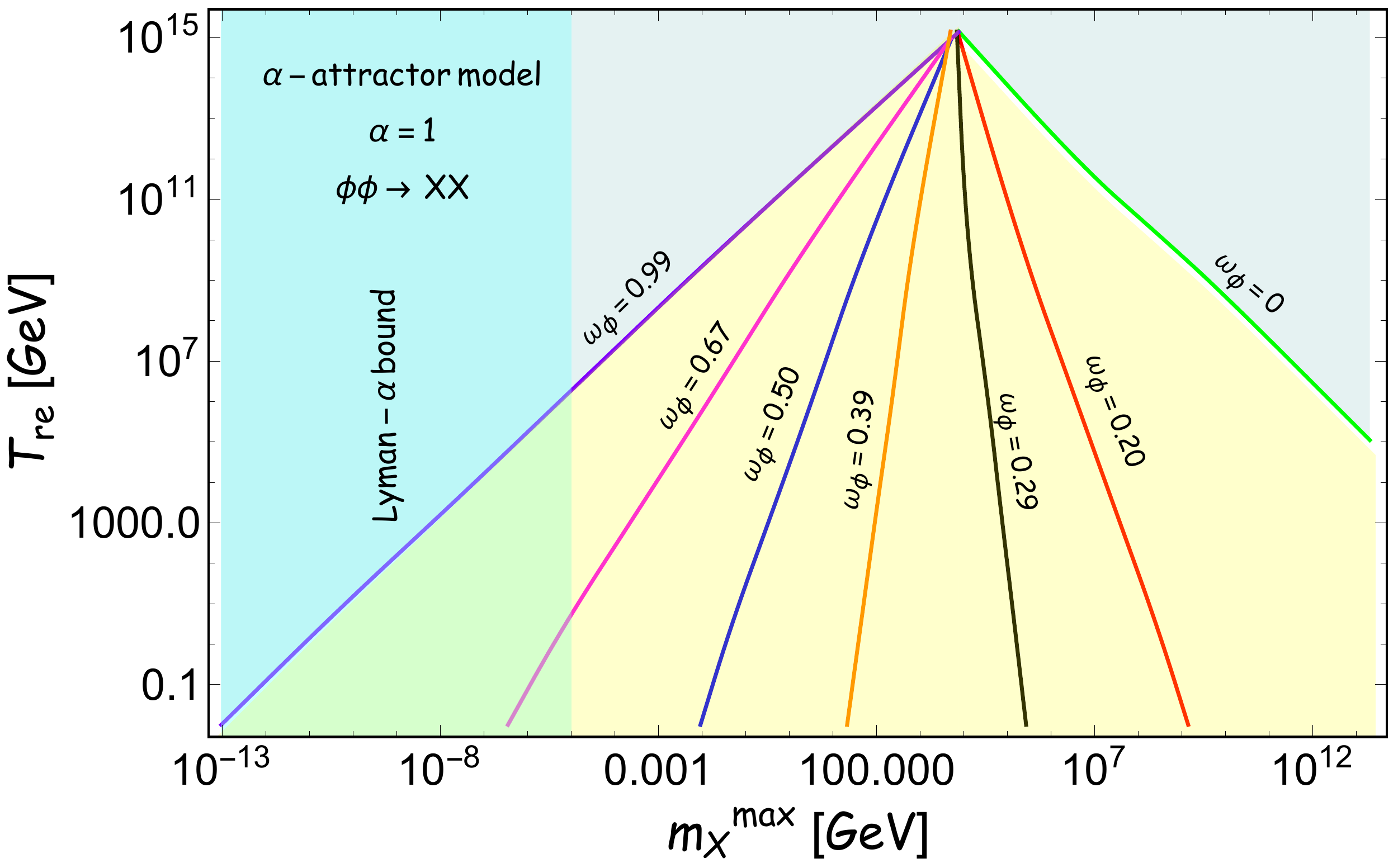}

\caption{{\bf Upper panel:} The variation of the maximum allowed values of the dark matter mass ($m_{Y}^{max}$) as a function of the scalar spectral index ($n_s$) corresponding to the fixed value of the dark matter abundance $\Omega_{Y}h^2\simeq 0.12$ for the cases wherein $\omega_\phi=(0,\,0.2,\,0.29,\,0.39,\,0.5,\,0.67,\,0.99)$ (in green, red, black, orange, blue, magenta, and purple). We have considered the scenario where the $\alpha-$ attractor model describes the inflationary dynamics.  We have indicated the $1-\sigma$ range of spectral index $n_s$ (as the violet band) associated with the constraints from the Planck \cite{Planck:2018jri}.  Further, the sky blue band corresponds to the dark matter masses lighter than $10$ KeV, indicating the Lyman-$\alpha$ bound \cite{Ballesteros:2020adh,DEramo:2020gpr}. {\bf Lower panel:} We have illustrated the variation of the reheating temperature as a function of the maximum allowed dark matter mass for seven different values of $\omega_\phi$ covering the entire possible range of $\omega_\phi \,(0,\,1)$.  Further, the yellow region shows the allowed parameters space, whereas the light green indicates the forbidden region.
} 
\label{mmax}
\end{center}
\end{figure}
\begin{figure}[t!]
\includegraphics[height=3.35cm,width=5.40cm]{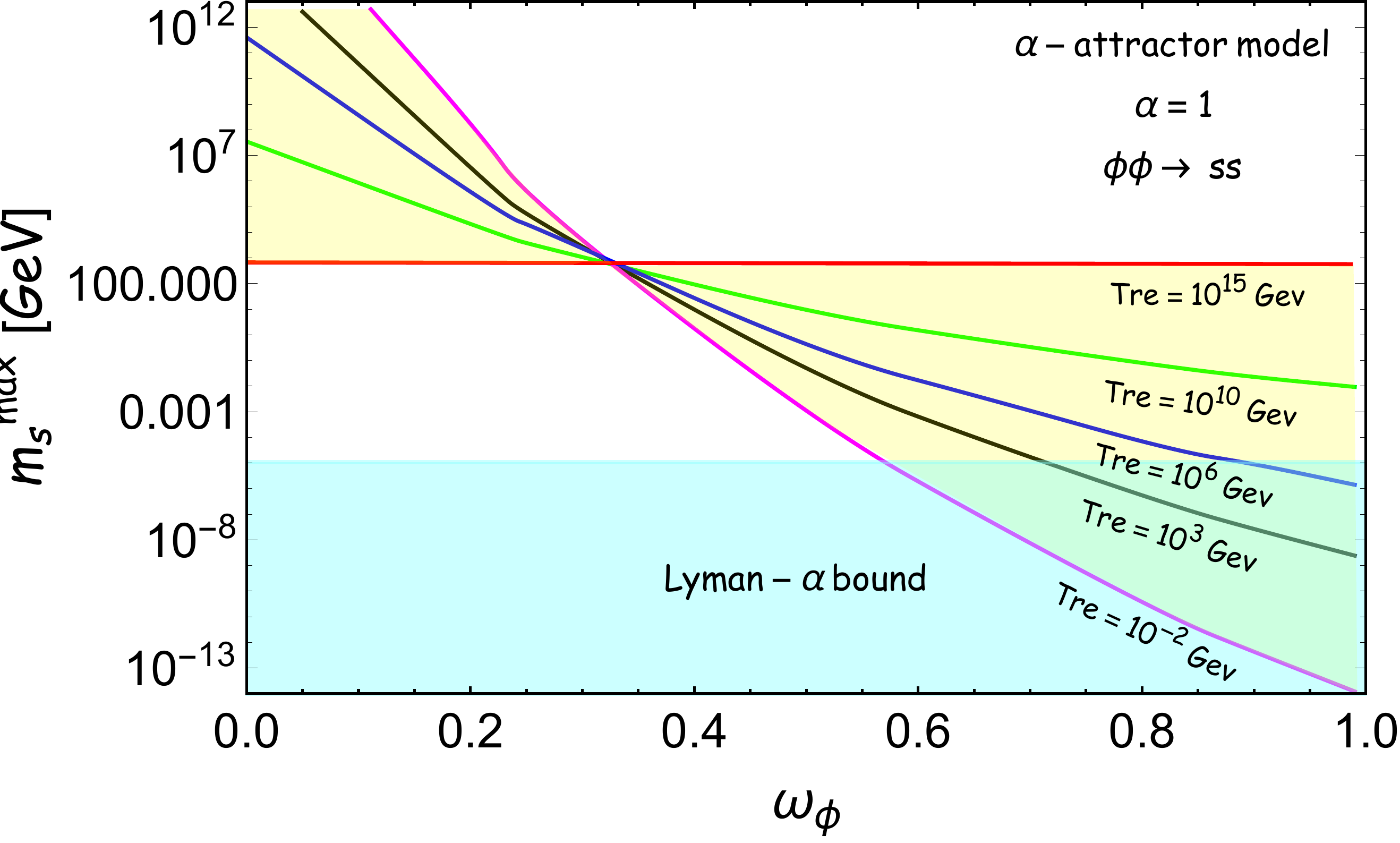}
\includegraphics[height=3.35cm,width=5.40cm]{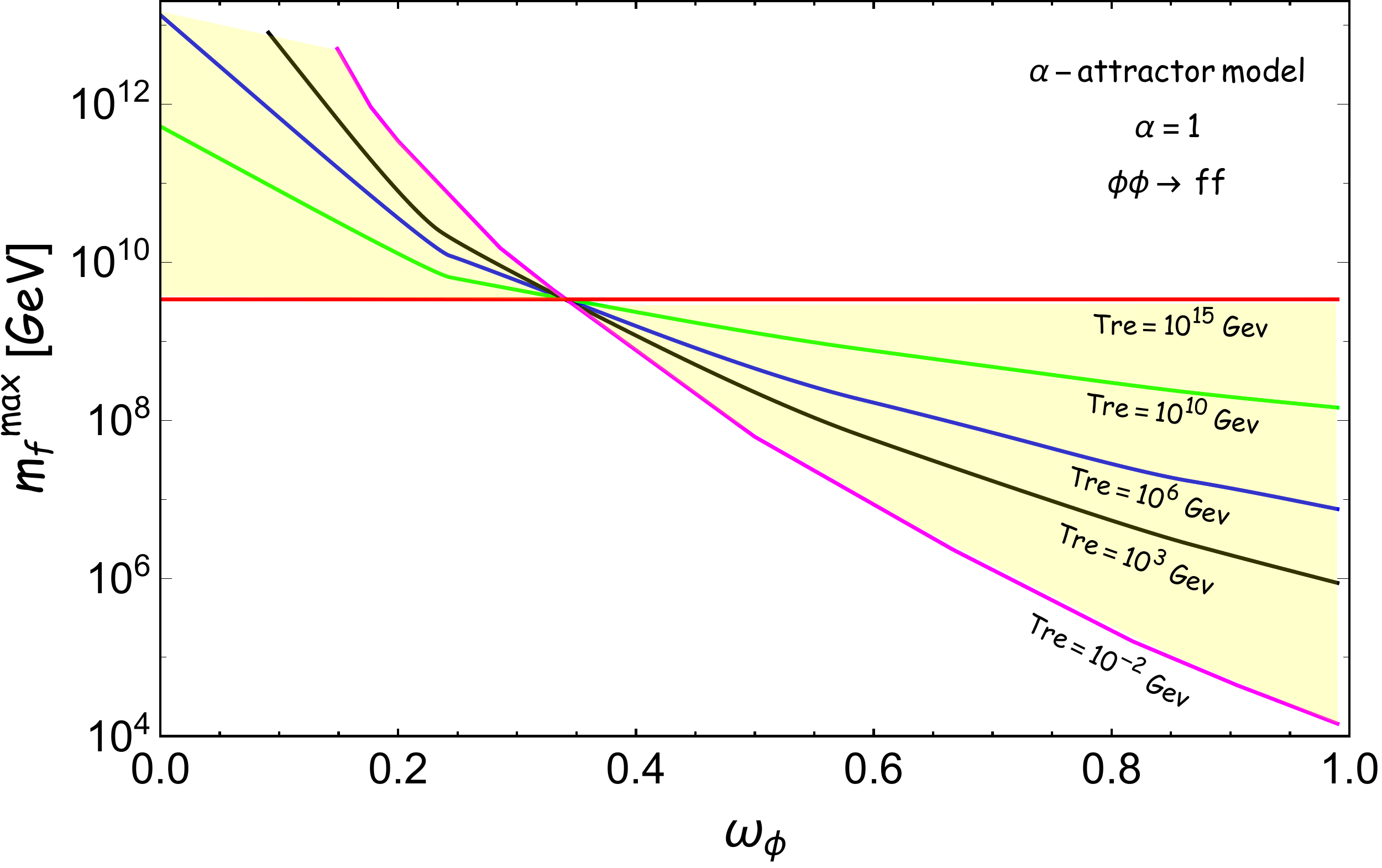}
\includegraphics[height=3.35cm,width=5.40cm]{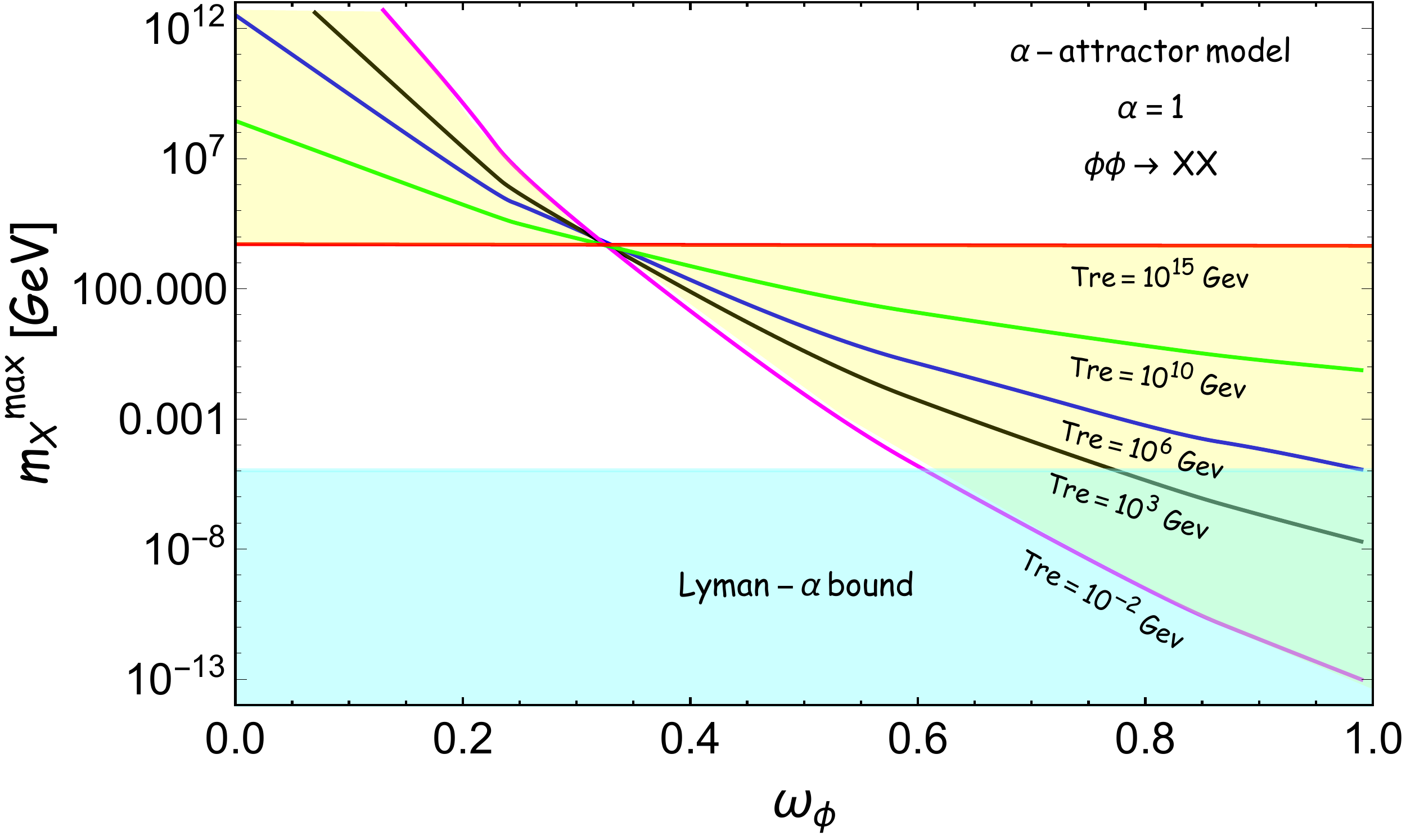}
\caption{The variation of the maximum allowed dark matter mass $m_{Y}^{max}$ over a range of inflaton equation of state $\omega_\phi=(0,\,1)$ for five different values of the reheating temperature $T_{re}=(10^{-2},\,10^3,\,10^6,\,10^{10},\,10^{15})$ GeV (showed in magenta, black, blue, green, and red). The sky blue band indicates restriction from Lyman-$\alpha$ observations, and the yellow shaded region indicates the allowed parameters in $m_{Y}^{max}-\omega_\phi$ plane.} 
\label{mfmaxomega}
\end{figure}
\begin{table}[t!]
	\caption{Different inflaton equation of state, associated bound on scalar spectral index ($n_s$), reheating temperature $T_{re}$ (measured in units of GeV) and dark matter mass $m_{Y}$ (measured in units of GeV), considering purely gravitational production of dark matter.\\}
  \begin{tabular}{|p{1.75cm}|p{1.5cm}|p{1.5cm}|p{1.5cm}|p{1.5cm}|p{1.5cm}|p{1.5cm}|p{1.5cm}|p{1.5cm}|p{1.5cm}}
\hline
Parameters &\multicolumn{3}{c|}{$\omega_\phi=0$} &\multicolumn{3}{c|}{$\omega_\phi=0.5$}&\multicolumn{3}{c|}{$\omega_\phi=0.99$}\\
\cline{2-10}
~& \multicolumn{1}{c|}{$\phi\phi\to SS$} & \multicolumn{1}{c|}{$\phi\phi\to ff$} & \multicolumn{1}{c|}{$\phi\phi\to XX$} &  \multicolumn{1}{c|}{ $\phi\phi\to SS$}&\multicolumn{1}{c|}{$\phi\phi\to ff$}& \multicolumn{1}{c|}{$\phi\phi\to XX$} & \multicolumn{1}{c|}{$\phi\phi\to SS$}&\multicolumn{1}{c|}{$\phi\phi\to ff$}& \multicolumn{1}{c|}{$\phi\phi\to XX$}\\
\hline
\quad$n_s^{min}$& \multicolumn{1}{c|}{0.9596}& \multicolumn{1}{c|}{0.9604}& \multicolumn{1}{c|}{0.9601}& \multicolumn{1}{c|}{0.9648}& \multicolumn{1}{c|}{0.9648}& \multicolumn{1}{c|}{0.9648}&  \multicolumn{1}{c|}{0.9645}& \multicolumn{1}{c|}{0.9645}& \multicolumn{1}{c|}{0.9645}\\
\hline
\quad$n_s^{max}$& \multicolumn{1}{c|}{0.9656}& \multicolumn{1}{c|}{0.9656}& \multicolumn{1}{c|}{0.9656}& \multicolumn{1}{c|}{0.9672}& \multicolumn{1}{c|}{0.9672}& \multicolumn{1}{c|}{0.9672}& \multicolumn{1}{c|}{0.9676}& \multicolumn{1}{c|}{0.9700}& \multicolumn{1}{c|}{0.9680}\\
\hline
\quad $T_{re}^{min}$& \multicolumn{1}{c|}{$1.8\times 10^4$}& \multicolumn{1}{c|}{$3.5\times 10^5$}& \multicolumn{1}{c|}{$1.1\times 10^5$}& \multicolumn{1}{c|}{$10^{-2}$}& \multicolumn{1}{c|}{$10^{-2}$}& \multicolumn{1}{c|}{$10^{-2}$}& \multicolumn{1}{c|}{$1.4 \times 10^7$}& \multicolumn{1}{c|}{$10^{-2}$}& \multicolumn{1}{c|}{$1.5 \times 10^6$}\\
\hline
\quad $T_{re}^{max}$& \multicolumn{1}{c|}{$10^{15}$}& \multicolumn{1}{c|}{$10^{15}$}& \multicolumn{1}{c|}{$10^{15}$}& \multicolumn{1}{c|}{$10^{15}$}& \multicolumn{1}{c|}{$10^{15}$}& \multicolumn{1}{c|}{$10^{15}$}& \multicolumn{1}{c|}{$10^{15}$}& \multicolumn{1}{c|}{$10^{15}$}& \multicolumn{1}{c|}{$10^{15}$}\\
\hline
$m_{Y}^{max}$(min)& \multicolumn{1}{c|}{$960$}& \multicolumn{1}{c|}{$8.0\times 10^{9}$}& \multicolumn{1}{c|}{$7.7\times 10^{3}$}& \multicolumn{1}{c|}{$1.1\times10^{-3}$}& \multicolumn{1}{c|}{$6.1\times 10^{7}$}& \multicolumn{1}{c|}{$9.0\times 10^{-3}$}&  \multicolumn{1}{c|}{$10^{-5}$}& \multicolumn{1}{c|}{$1.4\times10^{4}$}&  \multicolumn{1}{c|}{$10^{-5}$}\\
\hline
$m_{Y}^{max}$(max)& \multicolumn{1}{c|}{$m_\phi^{end}$}& \multicolumn{1}{c|}{$m_\phi^{end}$}& \multicolumn{1}{c|}{$m_\phi^{end}$}& \multicolumn{1}{c|}{$600$}& \multicolumn{1}{c|}{$5.0\times 10^{9}$}& \multicolumn{1}{c|}{$5.0\times 10^{3}$}& \multicolumn{1}{c|}{$640$}& \multicolumn{1}{c|}{$6.0\times10^{9}$}& \multicolumn{1}{c|}{$7.0\times10^{3}$}\\
\hline

  \end{tabular}\\[.2cm]
  
  \label{maxnstre1}
\end{table}
\begin{table}[t!]
	\caption{Different reheating temperatures, associated bound on inflaton equation state $\omega_\phi$, and dark matter mass $m_{Y}$ (measured in units of GeV), considering only gravitationally produced dark matter.\\}
  \begin{tabular}{|p{2cm}|p{1.4cm}|p{1.4cm}|p{1.4cm}|p{1.4cm}|p{1.4cm}|p{1.4cm}|p{1.4cm}|p{1.4cm}|p{1.4cm}|}
\hline
~Parameters &\multicolumn{3}{c|}{$T_{re}=10^{-2}$ GeV} &\multicolumn{3}{c|}{$T_{re}=10^{3}$ GeV}&\multicolumn{3}{c|}{$T_{re}=10^{10}$ GeV}\\
\cline{2-10}
~& \multicolumn{1}{c|}{$\phi\phi\to SS$} & \multicolumn{1}{c|}{$\phi\phi\to ff$}& \multicolumn{1}{c|}{$\phi\phi\to XX$} & \multicolumn{1}{c|}{$\phi\phi\to SS$} &  \multicolumn{1}{c|}{ $\phi\phi\to ff$}& \multicolumn{1}{c|}{$\phi\phi\to XX$}&\multicolumn{1}{c|}{$\phi\phi\to SS$}& \multicolumn{1}{c|}{$\phi\phi\to ff$} & \multicolumn{1}{c|}{$\phi\phi\to XX$}\\
\hline
\qquad$\omega_\phi^{min}$& \multicolumn{1}{c|}{0.11}& \multicolumn{1}{c|}{0.15}&
\multicolumn{1}{c|}{0.13}& \multicolumn{1}{c|}{0.05}& \multicolumn{1}{c|}{0.09}& \multicolumn{1}{c|}{0.07}& \multicolumn{1}{c|}{0.0}& \multicolumn{1}{c|}{0.0}& \multicolumn{1}{c|}{0.0}\\
\hline
\qquad$\omega_\phi^{max}$& \multicolumn{1}{c|}{0.56}& \multicolumn{1}{c|}{1.0}&
\multicolumn{1}{c|}{0.60}& \multicolumn{1}{c|}{0.71}& \multicolumn{1}{c|}{1.0}& \multicolumn{1}{c|}{0.77}& \multicolumn{1}{c|}{1.0}& \multicolumn{1}{c|}{1.0}& \multicolumn{1}{c|}{1.0}\\
\hline
~$m_{Y}^{max}$ (min)& \multicolumn{1}{c|}{$10^{-5}$}& \multicolumn{1}{c|}{$1.4\times 10^4$}&\multicolumn{1}{c|}{$10^{-5}$}& \multicolumn{1}{c|}{$10^{-5}$}& \multicolumn{1}{c|}{$8.7\times 10^5$}& \multicolumn{1}{c|}{$10^{-5}$}& \multicolumn{1}{c|}{$9.2\times 10^{-3}$}& \multicolumn{1}{c|}{$1.4\times 10^8$}& \multicolumn{1}{c|}{$7.0\times 10^{-2}$}\\
\hline
~$m_{Y}^{max}$ (max)& \multicolumn{1}{c|}{$m_\phi^{end}$}& \multicolumn{1}{c|}{$m_\phi^{end}$}& \multicolumn{1}{c|}{$m_\phi^{end}$}& \multicolumn{1}{c|}{$m_\phi^{end}$}&\multicolumn{1}{c|}{$m_\phi^{end}$}&\multicolumn{1}{c|}{$m_\phi^{end}$}&  \multicolumn{1}{c|}{$3.5\times 10^{7}$}& \multicolumn{1}{c|}{$5.2\times 10^{11}$}& \multicolumn{1}{c|}{$2.8\times 10^{8}$}\\
\hline
  \end{tabular}\\[.2cm]
  
  \label{maxntre}
\end{table}
\begin{figure}[t!]
\includegraphics[height=8.7cm,width=12.70cm]{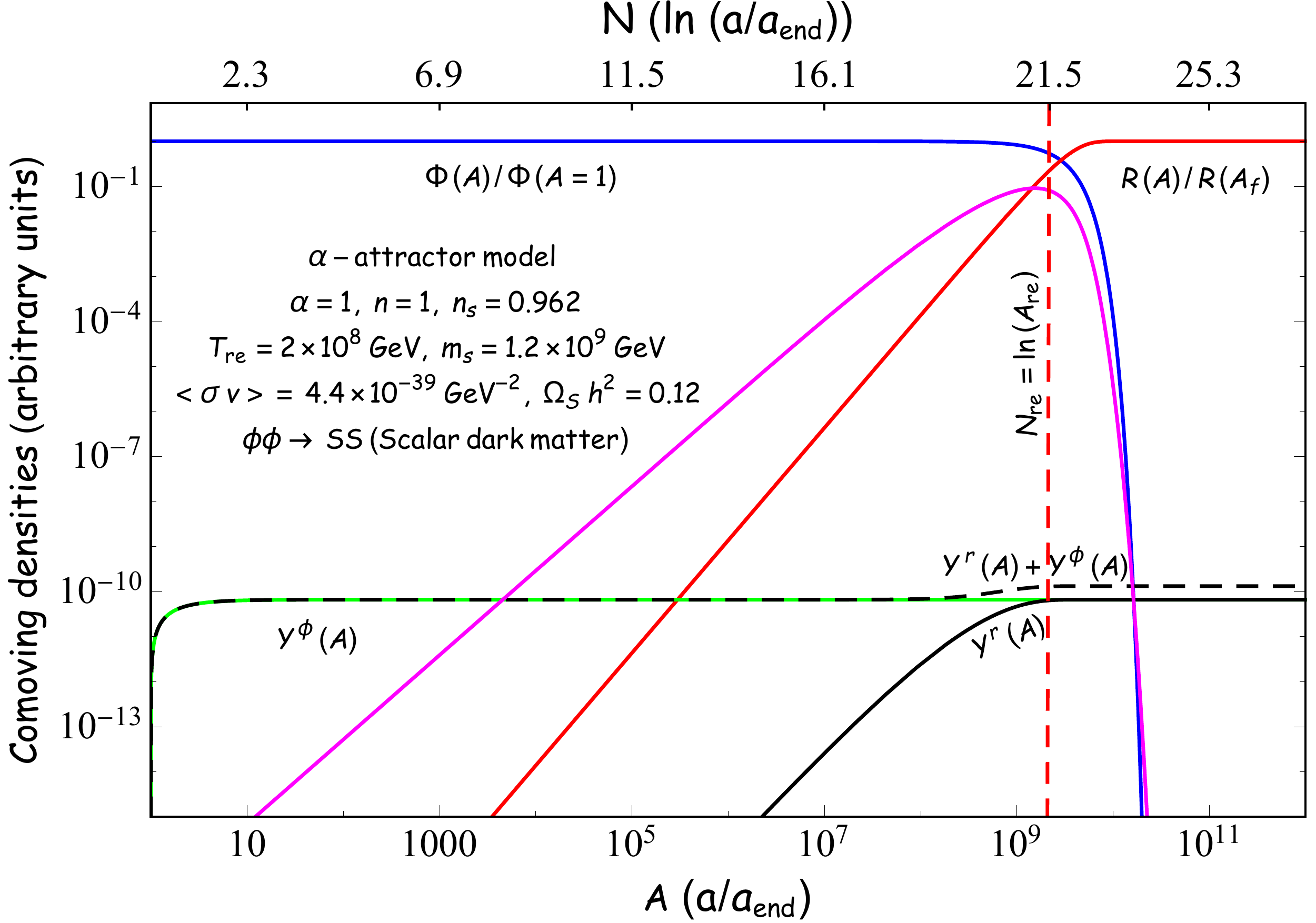}

\caption{We have plotted the evolution of the different energy components (inflaton, radiation) and the number density of dark matter as a function of the normalized scale factor (alternatively, the e-folding number is counting after the inflation) for $\alpha-$ attractor model with $\alpha=1$.  The blue and red curve indicates the variation of the comoving densities of inflaton and radiation, respectively.  Further, the black and green curve shows the evolution of the comoving number densities $(\,Y^r,\,Y^\phi)$ in arbitrary units produced from the radiation bath and the inflaton (mediated by gravity), accordingly.  Moreover, the dotted black curve shows the evolution of the total comoving dark matter number density $(\,Y^r+\,Y^\phi)$, where we are taking into account both possibilities of the dark matter production.} 
\label{density component}
\end{figure}

\begin{figure}[t!]
\includegraphics[height=3.6cm,width=5.40cm]{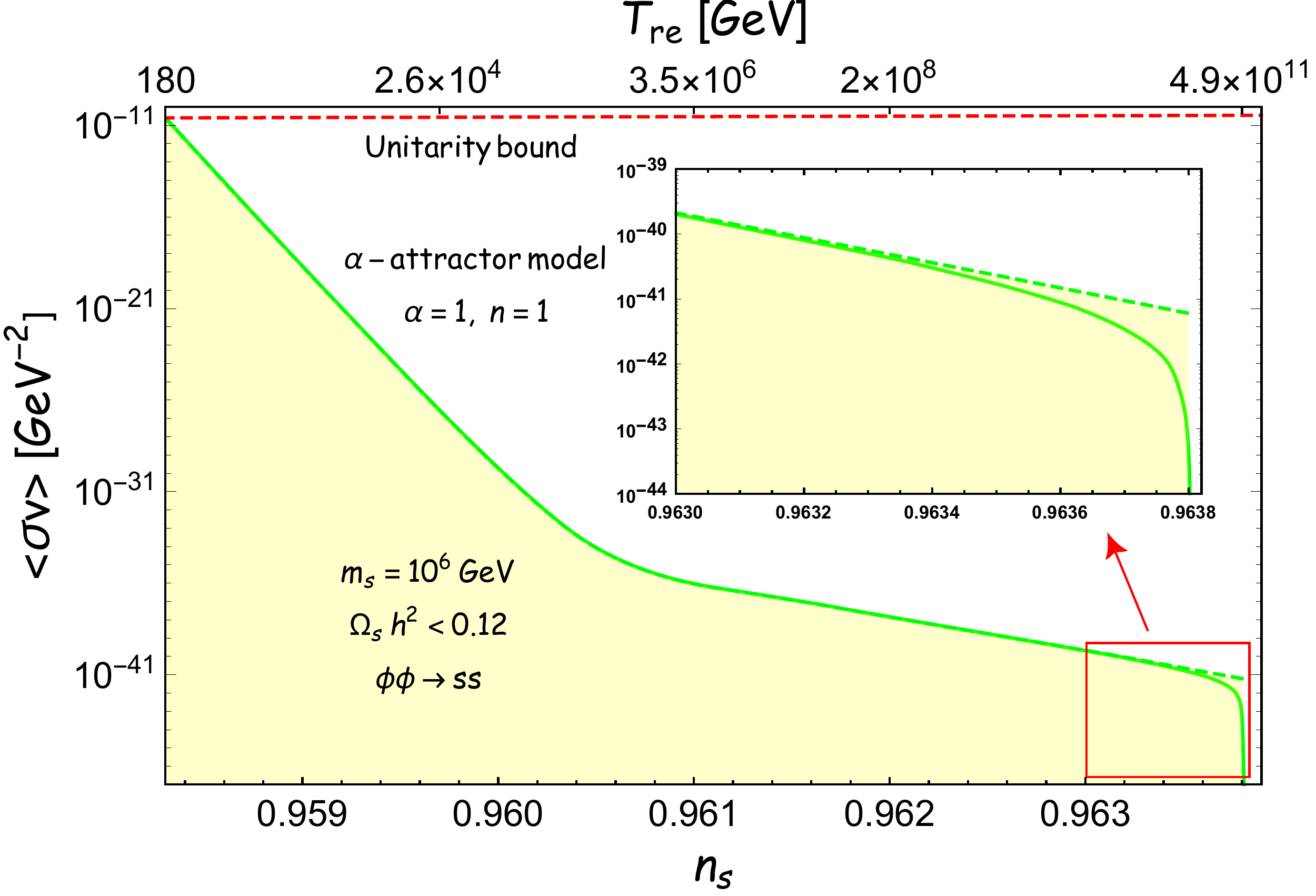}
\includegraphics[height=3.60cm,width=5.40cm]{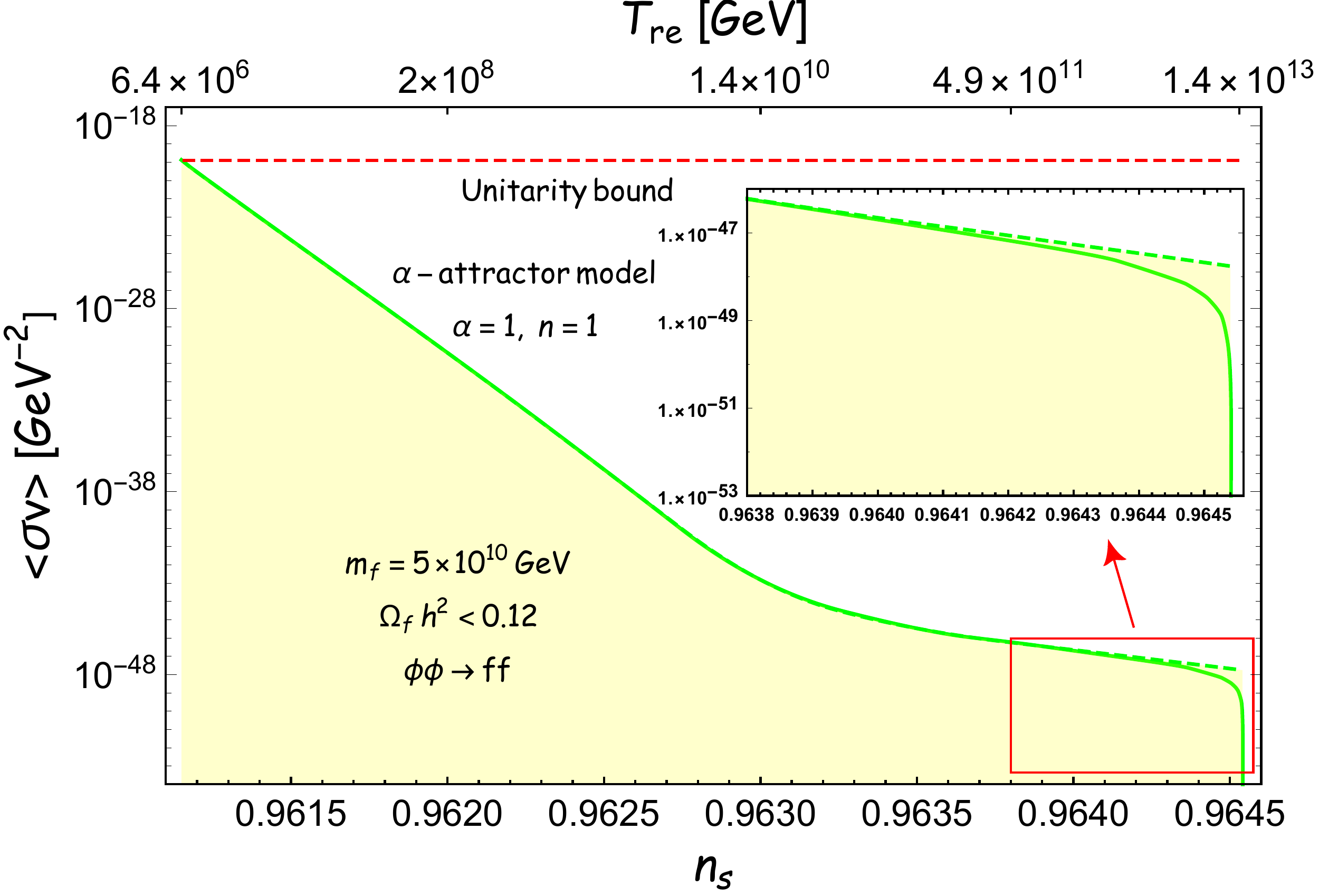}
\includegraphics[height=3.6cm,width=5.40cm]{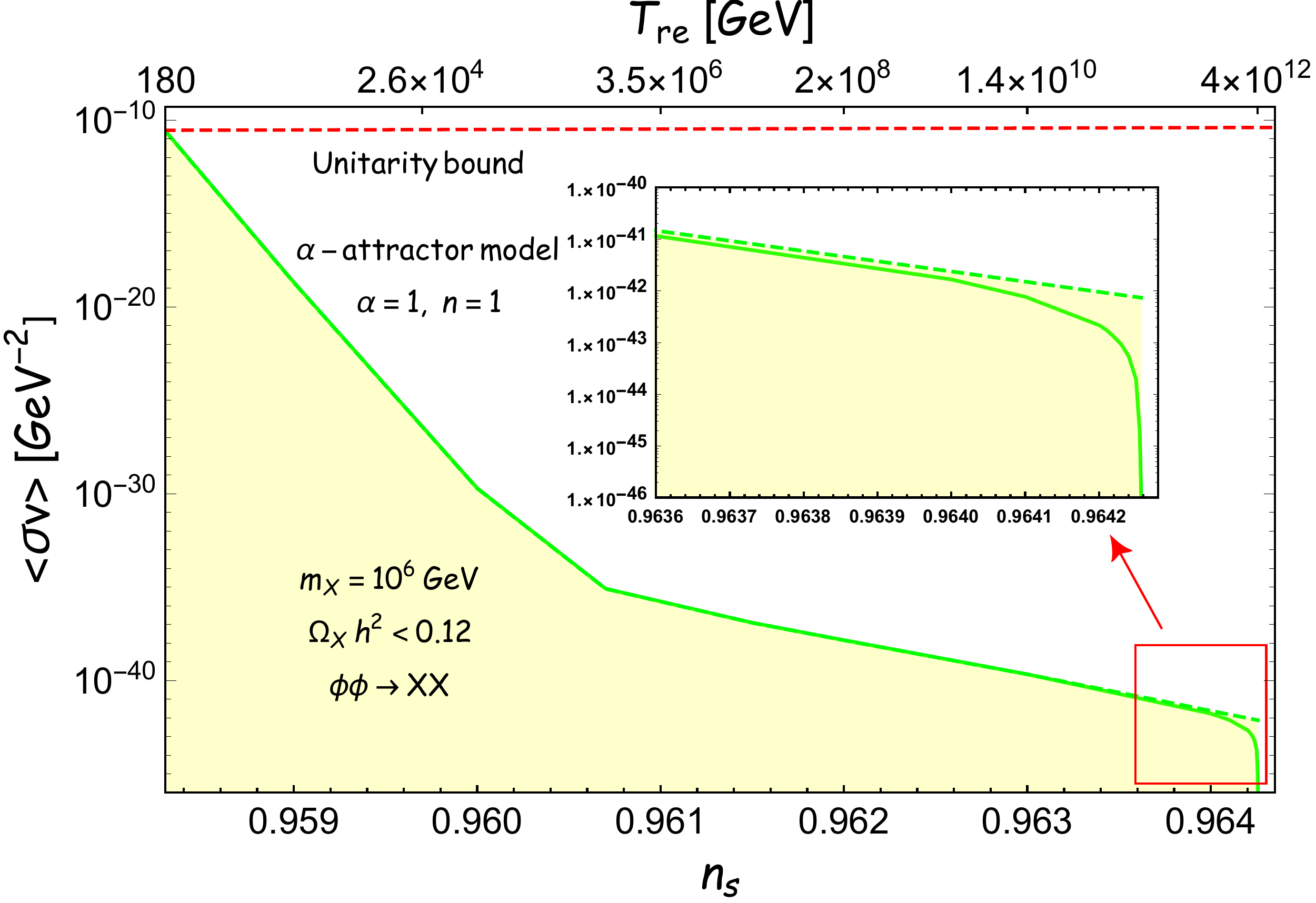}
\includegraphics[height=3.6cm,width=5.40cm]{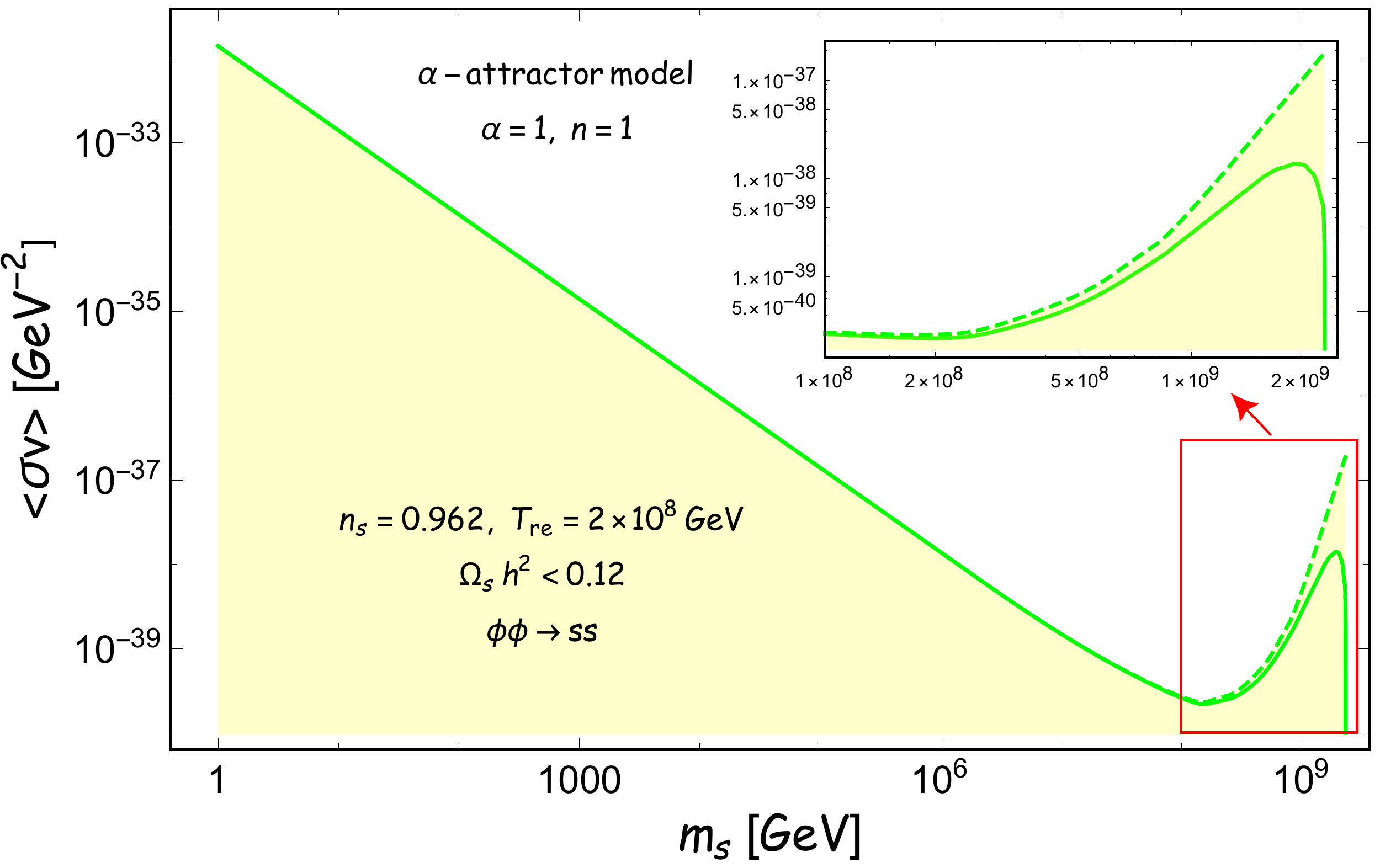}
\includegraphics[height=3.6cm,width=5.40cm]{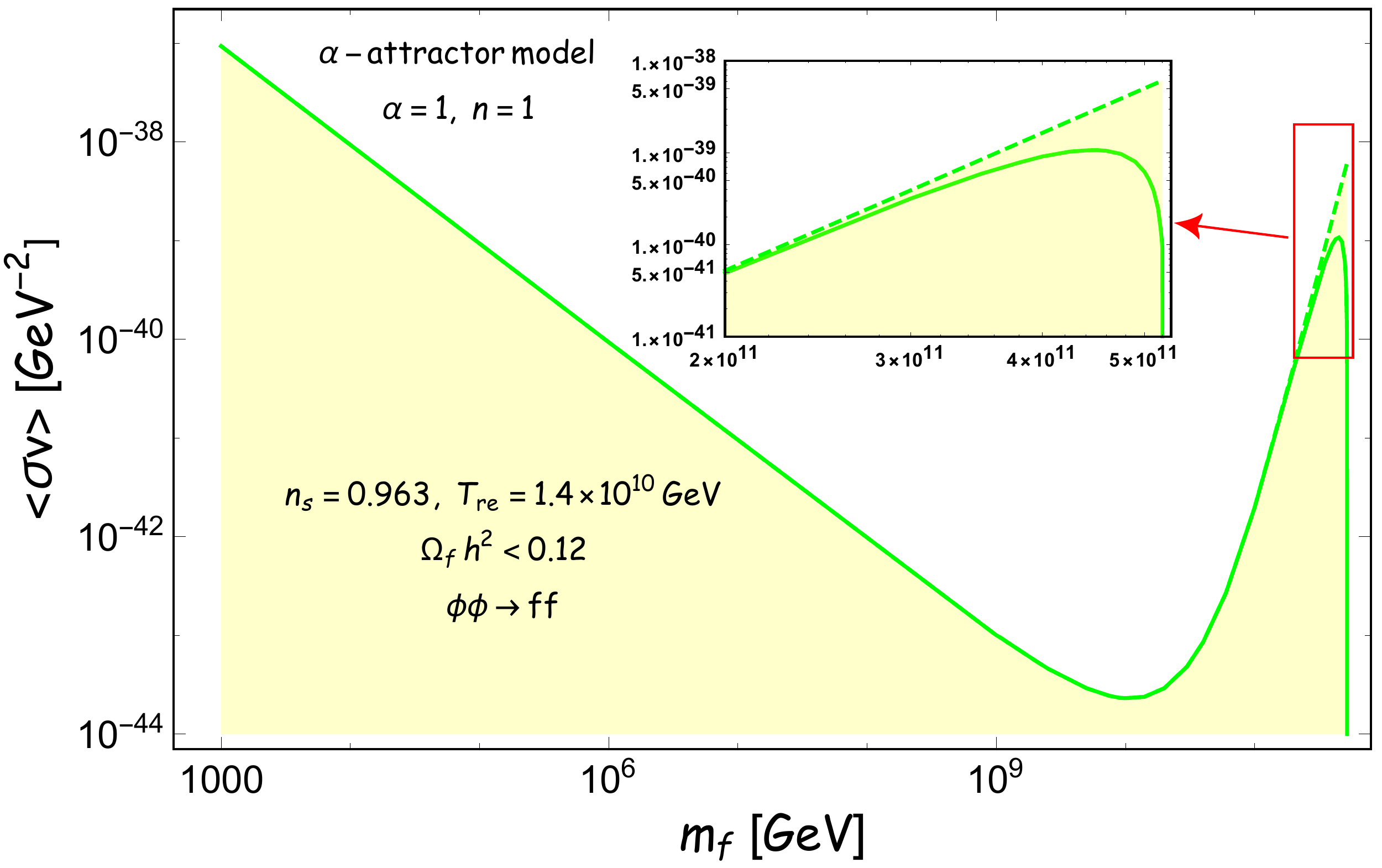}
\includegraphics[height=3.6cm,width=5.40cm]{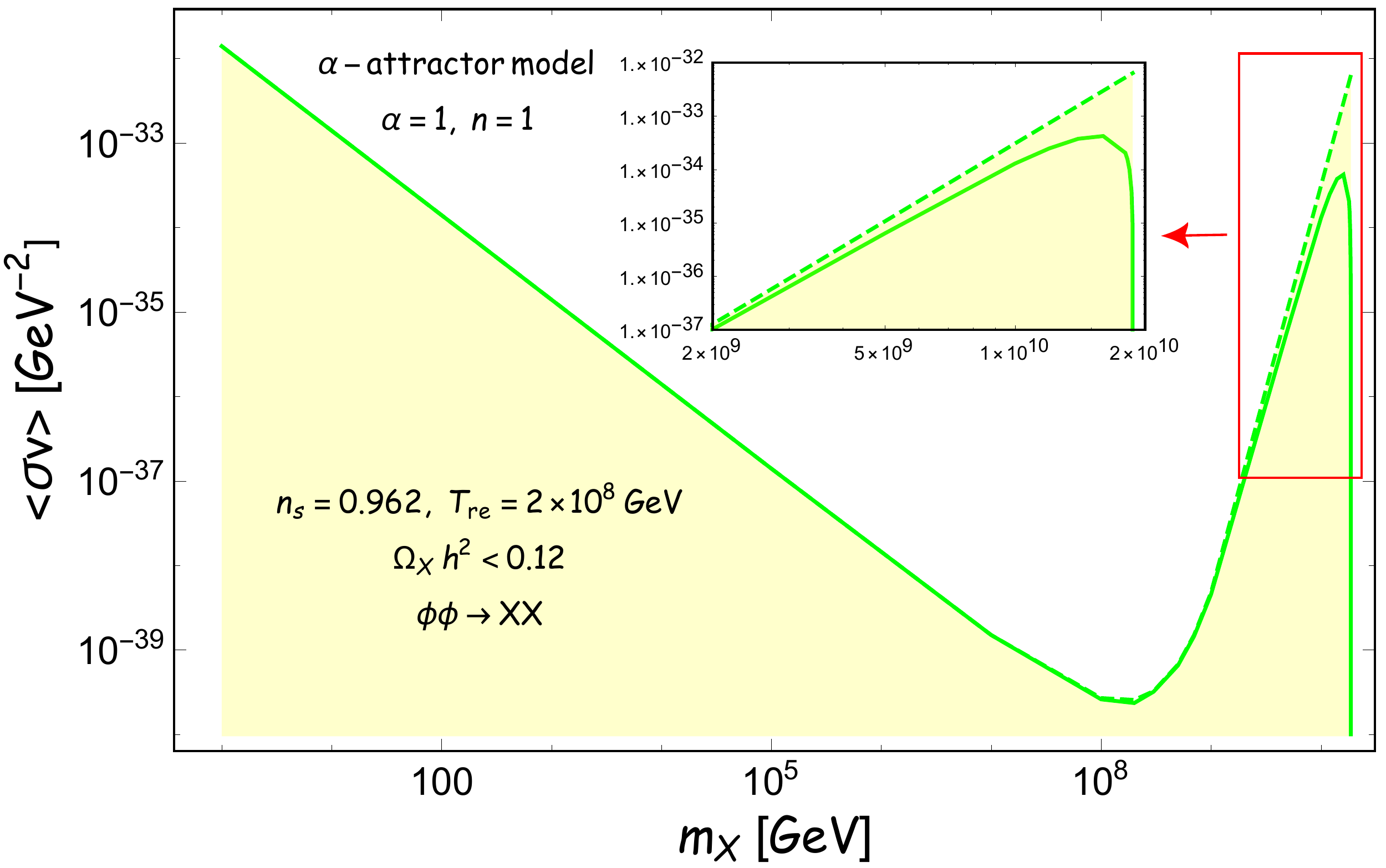}
\caption{{\bf Upper panel :} $\langle\, \sigma\, v\, \rangle$ vs. $n_s$  plotted for three different gravitationally produced dark matter scenarios: $\phi\phi \to SS$ (scalar dark matter), $\phi \phi\to ff$ (fermionic dark matter) and $\phi \phi\to XX$ (vector dark matter) considering $\alpha-$attractor model with $\alpha=1,\,n=1$ (Higgs-Starobinsky model). The yellow shaded region corresponds to the dark matter abundance $\Omega_{Y} h^2\leq 0.12$. The dashed green line implied the results when we took into account one possibility: dark matter production from radiation bath and the solid green line correspond to both possibilities:  dark matter can be produced from the decay of inflaton as well as from the radiation bath. The lower limit on the spectral index is given by the perturbative unitarity limit of cross-section  $\langle \sigma v \rangle_{max}=\frac{8\pi}{m_{Y}^2}$  (shown by the red dashed line). Whereas the upper limit is associated with that particular value of the spectral index or reheating temperature when only gravitational production of the dark matter is sufficient to produce the correct relic of the dark matter. So any value of $n_s$ above this is excluded because this leads to an overabundance. {\bf Lower panel :} Variation of $\langle\,\sigma\,v\,\rangle$ as function of dark matter mass $m_{Y}$. The upper limit on dark matter mass is associated with that particular value of the dark matter mass $m_{Y}^{max}$ when only gravitational production of the dark matter is sufficient to produce the correct relic.} 
\label{alpha1}
\end{figure}
\begin{figure}[t!]
\includegraphics[height=3.6cm,width=5.40cm]{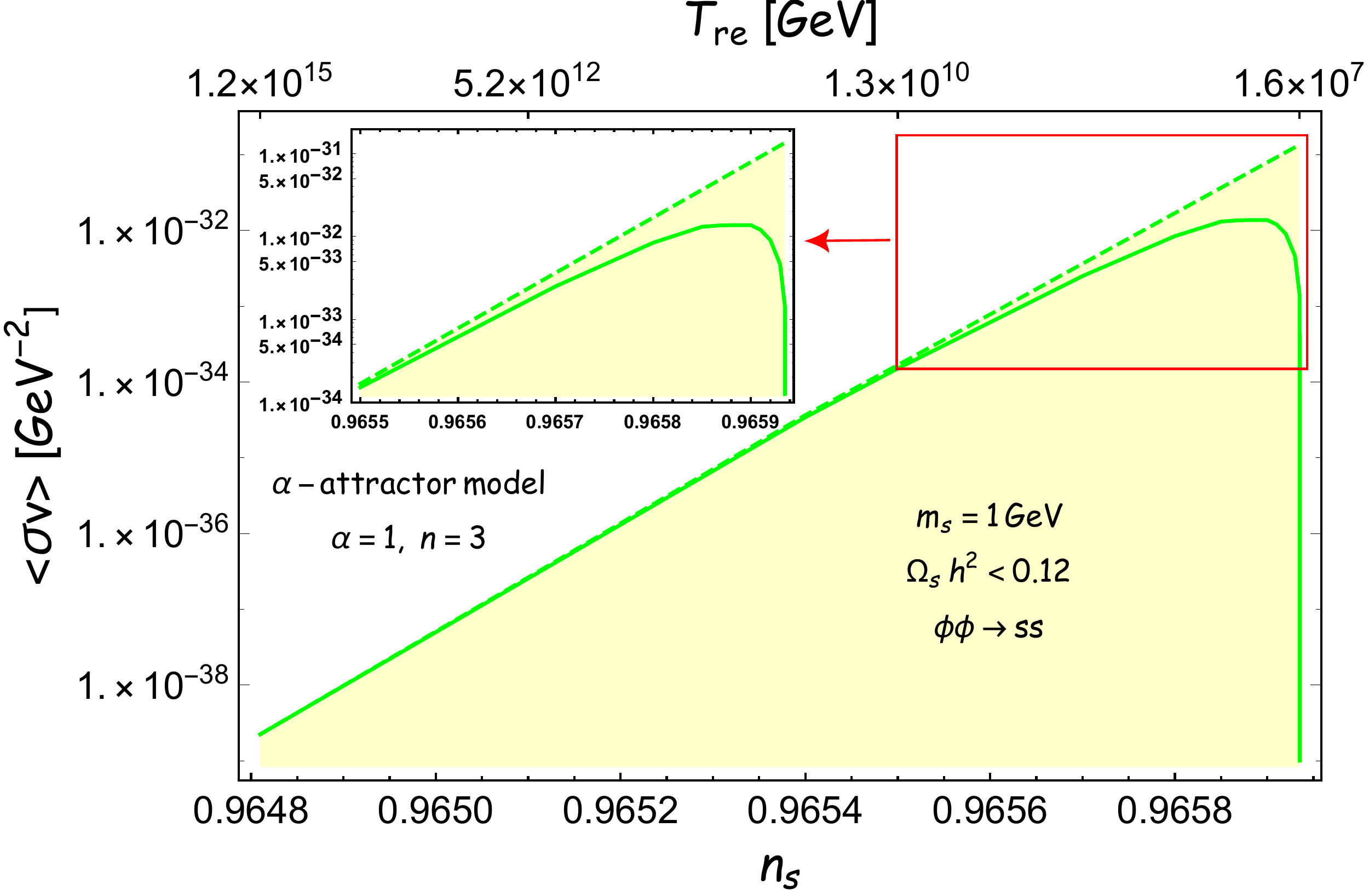}
\includegraphics[height=3.6cm,width=5.40cm]{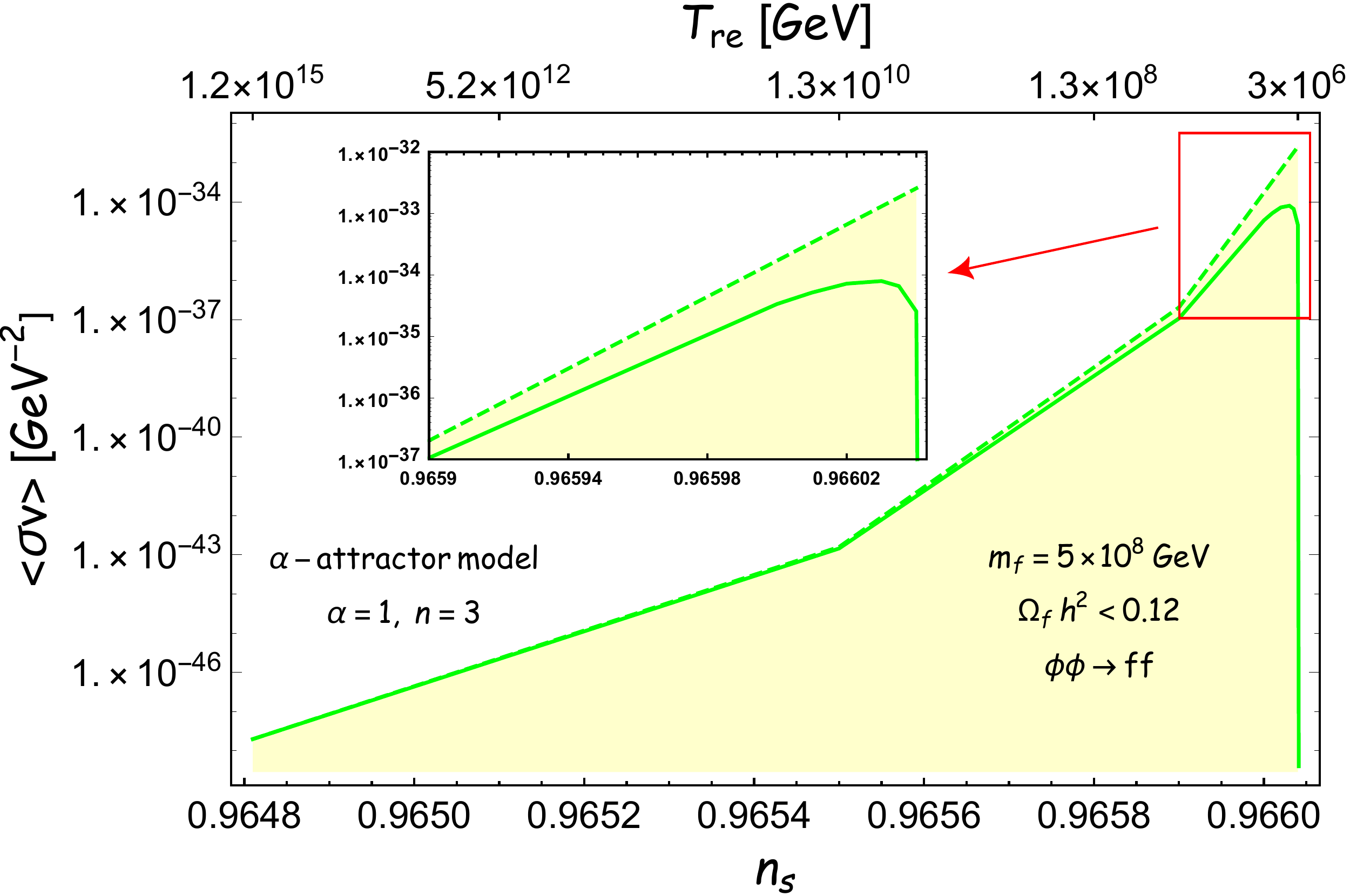}
\includegraphics[height=3.6cm,width=5.40cm]{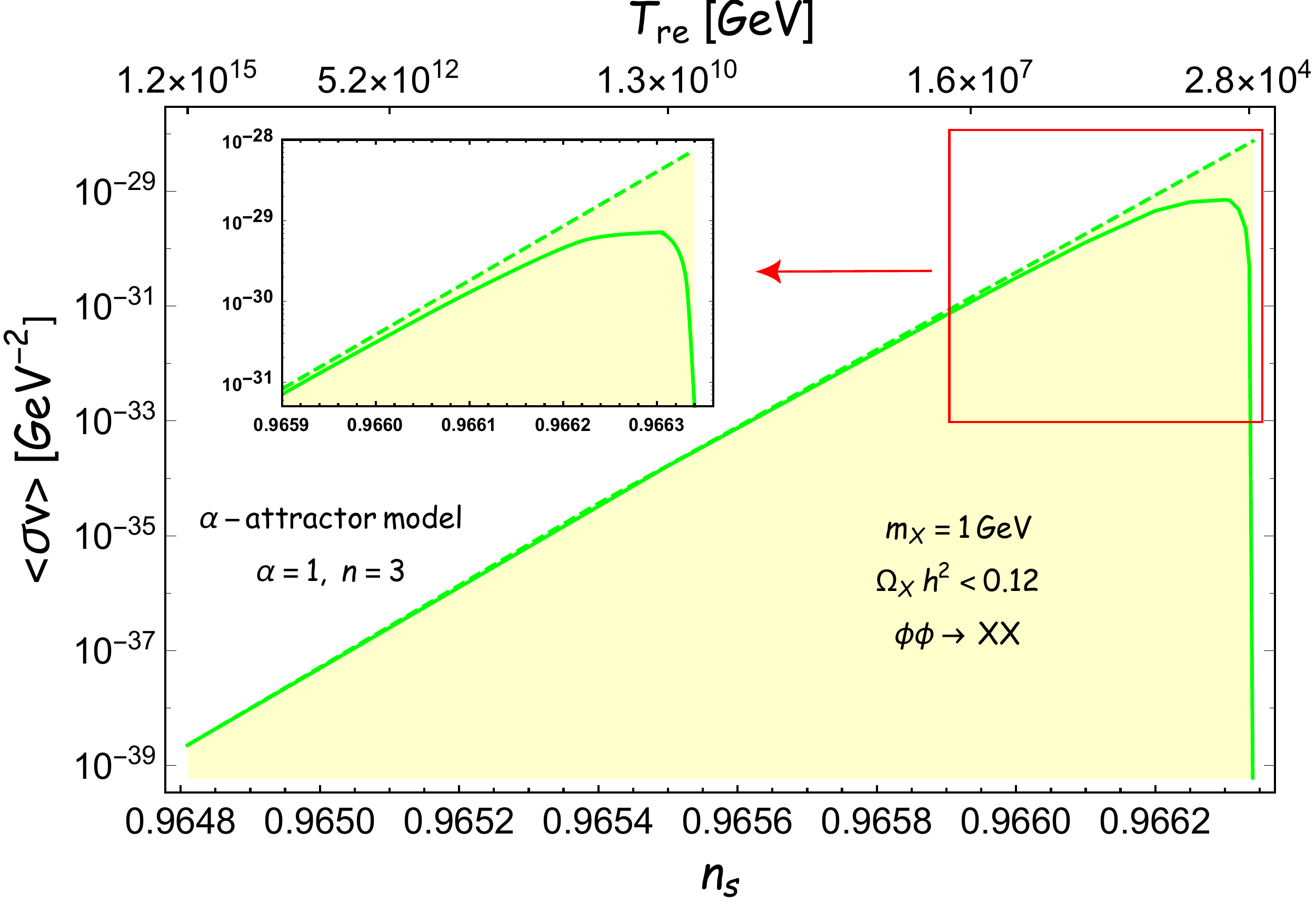}
\includegraphics[height=3.6cm,width=5.40cm]{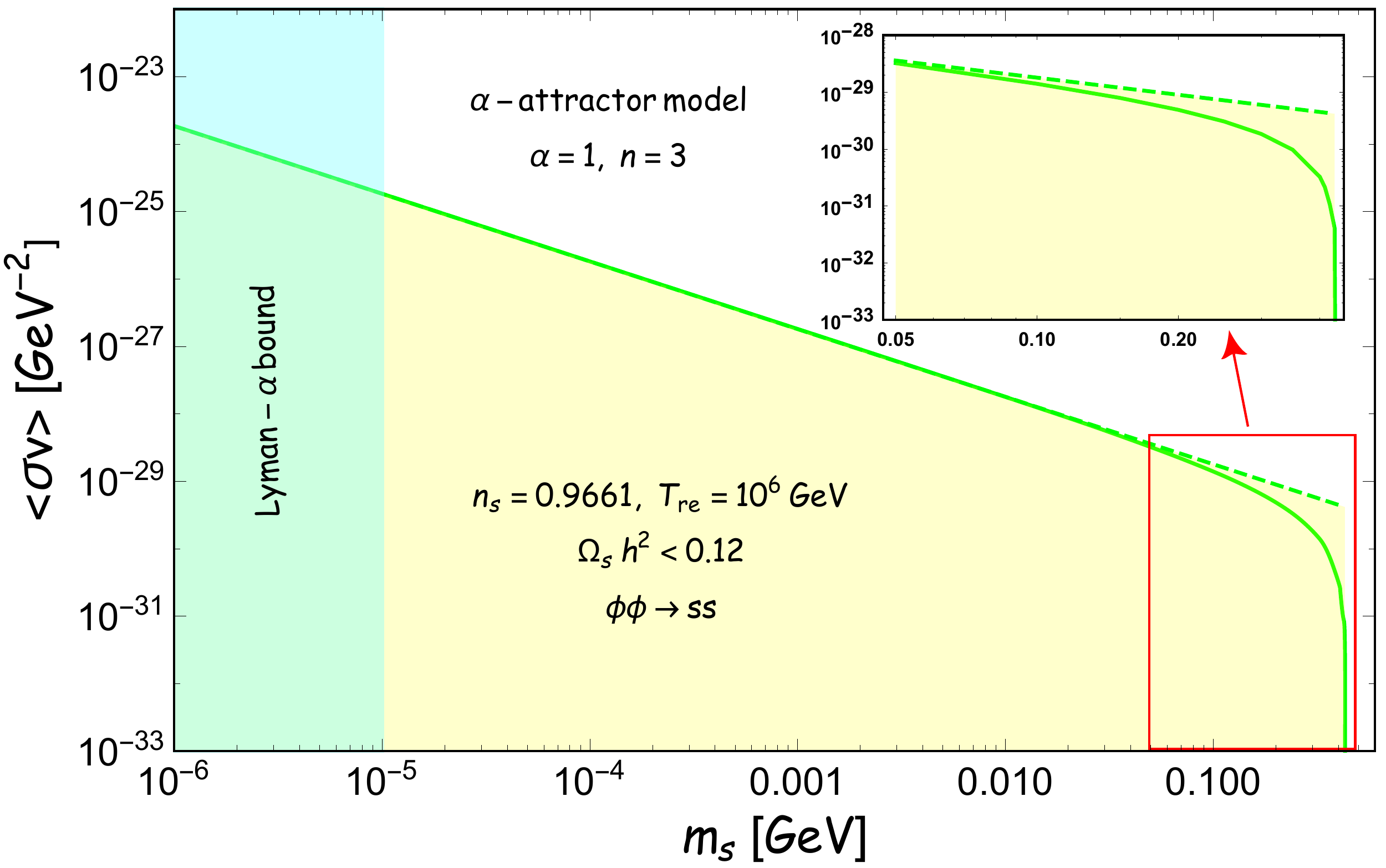}
\includegraphics[height=3.6cm,width=5.40cm]{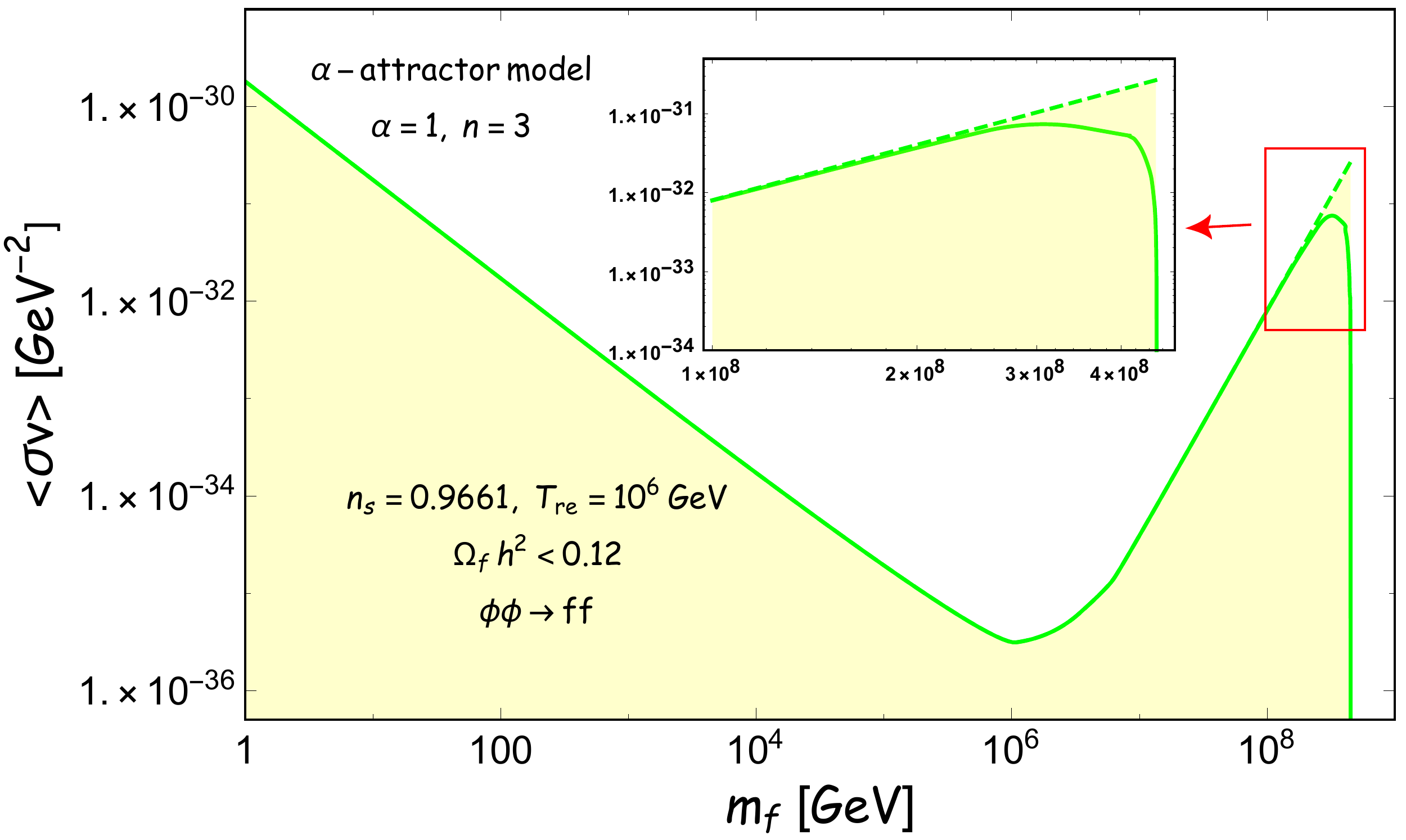}
\includegraphics[height=3.6cm,width=5.40cm]{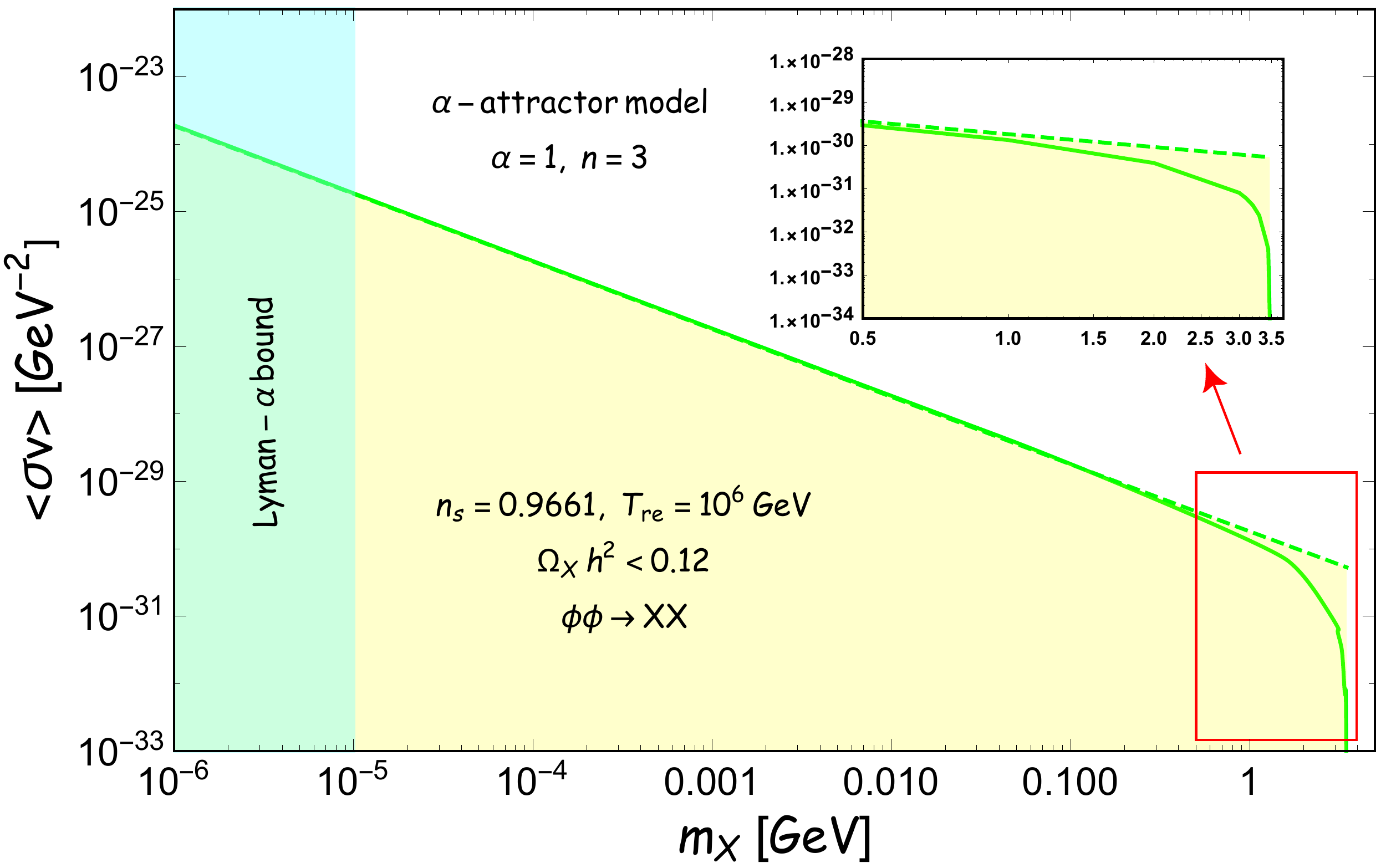}

\caption{{\bf Upper panel :} $\langle\, \sigma\, v\, \rangle$ vs. $n_s$, the description of this figure is the same as previous fig.\ref{alpha1}; the main difference is that here we have plotted for $\alpha-$attractor model with $\alpha=1$ and $n=3$. In addition to that, the lower limit on the spectral index corresponds to instantaneous reheating ($N_{re}\to 0$). {\bf Lower panel :} $\langle\, \sigma\, v\, \rangle$ vs. $m_{Y}$, the description of this figure is the same as previous fig.\ref{alpha1}. The sky blue band indicates restriction from Lyman-$\alpha$ observations.}
\label{scalarcrosssection3}
\end{figure}
As already emphasized in the beginning, the production of gravitational dark matter is an interesting, physically motivated scenario that needs detailed exploration. 
Following the references on gravitational dark matter \cite{Mambrini:2021zpp,Barman:2021ugy,Donoghue:1994dn,Choi:1994ax,Holstein:2006bh}, in this paper we explore the observationally viable dark matter scenarios in terms of different inflationary models. Important reheating parameters such as the equation of state $\omega_\phi$ associated with the inflaton potential and reheating temperature $T_{re}$ will play an important role in constraining the parameters such as the maximum possible mass of the dark matter.  Moreover, since we consider the dark matter production from the radiation bath, we also place constraints on the average cross-section times velocity $\langle\,\sigma\,v\,\rangle$.  The dark sector may have different possibilities in terms of the nature of the dark matter and the number of components.
\subsection{Single component dark matter}
In our present framework, we have two different underlying production mechanisms. To understand the construction from each, one
examines the evolution of different density components during reheating, as shown in Fig.\ref{density component}. 
The production of dark matter components due to gravity mediation will naturally occur at the very beginning of the reheating phase when the inflation energy density is maximum, and this is depicted by the green curve. On the other hand, the dark matter production from the radiation bath will follow the evolution of radiation itself which is depicted by a solid red curve. Therefore, maximum production will naturally occur near the end of reheating, as depicted by the solid black line. Finally, combining both the gravitational production and production from radiation bath will contribute to the current dark matter abundance. An interesting aspect of such products of the same type of dark matter from two different mechanisms is that it will lead to a mixture of components with different velocity distribution, whose density perturbation may grow differently and provide distinct signatures in the small scale structure, which will be studied in our future publication. 

Anyway, for the case of a purely gravity-mediated scenario, the mass of the dark matter is the only parameter. Therefore, in this scenario reheating dynamics is controlled by two parameters $(\Gamma_{\phi}, m_Y)$ and two constraints relations Eq.\ref{eqtre}, \ref{darkmatter relic}. Hence, the dynamics are determined completely by the inflation model under consideration instead of the non-gravitational dark matter production scenario, which contains annihilation cross-section as an additional parameter. A large class of inflationary models such as $\alpha$-attractor endows with a degenerate prediction of large scale observables, namely, scalar spectral index $(n_s)$ and tensor to scalar ratio ($r$) but with distinguishing properties in terms of their effective inflaton equation of state $\omega_{\phi}$ during reheating. Such degeneracy can be lifted during reheating, considering various other observables. For example, primordial gravitational waves encode distinct signatures depending on the reheating equation of state \cite{Haque:2021dha,Mishra:2021wkm}. In our present analysis also for a given equation of state $\omega_{\phi}$, solely gravitationally produced dark matter 
will assume distinct value of its mass $m_Y^{max}$ in compatible with dark matter abundance as can be seen in the Fig.\ref{mmax}, \ref{mfmaxomega} and the shaded yellow regions are the only allowed parameter plane which are either bounded by the value of $\omega_{\phi} \sim (0,1)$ or by the minimum $T_{re}^{min}\simeq 10^{-2} \mbox{GeV}$ set by the BBN and maximum $T_{re}^{max} \simeq 10^{15}$ GeV set by the instantaneous reheating. Therefore, simple dark matter mass produced gravitationally during reheating can give valuable information about inflaton potential. An important point we should remember is that the condition $H = \Gamma_{\phi}$ leads to unique reheating temperature $T_{re}$ for a given $n_s$ and this is precisely the reason the present dark matter abundance is satisfied for a fixed dark matter mass. However, the suffix "max" in $m_Y^{max}$ is due to the reason that this is the maximum possible dark matter mass in $(\langle\sigma v\rangle~\mbox{Vs}~m_Y)$ plane satisfies the abundance $\Omega_Y h^2 = 0.12$, when finite dark matter coupling with the radiation bath is included in the process; and it is in the limit $\langle \sigma v\rangle \rightarrow 0$, when $m_Y \rightarrow m_Y^{max}$ as shown in Figs. \ref{alpha1}, \ref{scalarcrosssection3}. The generic expression of $m_Y^{max}$ is assumed as,  
\bea \label{maxmass}
m_{Y}^{max}=\frac{\mathcal{G}\,\beta\,T_{now}}{ n_{Y}^{re}\,A_{re}^3} \left(\frac{\Omega_{Y}\,h^2}{\Omega_r\,h^2}\right)_{now},~\mbox{where} ~~A_{re}=\left(\,\frac{12\,M_p^2\,H_{end}^2\,(\,1+\,\omega_\phi\,)^2}{\mathcal{G}^4\,\beta\,(\,5-\,3\,\omega_\phi\,)^2}\,\right)^{\frac{-1}{(1-\,3\,\omega_\phi)}}.
\eea
The number density $n_Y^{re}$ is calculated at the end of reheating with normalized scale factor $A_{re}$ and associated expressions for each component are 
\bea  \label{mymax}
&&n^{re}_{X}\approx \,\frac{3}{4096\,\pi}\,\frac{(\,1+\,\omega_\phi\,)}{(\,1+\,3\,\omega_\phi\,)}\left(\frac{H_{end}}{A_{re}}\right)^3~~,\\
&&n^{re}_{s} \approx \frac{3}{512\,\pi}\,\frac{(\,1+\,\omega_\phi\,)}{(\,1+\,3\,\omega_\phi\,)}\left(\frac{H_{end}}{A_{re}}\right)^3,\nno\\
&&n^{re}_{f}\approx\frac{m_f^2\,\lambda^{\frac{\omega_\phi-1}{\omega_\phi+1}}\,\nu(\omega_\phi)}{4096\,\pi\left(1+3\,\omega_\phi\right)\left(H_{end}^2M_p^2\right)^{\frac{2\omega_\phi}{1+\omega_\phi}}}\,\left(\frac{H_{end}}{A_{re}}\right)^3=\frac{3}{2048\,\pi}\,\frac{1+\omega_\phi}{1-\omega_\phi}\,\left(\frac{m_f}{m_\phi^{end}}\right)^2\,\left(\frac{H_{end}}{A_{re}}\right)^3\,.\nno
\eea
We should emphasize at this point that the above expression for the dark matter mass is sensitive to $(H_{end}, \omega_{\phi})$.  
Detailed derivation and the associated symbols of the above expressions are given in the appendix \ref{maxdark matter}.
Any value of the dark matter mass above $m_{Y}^{max}$ is excluded because of overabundance. 
In fig.\ref{mmax}, we have shown the allowed dark matter masses $m_Y^{max}$ as a function of the spectral index and reheating temperature  for different inflaton equations of state $\omega_\phi=(\,0,\,0.2,\,0.29,\,0.39,\,0.50,\,0.67,\,0.99\,)$ and assumed different single component dark matter species namely, scalar, fermion and vector. Therefore, we cover the whole possible range of inflaton equation of state 
$\omega_\phi=(\,0,1\,)$ and the allowed parameter space is shown by the shaded yellow region in the $(T_{re}-m_{Y}^{max})$ plane. It suggests that for the entire range of inflaton equation of state between $(0,1)$, the allowed mass of the scalar and vector dark matter must lie between ($10^{-8}$,\,$10^{13}$) GeV. And for the fermionic dark matter, the possible range turns out to be $(10^4,\,10^{13})$ GeV. Here, one should notice the distinct mass range allowed for the dark matter for boson and fermion. Bosonic dark matter mass can be as low as in the eV range, which can be identified as an axion-like particle
. It would be interesting to study in detail along this direction. Anyway, as has already been pointed out, there is one to one correspondence between the dark matter mass and the reheating temperature, we provide possible constraints on the value of (\,$n_s,\,T_{re},\,m_{Y}^{max}$\,) in terms of different inflaton equations of states in Table-\ref{maxnstre1}. To determine the possible bound on the minimum value of the dark matter mass, we use the additional constraints arising from the Lyman-$\alpha$ forest data set, which in turn impose further restrictions on the inflationary and reheating parameters ($n_s,\, T_{re}$). Additionally, in Fig.\ref{mfmaxomega}, we have shown the allowed dark matter mass as a function of the inflaton equation of state for different sets of reheating temperature. 
Interestingly, depending upon the inflaton equation of state, the allowed DM mass range changes, and it shrinks to a point as $\omega_{\phi}$ approaches towards 1/3. The analytic expression of that specific mass (cf. Eq.\ref{maxmass} and \ref{mymax}) turned out to be dependent on the two factors, inflation energy scale $H_{end}$ and inflaton equation of state $\omega_\phi$. Moreover, the possible bound on the inflaton equation of state and the mass of the dark matter for different sample values of reheating temperatures are shown in Table-\ref{maxntre}. 

When dark matter production from radiation bath is included in the reheating process, Figs.\ref{alpha1}, \ref{scalarcrosssection3} depict the region of allowed cross-section $\langle \sigma v \rangle $ in terms of $n_s$ for two distinct values of the inflaton equation of state $\omega_\phi=(\,0,\,0.5\,)$ respectively. It is clear from the figures that for finite cross-sections with $m_Y < m_Y^{max}$, the production from radiation bath is always dominating compared to that of gravitational production. However, as one approach towards $m_Y^{max}$, gravity mediated dark matter production is increasingly dominated considering the fixed value of $\Omega_{Y}\,h^2\simeq 0.12$. This fact entails the value of $\langle \sigma v\rangle$ approaching towards zero for not to overproduce the dark matter. Another important point is to note that for $m_Y>T_{re}$, there always exists a maximum cross-section for a given temperature once we fixed $\omega_\phi$. In the in-set of all the figures, we show how the cross-section is approaching zero, and gravitational dark matter contributes to the abundance. Last three plots of fig.\ref{alpha1} and \ref{scalarcrosssection3} also show the similar behavior in $(\langle \sigma v \rangle ~\mbox{Vs}~m_Y)$ plane near the maximum possible dark matter mass.

Due to entropy conservation constraint, we generally observed that reheating temperature is sensitive to the inflationary scalar spectral index $n_s$. The spectral index $n_s$ is
observationally bounded with a central value \cite{Planck:2018jri,BICEP:2021xfz}. Because of this bounded region, one naturally obtains a limit on the reheating temperature. Furthermore, we get different bound on this reheating temperature for different dark matter masses as all are intertwined through the reheating dynamics and inflationary dynamics.
For example from fig.\ref{alpha1}, for $\omega_\phi=0$ ($\omega_\phi<\omega_r$), the upper bound on the reheating temperature turns out as  $T_{re}^{max}\simeq(\,4.9\times10^{11},\,4.0\times 10^{12}\,)$ GeV for scalar and vector dark matter respectively with $m_{s/X}=10^6$ GeV, and $T_{re}^{max}\simeq 1.4\times10^{13}$ GeV for fermionic dark matter with $m_f=5\times 10^{10}$ GeV. However, for $\omega_\phi=0.5 > 1/3$, one obtains $T_{re}^{min}\simeq(\,1.6\times10^7,\,2.8\times10^4\,)$ GeV for scalar and vector dark matter with $m_{s/X}=1$ GeV and $T_{re}^{min}\simeq 3\times 10^6$ GeV for fermionic dark matter with $m_f=5\times10^8$ GeV. In addition to that, the lower limit on the scalar spectral index is set by the BBN temperature for those models where $\omega_\phi<1/3$ and instantaneous reheating for $\omega_\phi>1/3$.  In the allowed range of $n_s$, the cross-section can not be arbitrarily large due to unitarity limit on the cross-section $\langle\,\sigma\,v\,\rangle_{max}={8\pi}/{m_{Y}^2}$. This will further constraint $n_s$ and $T_{re}$.  For $n=1$ model, the lower limit on the scalar spectral index is modified due to the perturbative unitarity limit on the cross-section. Moreover, the modification on the lower limit of $n_s$ changes the minimum allowed value of the reheating temperature. Such as for $\omega_\phi=0$,  $T_{re}^{min}\simeq (\,180,\,6.4\times 10^6\,)$ GeV for scalar/vector ($m_{s/X}=10^6$ GeV) and fermionic dark matter ($m_f=5\times 10^{10}$) respectively.

\begin{figure}[t!]
\includegraphics[height=3.6cm,width=5.4cm]{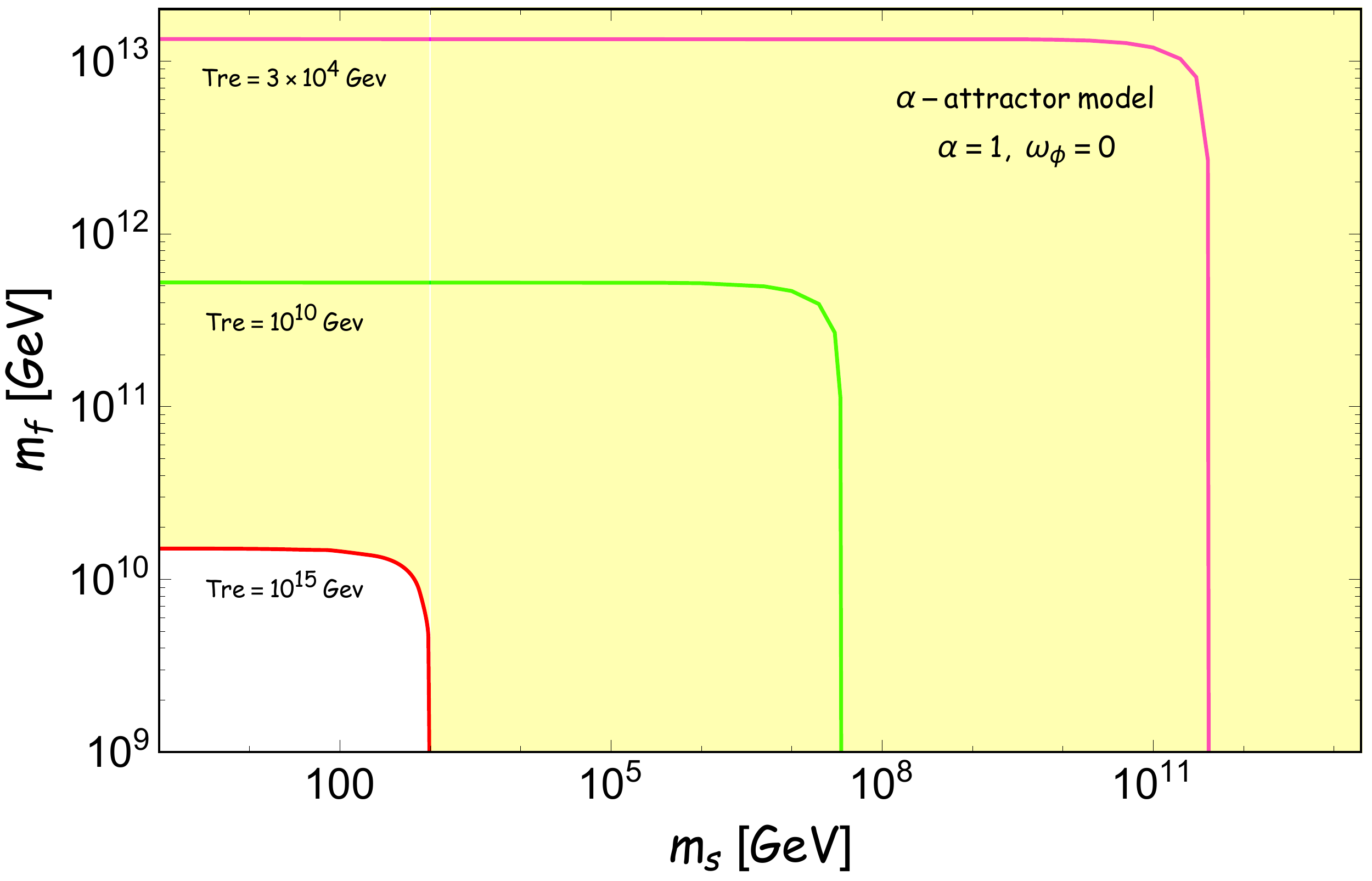}
\includegraphics[height=3.6cm,width=5.4cm]{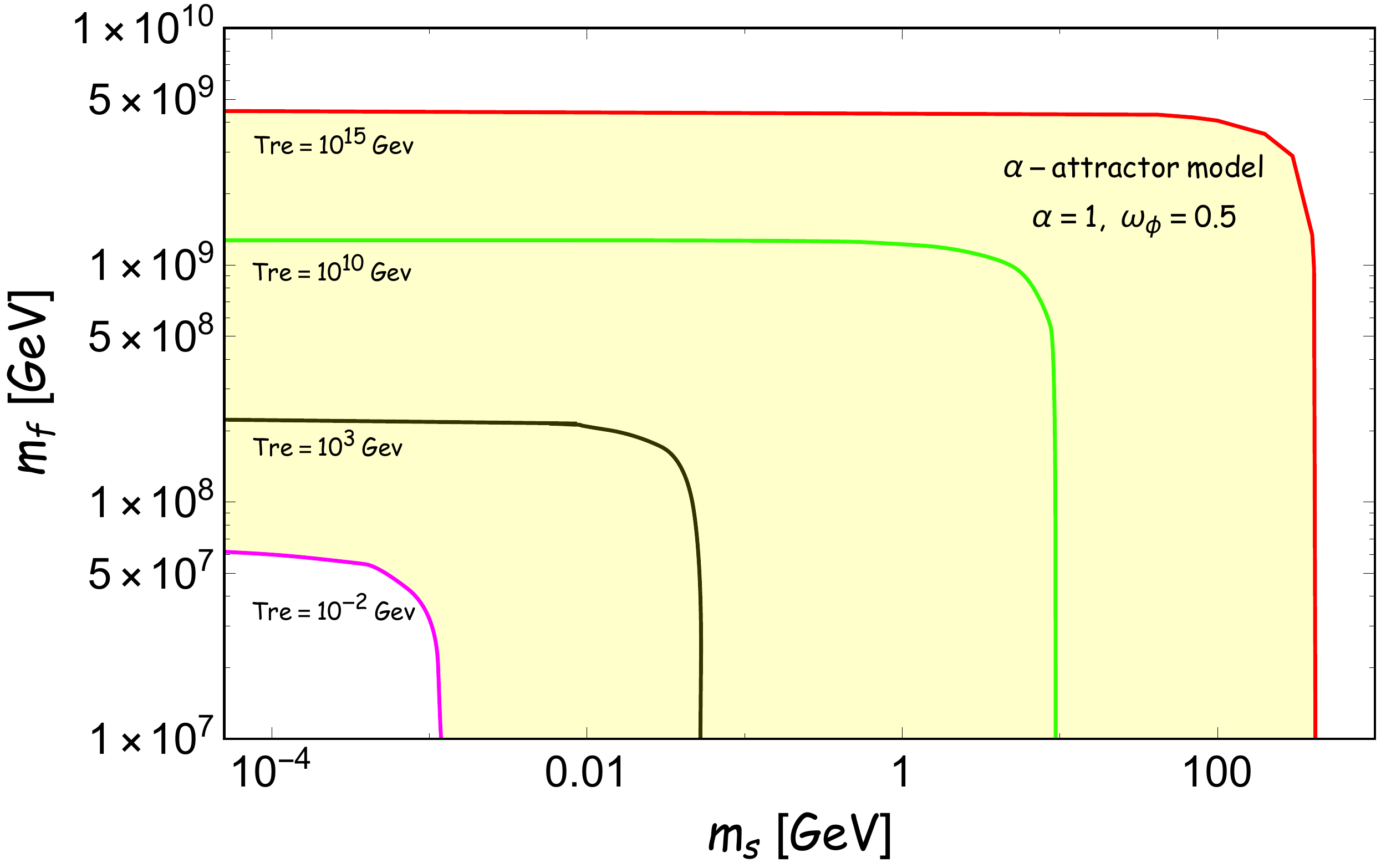}
\includegraphics[height=3.6cm,width=5.4cm]{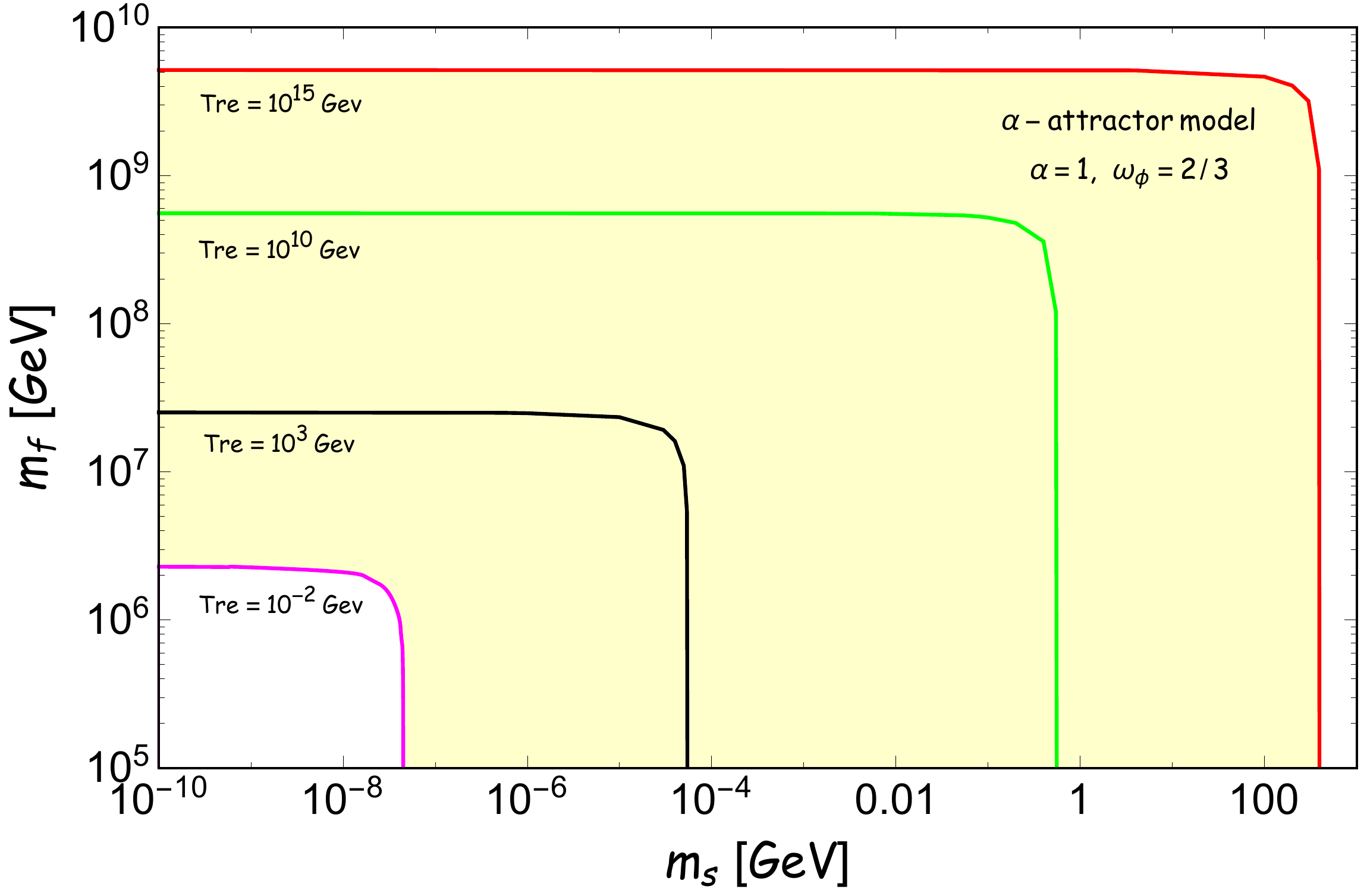}

\caption{{\bf Two-component dark matter scenario}: $m_f$ vs. $m_s$ were plotted for three different values of inflaton equation of state $\omega_\phi=(\,0,\,0.5,\,2/3\,)$ considering different reheating temperatures (shown by different colored line). Those lines corresponds to the fixed value of the present dark matter abundance $\Omega_{X(s+f)}\,h^2\simeq\,0.12$. For all the cases, we consider purely gravitationally produced dark matter. The dark matter sector consists of two sectors, one for scalar and another one for fermionic dark matter. Here the $\alpha-$ attractor model with $\alpha=1$ describes the inflationary dynamics, and the yellow shaded region shows the allowed dark matter masses.
 }
\label{Two-sector dark matter}
\end{figure}
\section{Two-component dark matter}
For the sake of completeness, in this section, we will briefly discuss the two-component dark matter scenario and the constraints on the parameter space. We explore possible allowed mass ranges when it is produced gravitationally. Since the behavior of scalar and vector dark matter is qualitative same, we assume the present-day abundance of total dark matter is composed of scalar and fermionic type particles. The dynamical equation will be the same as previously discussed in Eqs.\ref{boltzmann}-\ref{B2}, with no production from the radiation bath. From Fig.\ref{Two-sector dark matter}, it is clear that not all the range of mass is allowed, and as expected, it is explicitly dependent upon the reheating equation of state or rather types of the inflaton potential near its minimum. For each plot, the yellow shaded region is the allowed parameter space if we include all possibilities of reheating temperature. The region is either bounded by the maximum reheating temperature $\sim 10^{15}$ GeV, and the BBN Bound $10^{-2}$ GeV, or by $m_Y^{max}$ discussed in the previous section.

An interesting observation of this analysis is that there exists a one-to-one correspondence between scalar and fermionic dark matter masses. For a fixed combination of ($T_{re},\omega_\phi$), we can uniquely determine the mass of one component once another component is fixed. The maximum allowed mass for any one component is associated with the single component dark matter scenario, which we already discussed earlier. However, the minimum value of the mass approaches zero as the system starts dominating by only one component, either scalar or fermionic dark matter.
\begin{figure}[t!]
	\includegraphics[height=8cm,width=12.70cm]{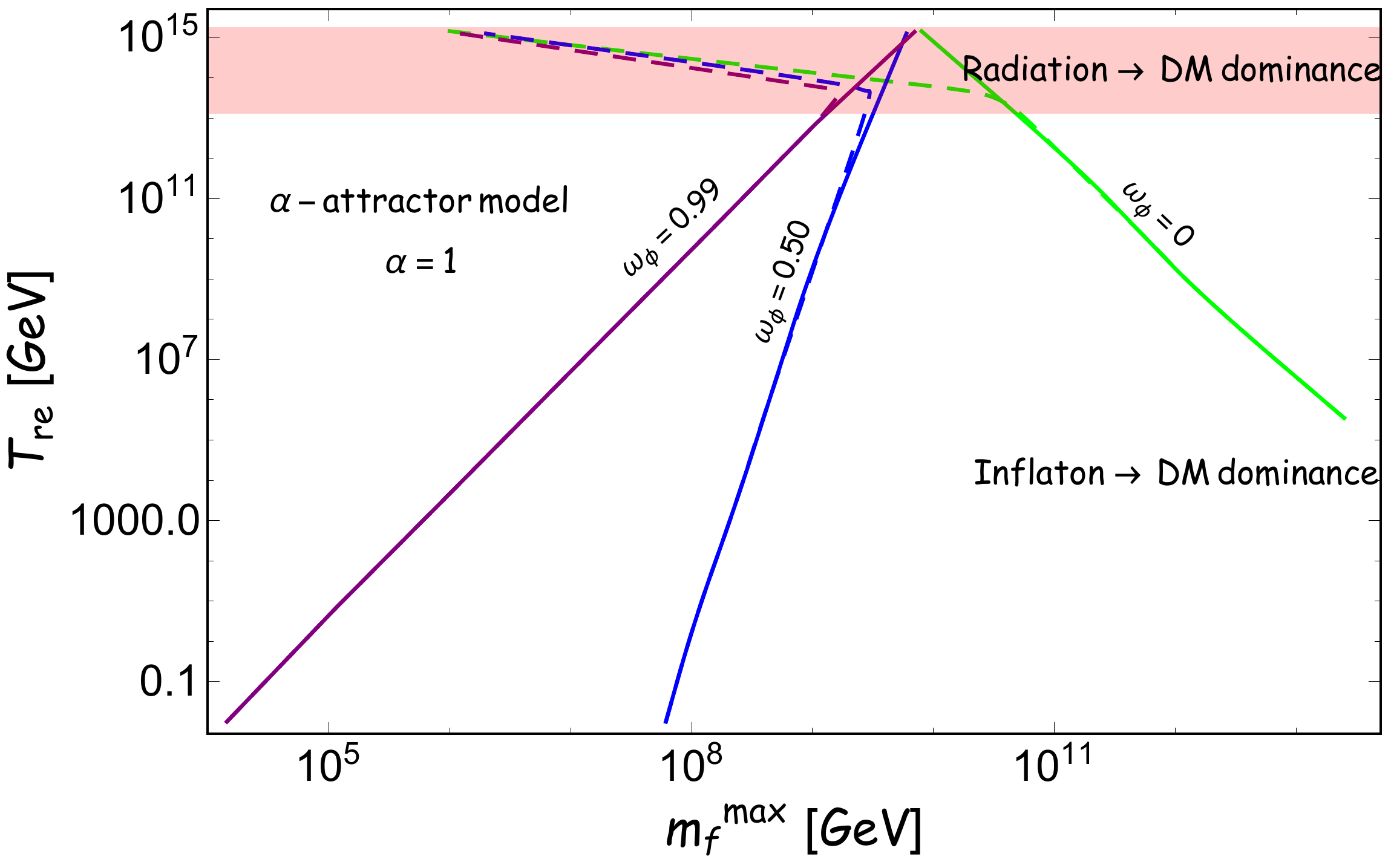}
	
	\caption{Variation of reheating temperature as a function of dark matter mass for two different gravitationally produced dark matter scenarios: 1) dark matter generated only from inflaton scattering (shown in solid line) 2) We took the contribution from inflaton as well as SM scattering (shown in dashed line). These results are for fermionic dark matter with three different inflaton equations of state $\omega_\phi=(0,\,0.5,\,0.99)$. Furthermore, the light red band indicates the dominating contribution in the dark matter relic from thermal bath over inflaton scattering.} 
	\label{purely gravitational dark matter}
\end{figure}
\section{Comparison on gravitational DM production from inflaton and radiation bath}\label{comparison rad-inflaton}
In our discussions so far, we considered gravitational dark matter production purely from the inflaton annihilation. 
However, in principle, gravitational production from the radiation bath will contribute, which we mentioned before, to be sub-leading compared to the production from inflaton.  This section will show through an explicit calculation that this is indeed the case. 
For the case of s-channel DM production from inflaton we have decay rates in Eqs.\ref{decaywidthscalar}-\ref{gauge}.  For the production of gravitational dark matter from the radiation bath during reheating has already been studied \cite{Bernal:2018qlk,Garny:2015sjg,Garny:2017kha,Tang:2017hvq}, and the  decay rate per unit physical volume is expressed as 
\bea \label{SMgrav}
R(T)=\gamma\,\frac{T^8}{M_p^4}\,,
\eea
where $\gamma=1.9\times 10^{-4}$ for scalar dark matter, $\gamma=1.1\times 10^{-3}$ for fermionic dark matter or $\gamma=2.3\times10^{-3}$ for vector dark matter. In addition to usual inflaton and the DM component from inflation we have modified radiation dynamics and an additional dark matter production channel from radiation bath as follows: 
\begin{eqnarray}\label{grav1}
\dot{\rho}_r+4\,H\,\rho_r-\Gamma_\phi\,\rho_\phi\left(\,1+\omega_\phi\,\right)+\,R(T)\,\langle\,E_Y\,\rangle^r=0~~, \label{grav1}
\\
\dot{n}_{Y(R)}+3\,H\,n_{Y(R)}-\,R(T)=0~~,
\label{grav2}
\end{eqnarray}
where $n_{Y(R)}$  is the DM number density produced from the radiation bath due to gravitational interaction.\\
Now let us compare the results for dark matter production from radiation bath mediated by gravity with the production from inflaton. The associated expressions for comoving dark matter number density in terms of reheating temperature calculated at the end of reheating for different types of dark matter, produced from either inflaton or radiation bath are (see appendix \ref{maxdark matter} and \ref{smcal})
\bea  \label{numbertre}
&&n^{re}_{s}\,A_{re}^3 \approx 8\,n^{re}_{X}\,A_{re}^3\approx \frac{3}{512\,\pi}\,\frac{(\,1+\,\omega_\phi\,)}{(\,1+\,3\,\omega_\phi\,)}\,\frac{\beta^2\,T_{re}^8\,e^{\,6\,N_{re}\,(1+\omega_\phi)}}{9\,M_p^4\,H_{end}},\nno\\
&&n^{re}_{f}\,A_{re}^3\approx \frac{3}{2048\,\pi}\,\frac{1+\omega_\phi}{1-\omega_\phi}\,\left(\frac{m_f}{m_\phi^{end}}\right)^2\,\frac{\beta^2\,T_{re}^8\,e^{\,6\,N_{re}\,(1+\omega_\phi)}}{9\,M_p^4\,H_{end}}\,,\\
&&n_{Y(R)}^{re}\,A_{re}^3\approx \frac{2\gamma}{3\,(\,1-\omega_\phi\,)}\frac{e^{\frac{3}{2}\,N_{re}\,(\,3+\,\omega_\phi\,)}\,T_{re}^8}{M_p^4\,H_{end}}\,.\nno
\eea
To derive above equation, we use the following approximate relation, $H_{re}^2={\rho_R^{re}}/{3\,M_p^2}={\beta\,T_{re}^8}/({3\,M_p^2})=H_{end}^2\,A_{re}^{-3\,(1+\omega_\phi)}$, which indicates that at the end of reheating the universe is dominated by radiation. The dark matter production from radiation bath is maximum when radiation temperature is maximum which is approximately equivalent to taking  $N_{re} = 0$. Therefore, it would be sufficient to compare the above comoving densities for different production channel at the point of instantaneous reheating;
\begin{eqnarray}
&&n^s_R =\frac{n^{re}_{s}\,A_{re}^3} {n_{Y(R)}^{re}\,A_{re}^3} = \left(\frac{3}{512\,\pi}\,\frac{(\,1+\,\omega_\phi\,)}{(\,1+\,3\,\omega_\phi\,)}\,\frac{\beta^2}{9}\right) \times \left(\frac{3\,(\,1-\,\omega_\phi\,)}{2\,\gamma}\right)  \\
&&n^f_R=\frac{n^{re}_{f}\,A_{re}^3} {n_{Y(R)}^{re}\,A_{re}^3} = \left(\frac{3}{2048\,\pi}\,\frac{(\,1+\,\omega_\phi\,)}{(\,1-\,\omega_\phi\,)}\,\frac{\beta^2}{9}\,\left(\frac{m_f}{m_\phi^{end}}\right)^2\right) \times \left(\frac{3\,(\,1-\,\omega_\phi\,)}{2\,\gamma}\right)
\end{eqnarray} 
From the above two equations it can be checked that for any $\omega_{\phi}$, $n^s_R >> 1$  (cf. Eq.\ref{numbertre}). Hence, comoving dark matter number density for scalar/vector produced from inflaton always dominates over the production from the radiation bath. However, for fermionic dark matter dominating production channel is crucially dependent on $(m_f/m_{\phi}^{end})$. For example, if the reheating is  instantaneous and the value of the fermionic dark matter mass produced from inflaton assumes $m_f\simeq 10^{-3}\,m_\phi^{end}$, then $n^f_R << 1$ which makes $n^{re}_{f}$ sub-dominant compared to $n_{Y(R)}^{re}$. If we convert this into reheating temperatures, it can be easily shown that above $T_{re} \gtrapprox 10^{13}\,\mbox{GeV}$, the production of fermionic dark matter from radiation bath will always dominate over the production from inflaton field and it is less sensitive to the inflaton equation of state (see Fig.\ref{purely gravitational dark matter}). 
 In Fig.\ref{purely gravitational dark matter}, solid lines correspond to gravitational dark matter production from inflaton scattering, and dotted lines correspond to dark matter production from both inflaton as well as radiation bath.  The light red shaded region within $10^{15} \gtrsim T_{re} \gtrsim 10^{13}$ GeV clearly shows that the production from the inflaton field is sub-leading compared to that from the radiation bath. Depending upon the reheating equation of state, the mass range of the fermionic dark matte is observed to be slightly different. 

So far, we have discussed dark matter production and its intimate connection with the inflationary and reheating phase.  However, dark matter abundance does not contain much information about the nature of dark matter and its underlying production mechanism.  In the subsequent discussions, we will focus more on the microscopic properties of dark matter, such as its phase-space distribution, free streaming lengths, etc.  These properties play a significant role in the subsequent cosmological evolution of DM perturbation, which is deeply connected with the large-scale structure formation.   
\section{Phase space distribution of gravitationally produced dark matter}
In this section we study the evolution of phase space distribution of dark matter which will be observed to encode not only the underlying production mechanism but also the very nature of the DM itself. The DM production is purely gravitational and produced from inflaton through the process $\phi\phi\to SS/XX/ff$ for scalar ($S$), vector ($X$), and fermion ($f$), mediated by gravity. Gravitational production from radiation bath will not be considered  unless otherwise stated. The phase-space distribution $(f_Y)$ of DM is evolved by the Boltzmann transport equation as,
\begin{eqnarray} \label{Boltz-transport}
\frac{\partial{f_Y}}{\partial{t}}-\,H\,|{\bf p}_Y|\,\frac{\partial{f_Y}}{\partial{|{\bf p}_{Y}|}}=\,c\left[f_Y(\,|{\bf p}_Y|,\,t)\,\right]\,
\end{eqnarray}
where $c\left[f_Y(\,|{\bf p}|,\,t)\,\right]$ symbolizes the collision term, determined through inflaton-DM interaction.\\
Let us first calculate the collision term for this process. To calculate collision term, one of the important quantities is the phase space distribution of inflaton. The inflaton field is homogeneous in nature, and the phase space distribution of the inflaton field can be effectively written as $f_\phi\,(\,k,\,t\,)=(2\,\pi)^3\,n_\phi(t)\,\delta^{(3)}({\bf k})$. Where, $n_{\phi}$ is the number density of the zero momentum inflaton particles. The required collision term for the transport equation is given by
\bea \label{collision term}
\begin{split}
	c\left[f_Y(\,|{\bf p}_Y|,\,t)\,\right] & =\,\frac{1}{2\,p_{Y_0}}\,\int\,\frac{d^3\,{\bf k}}{(2\,\pi)^3\,2\,k_0}\,\frac{d^3\,{\bf k'}}{(2\,\pi)^3\,2\,k'_0}\,\frac{g_{Y'}\,d^3\,{\bf p}_{Y'}}{(2\,\pi)^3\,2\,p_{Y'_0}}\,(2\,\pi)^4\,\delta^{(4)}\,(k+k'-p_Y-p_{Y'})\\\ 
	& \qquad \qquad \qquad \qquad\qquad|\mathcal{M}|^2_{\,\phi\phi\to YY'}\,f_\phi\,(k)\,f_\phi\,(k')\,\left[\,1\pm f_Y(p_Y)\pm\,f_{Y'}(p_{Y'})\,\right]\\
	& =\frac{\pi \,n_\phi(t)^2}{p_{Y_0}}\,\int\,\frac{1}{\,4\,m_\phi^2}\,\frac{g_{Y'}\,d^3\,{\bf p}_{Y'}}{2\,p_{Y'_0}}\,\delta\,(2\,m_\phi-\,p_{Y_0}-\,p_{Y'_0})~ \delta^{(3)}({\bf p}_{Y'}+{\bf p}_{Y})\\
	& \qquad \qquad \qquad \qquad \qquad \qquad \qquad \,|\mathcal{M}|^2_{\,\phi\phi\to YY'} \left[\,1\pm f_Y(p_Y)\pm\,f_{Y'}(p_{Y'})\,\right]\\
	& =\frac{\pi\,n_\phi (t)^2}{8\,g_Y\,p_{Y_0}\,p_{Y'_0}}\,\frac{g_Y\,g_{Y'}\,|\mathcal{M}|^2_{\,\phi\phi\to YY'}}{m_\phi^2}\,\,\delta\,(2\,m_\phi-\,p_{Y_0}-\,p_{Y'_0})\,\left[\,1\pm f_Y(p_Y)\pm\,f_{Y'}(p_{Y'})\,\right]\,,
\end{split}
\eea
where $(+)$ and $(-)$ sign, in the third bracket, are for bosonic and fermionic dark matter respectively. $g_y$, $g_{Y'}$ represents the number of internal degrees of freedom for $Y$ and $Y'$. Moreover, in the absence of Bose condensation or fermionic degeneracy, one may approximate the blocking and stimulated emission factor as $\left[\,1\pm f_Y(p_Y)\pm\,f_{Y'}(p_{Y'})\,\right]\simeq 1$. Furthermore, the corresponding gravitational DM production rate for the process $\phi\phi\to YY$ can be related with spin-averaged squared amplitude $|\mathcal{{M}}|^2_{\,\phi\phi\to YY}\to |\mathcal{{\bar{M}}}|^2_{\,\phi\phi\to YY}=\,\sum_{avg\, over \,initial\, pol.}\sum_{sum \,over \,final \,pol.}\,\mathcal{{M}}|^2_{\,\phi\phi\to YY}$, (sum over the polarizations (spins) of the final particles and average over the polarizations (spins) of the initial ones) as 
\bea \label{decay-amplitude compare}
\Gamma_{\phi\phi\to YY}=\,n_\phi\,\frac{g_Y^2\,|\mathcal{M}|^2_{\,\phi\phi\to YY}}{32\,\pi\,m_\phi^2}\,\sqrt{1-\frac{m_Y^2}{m_\phi^2}}\simeq\,\,n_\phi\,\frac{g_Y^2\,|\mathcal{M}|^2_{\,\phi\phi\to YY}}{32\,\pi\,m_\phi^2}\,,
\eea
where we use the approximation $m_Y <m_\phi$. Therefore, combining equations (\ref{collision term}) and (\ref{decay-amplitude compare}) and acknowledging the approximations mentioned above, the collision term takes the form
\bea \label{collision-term final}
c\left[f_Y(p_Y,t) \right]\simeq \frac{2\pi^2  n_\phi(t)}{g_Y\,p_{Y_0}^2}\,\Gamma_{\phi\phi\to YY}\,\delta(m_\phi-\,p_{Y_0}) =  \frac{2\pi^2  n_\phi(t)}{g_Y\,p_{Y_0}^3 H} \Gamma_{\phi\phi\to YY}\,\delta(t-t')
\eea
were $t'$ is the cosmic time when $\textit{p}_Y$ is equal to the inflaton mass which satisfies the relation $
p_Y a(t)= {a\,(t')} {m_\phi}$. 
The energy associated with each dark matter particle is $p_{Y_0}=m_\phi$. Upon substituting the above collision term into the transport equation (\ref{Boltz-transport}), one obtains the dark matter phase space distribution as
\bea \label{momentum-distribution}
f_Y(\,p_Y,\,t)=\frac{2\,\pi^2\,n_\phi\,(t')}{g_Y\,m_\phi^3\,H\,(t')}\,\Gamma_{\phi\phi\to YY}\,(t')\,\theta\,(t-t')\,
\eea

\begin{figure}[t!]
	\includegraphics[height=5.1cm,width=7.5cm]{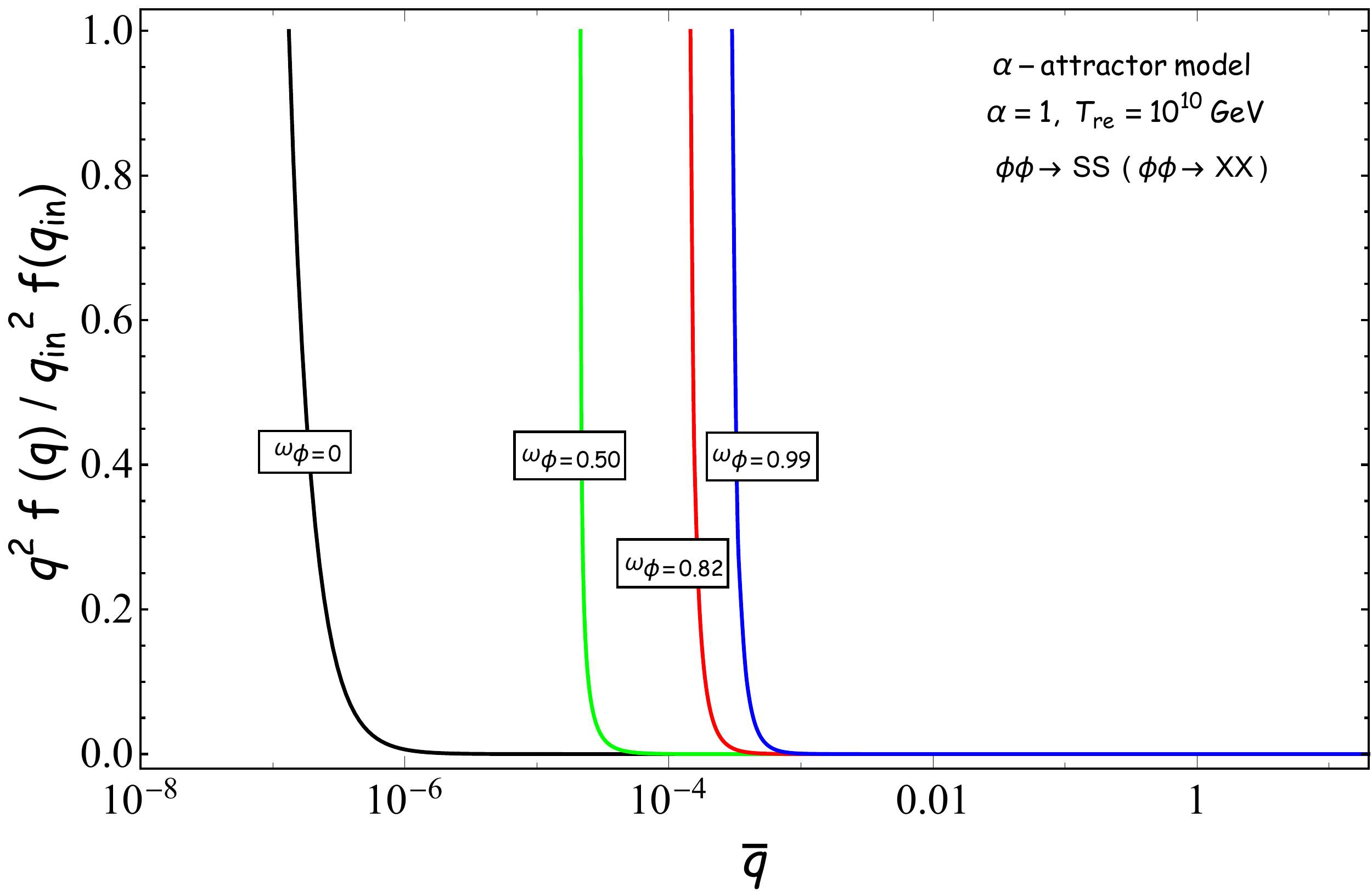}
	\includegraphics[height=5.1cm,width=7.5cm]{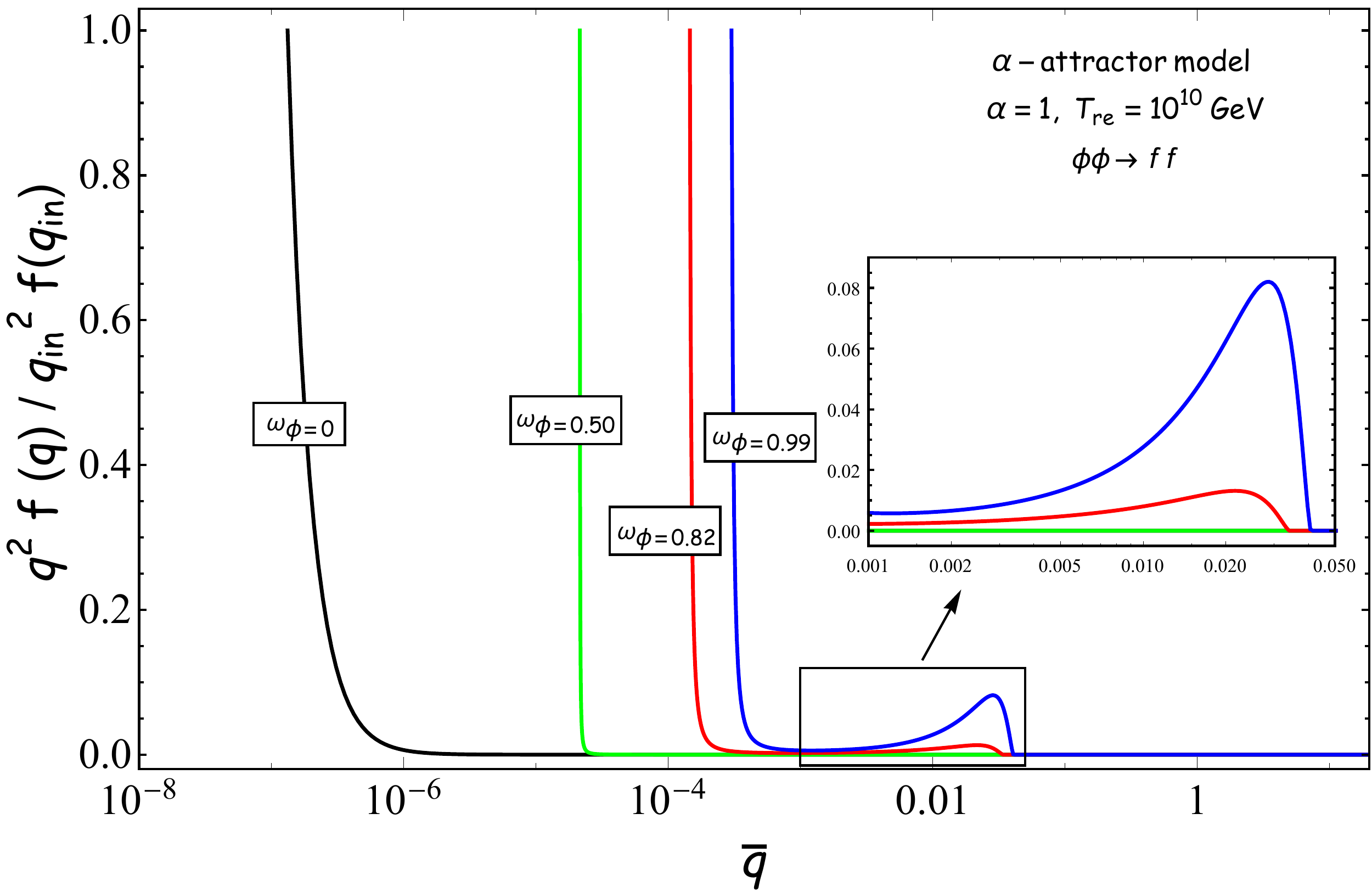}
	\includegraphics[height=5.1cm,width=7.5cm]{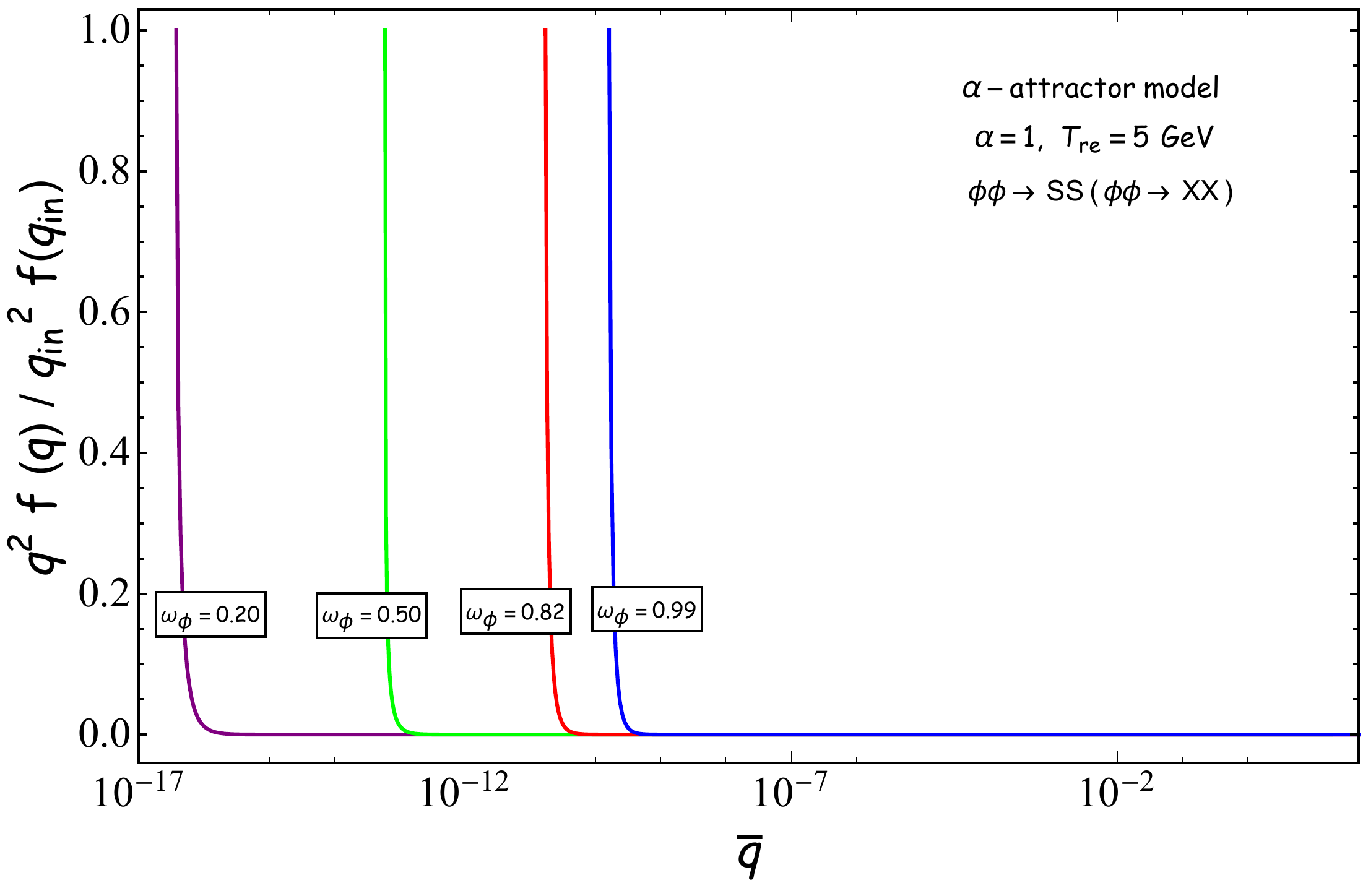}
	\includegraphics[height=5.1cm,width=7.5cm]{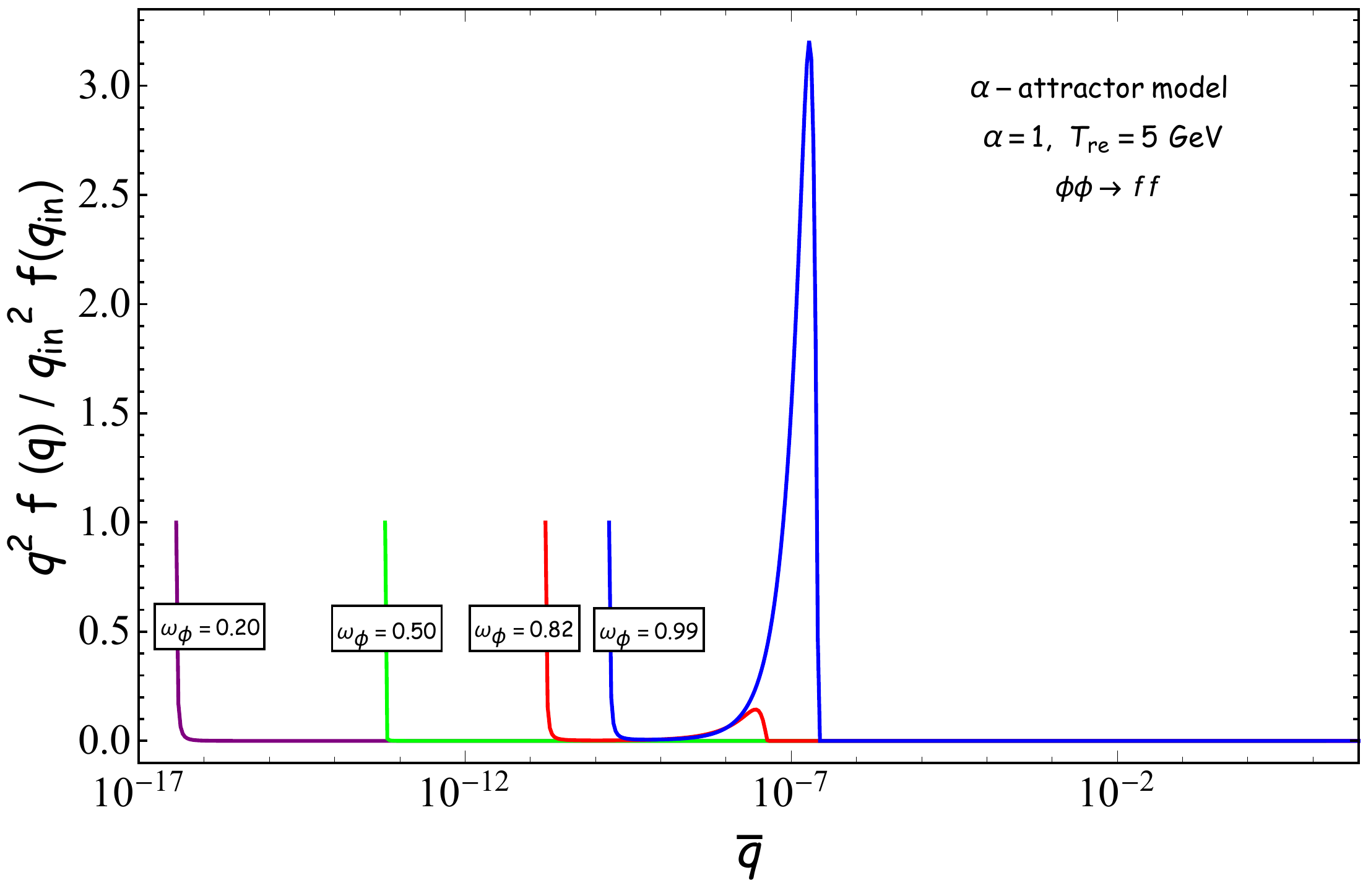}

	\caption{The rescaled momentum distribution function of dark matter as a function of $q$, defined in equation (\ref{comovingmomen}), indicates rescaled comoving momentum for time-independent inflaton mass ($V(\phi)\sim\phi^2$) with a few different inflaton equations of state $\omega_\phi=(\,0,\,0.2,\,0.5,\,0.82,\,0.99\,)$ (shown in different color) with two specific values of the reheating temperature $T_{re}=(\,10^{10},\,5)$ GeV. On the left, we have plotted results for gravitationally produced scalar $(\phi\phi\to SS)$ or vector dark matter ($\phi\phi\to XX$) and on the right for gravitationally produced fermionic dark matter ($\phi\phi\to ff$).} 
	\label{momentum distribution zero}
\end{figure}
Inflaton energy density  during reheating can be evaluated by integrating Eqn.\ref{boltzmann}, which leads to 
\bea \label{ied}
\rho_\phi(\,t'\,)=\rho_\phi^{end}\,A\, (\,t'\,)^{-3\,(\,1+\omega_\phi \,)}\,e^{-\Gamma_\phi\,(t'-t_{end})}\,,
\eea
Here, we ignore the effect of dark matter production on the inflaton energy density as it is negligible compared to the production of radiation. The subscript ``end" indicates the end of the inflation. We can write the above equation (\ref{ied}) for the inflaton energy density  in terms of the inflaton energy density at the end of the reheating ( $\rho_\phi^{re}$ ) as
\bea
\rho_\phi (t')=\rho_\phi^{re}\,\left( \frac{A\,(\,t'\,)}{A_{re}}\right)^{-3\,(\,1+\,\omega_\phi\,)}\,e^{-\,\Gamma_\phi\,(\,t'-\,t_{re}\,)}\,.
\eea
As most of the region during reheating is dominated by the inflaton EoS, we can approximate the scale factor as $a\propto\,t^ {2/3\,(\,1+\omega_\phi\,)}$ and $t'=\, t_{re}\,\left(\frac{A\,(t')}{A_{re}}\right)^{3\,(1+\,\omega_\phi)/2}$. Further, at the end of the reheating, when $t_{re}=\Gamma_\phi^{-1}$, the Hubble parameter $H_{re}\simeq \Gamma_\phi$ and the inflaton energy density approximately equals to the radiation energy density $\rho_\phi\simeq\,\rho_r=\,\frac{\pi^2}{30}\,g_{re}\,T_{re}^4$. Under these approximations, $\rho_\phi\,(\,t'\,)$ assumes following form,
\bea \label{rhore}
\rho_\phi (t')=\frac{\pi^2}{30}\,g_{re}\,T_{re}^4\,\left( \frac{A\,(\,t'\,)}{A_{re}}\right)^{-3\,(\,1+\,\omega_\phi\,)}\,e^{1-\left(\frac{A\,(t')}{A_{re}}\right)^{3\,(1+\,\omega_\phi)/2}}
\eea
In the same way Hubble parameter during reheating phase turns out as 
\bea \label{hubblere}
H\,(\,t'\,)\simeq \,H_{re}\,\left(\frac{A\,(t')}{A_{re}}\right)^{-\frac{3}{2}\,(\,1+\,\omega_\phi\,)}\simeq\,\Gamma_\phi\,\left(\frac{A\,(t')}{A_{re}}\right)^{-\frac{3}{2}\,(\,1+\,\omega_\phi\,)}
\eea
Substituting equations (\ref{rhore}) and (\ref{hubblere}) into the phase-space distribution equation (\ref{momentum-distribution}) one obtains the following form of the dark matter phase space distribution during reheating phase as,
\bea
\begin{split}
	f_Y(\,p_Y,\,t)=\,\frac{\pi^4\,g_{re}}{15\,g_Y\,\Gamma_\phi}\,\left(\,\frac{T_{re}}{m_\phi^{end}}\,\right)^4\,\,\left(\frac{m_\phi^{end}}{m_\phi\,(\,t'\,)}\right)^4\,\left(\frac{A\,(t')}{A_{re}}\right)^{-\frac{3}{2}\,(\,1+\,\omega_\phi\,)}\,e^{1-\left(\frac{A\,(t')}{A_{re}}\right)^{\frac{3}{2}\,(1+\,\omega_\phi)}}\\ \times\Gamma_{\phi\phi\to YY}\,(t')\,\theta\,(t-t')\,.
\end{split}
\eea
Instead of symbolizing the inflaton's mass by $m_\phi$, we use $m_\phi (t)$ as the effective mass of the inflaton being a function of time, and its evolution is followed by Eq.\ref{effectivemass}. 
To get a better approximation for the momentum distribution function $f_Y\,(p_Y,\,t)$, we have to calculate equation (\ref{momentum-distribution}) by solving the sets of Boltzmann equations [\ref{boltzmann} - \ref{boltzmann1}] numerically. The numerical solution of the rescaled momentum distribution function $f(q)$ is shown in fig.\ref{momentum distribution zero}; the form of $f(q)$ is defined in the following manner 
\bea \label{rescale-momentum distribution}
f_{Y}\,(p_Y,\,t)\,d^3p=\,\frac{\pi^4\,g_{re}}{15\,g_Y}\,\left(\,\frac{T_{re}}{m_\phi^{end}}\,\right)^4\,\left(\,\frac{T_*}{a}\,\right)^3\,f(\,q\,)\,d^3q\,,
\eea
where $q$ is the rescaled comoving momentum of the dark matter, which is defined as
\bea \label{comovingmomen}
q=\,\frac{p\,a\,(\,t\,)}{T_*}=\,\frac{A\,(\,t'\,)}{A_{re}}\,\frac{m_\phi\,(\,t'\,)}{m_\phi\,(\,t_{re}\,)}=\,\bar{q}\,\frac{m_\phi\,(\,t'\,)}{m_\phi\,(\,t_{re}\,)}\,.
\eea
Here $T_*$ is the time-independent quantity, defined as $T_*=\,m_\phi\,(\,t_{re}\,)\,a_{re}$. As can be observed from the Fig.\ref{momentum distribution zero}, the phase space distribution function naturally contains peaks at the initial time when the DM particles would be maximally produced from the inflaton decay, and the momentum of those produced particles should be around the mass of the inflaton.  The characteristics of the peak and location will certainly be dependent on the background dynamics determined by the inflaton equation of state $\omega_{\phi}$ and reheating temperature $T_{re}$ as one can imagine that this characteristic peak will naturally be imprinted on the subsequent evolution of DM structures.  In addition, the free streaming properties of DM will help understand the formation of the dark matter structure, and we will discuss this in detail in the following section. 
Furthermore, it can be observed that there exist a secondary peak in the fermionic distribution function at even higher momentum which is arising due to non-trivial mass dependence in the fermion decay width $\Gamma_{\phi\phi\rightarrow ff} \propto \rho_{\phi}/m_{\phi}$ and consequently, the phase space distribution $q^2 f_f (q,t)\propto ({a^2 \rho_\phi^2})/({m_\phi^3 H})\propto a^{\frac{1}{2}(5-9\omega_\phi\,)}$, as opposed to the bosonic phase-space distribution function $q^2 f_s (q,t)\propto ({a^2 \rho_\phi^2})/({m_\phi H})\propto a^{-\frac{1}{2}(5+3 \omega_\phi)}$. Therefore, in the case of fermionic dark matter, for $\omega_\phi>5/9$, the phase space distribution function increases till the point when inflaton mass is equal to the mass of the dark matter ($m_Y=m_\phi$), and after that point, the distribution function approaches zero as the dark matter production is kinematically forbidden ($\Gamma_{\phi\phi\to YY}\to0$) in the region where $m_\phi<m_Y$.  The important point is to note that the peak value associated with the secondary peak increases as we $T_{re}$ decreases, which increases the time elapsed to reach the point $m_\phi=m_Y$.  However, for bosonic DM, such secondary peak does not arise as $q^2 f_s(q,t)$ drops with scale factor during reheating for a viable range of $\omega_\phi$, $0\leq\omega_\phi\leq1$.  It would be interesting to look into this secondary peak and its physical significance in detail.

\section{Momentum, free streaming length, and constraints}
We have already observed the peak of the dark matter phase space distribution occurs near the beginning of reheating, where inflaton decay to DM will be maximized.  The momentum around that peak will also maximum, which is $\sim m_{\phi}^{end}$, which naturally depends on the inflaton equation of state.  The obvious physical effect of this large initial momentum of the dark matter would be on their free steaming properties, which will have a significant impact on the perturbation evolution at a small scale.  Large initial momentum will naturally suppress the structure formation at small scales.  In this section, we will study this in detail and evaluate the possible constraints on the present dark matter velocity from the well-known Lyman-$\alpha$ bound on the dark matter mass for warm dark matter (WDM) \cite{Viel:2013fqw, Narayanan:2000tp, Viel:2005qj, Baur:2015jsy,Irsic:2017ixq,Palanque-Delabrouille:2019iyz,Garzilli:2019qki}.  If the dark matter has no interaction with itself or with the SM particles, the momentum of the dark matter particles is redshifted by the expansion of the universe.  Therefore, we can relate the present momentum of the dark matter with the mean initial momentum at the time of its production as, $
\textit{p}_{now}=({a_{in}}/{a_{now}})\,\textit{p}_{in}.
$
For example, if the dark matter particles produced from the thermal bath, the mean initial momentum would assume $\textit{p}_{in}\sim 3\, T_{re}$ at scale factor $a_{in}=a_{re}$  and assuming the entropy being conserved between the end of reheating to today, the momentum at present would be calculated as 
\bea\label{pnowrad}
\textit{p}_{now}=\left(\,\frac{387}{11\,g_{s,re}}\right)^{1/3}\,T_{0}~~.
\eea 
Where the present CMB temperature $T_{0}=2.725\,k\,=2.3\times 10^{-13}$ GeV and the $g_{s,re}$ is the effective number of degrees of freedom for entropy at reheating temperature. Now, by using the various experimental constraints on the warm dark matter such as:  the MCMC analysis of the XQ-$100$ and HIRES/MIKE Lyman-$\alpha$ forest data sets, constraints the mass of the warm dark matter (WDM) particle $m_{wdm}> 5.3$ keV at $2\,\sigma$ range \cite{Irsic:2017ixq}. In references \cite{Viel:2013fqw,Hoof:2018hyn}, using the same  Lyman-$\alpha$ forest data set the authors obtained the bound on $m_{wdm}>3.3$ keV using HIRES/MIKE and $>3.95$ keV  using SDSSIII/BOSS. Considering the over all conservative estimate of $m_{wdm}>3.9$ keV and $g_{s,re}\sim 100$
, using Eq.\ref{pnowrad} one gets the lower bound on the present dark matter  velocity $v_{dm}<4.1\times10^{-8}$.\\
Now using the above bounds on the warm dark matter mass, we will first estimate the dark matter velocity for different production scenarios described so far. 

{\bf Production from inflaton}: For the gravitational dark matter produced from the inflaton, the initial momentum at production can be approximately taken to be $\textit{p}_{in}\sim m_\phi$ and the radiation temperature correspond to the scale factor $a=a_{in}$ can be taken as the maximum radiation temperature $T_{rad}^{max}$. The radiation energy density will evolve as $\rho_r\,\propto\,T_{rad}^4\,\propto\,a^{-\frac{3\,(\,1+\,\omega_\phi\,)}{2}}$ \cite{Haque:2020zco}. Accumulating all the above expressions, one can find the present value of the dark matter momentum as
\bea \label{pnowinflaton}
\textit{p}_{now}=\frac{a_{in}}{a_{re}}\,\frac{a_{re}}{a_{now}}\,m_\phi=\left(\,\frac{43}{11\,g_s\,(T_{re}\,)}\,\right)^{1/3}\,\frac{T_{now}}{T_{re}}\,\left(\,\frac{T_{re}}{T_{rad}^{max}}\,\right)^{\frac{8}{3\,(\,1+\,\omega_\phi\,)}}\,m_\phi~~.
\eea
Moreover, in the perturbative reheating scenario, the approximated analytical expression of the maximum radiation temperature $T_{rad}^{max}$ can be written as \cite{Haque:2020zco,Maity:2018exj} 
\begin{equation}
T_{rad}^{max}=\left(\,\frac{60\,\sqrt{3}\,M_p\,\Gamma_\phi}{g_{re}\,\pi^2}\,\frac{1+\,\omega_\phi}{5-3\,\omega_\phi}\right)^{\frac{1}{4}}\,\left(3\,M_p^2\,H_{end}^2\right)^{\frac{1}{8}}\left\{y^{-\frac{3\,(\,1+\,\omega_\phi)}{5-3\,\omega_\phi}}-\,y^{-\frac{8}{5-3\,\omega_\phi}}\right\},
\end{equation}
where $y={8}/(3+3\omega_\phi)$.
\begin{figure}[t!]
	\includegraphics[height=8cm,width=12.70cm]{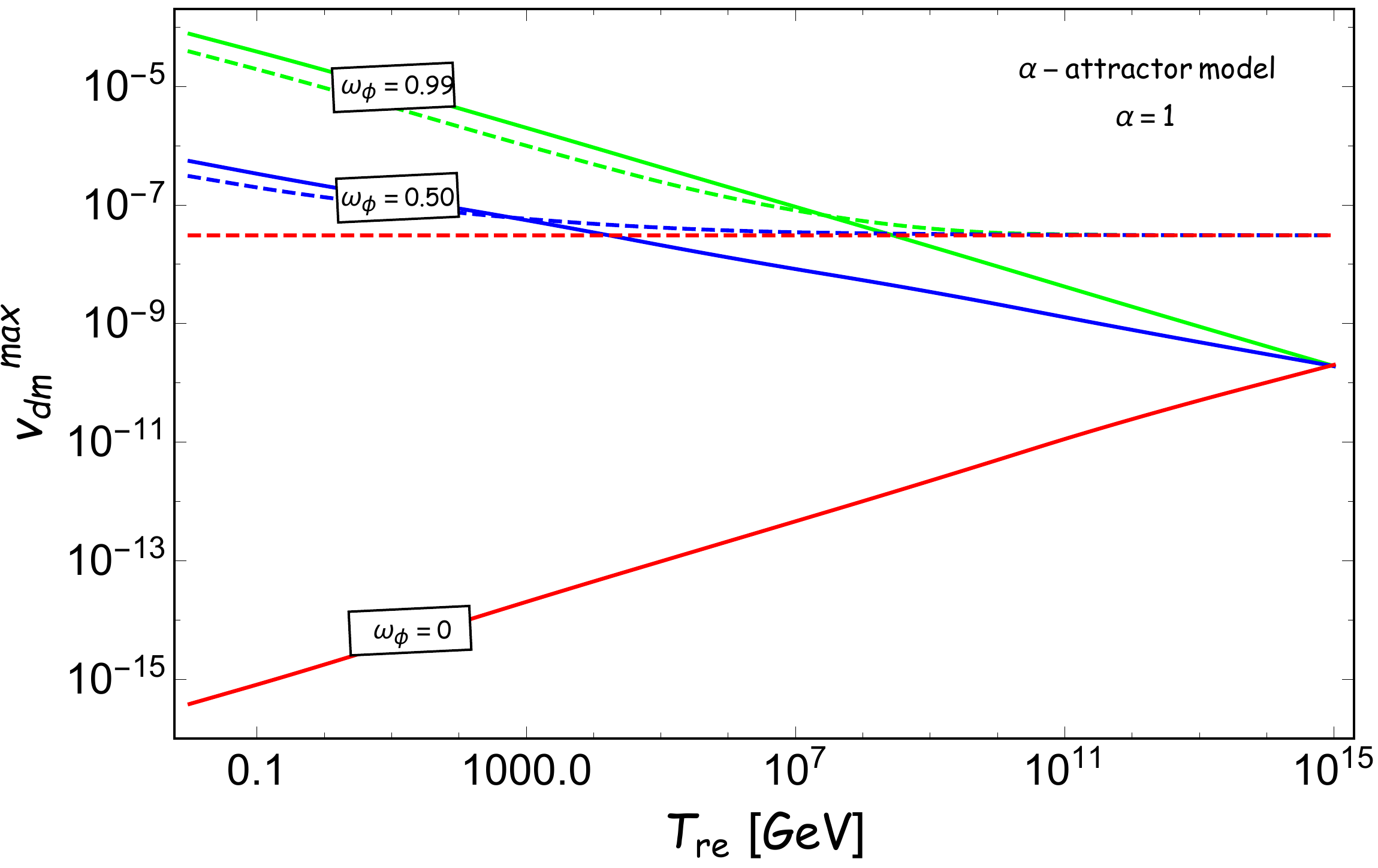}
	
	\caption{We have plotted the upper bound on the present dark matter velocity $v_{dm}^{max}$ as a function of reheating temperature for three distinct values of $\omega_\phi=(0,\,0.5,\,0.99)$. Here, solid lines indicate results for dark matter production from inflaton whereas, the dashed line is for production from both inflaton and radiation bath.  In the case of production from both inflaton and thermal bath, we choose $\xi=0.5$.  These bounds are estimated in choice of $m_{wdm}>3.9$ keV.} 
	\label{maxv}
\end{figure}\\
{\bf Production from inflaton and radiation bath:}  In this scenario, the fraction of the dark matter (say $\xi$) is produced from the inflaton through gravitational interaction with initial momentum $ p\sim m_{\phi}$ at the beginning of reheating, and remaining fraction, $(1-\xi)$ is produced from the radiation bath because of non-zero cross-section $\langle\sigma v\rangle$ near the end of reheating with momentum around $p \sim 3\, T_{re}$.  Detailed study of the evolution of dark matter perturbation will be interesting in such a scenario which we will study later.  For the present study, let us define an average momentum of the dark matter particles at the reheating end as
\bea \label{avgmom}
\langle\, \textit{p}\,\rangle_{re}=\xi \textit{p}_1(a_{re})+\,(1-\xi)\,\textit{p}_2(a_{re})~~,~~\xi=\frac{n_1(a_{re})}{n_1(a_{re})+n_2(a_{re})}~~,
\eea
where $n_1(a_{re})$ and $\textit{p}_1(a_{re})$ represent number density and momentum respectively for the gravitationally produced dark matter at reheating end. Whereas, $n_2(a_{re})$ and $\textit{p}_2(a_{re})$  represent corresponding number density and momentum at the end of reheating for the particles produced from radiation bath. As described before the dark matter particles produced gravitationally from inflaton redshifts due to expansion from $a_{in}=a_{max}$ (corresponding to maximum radiation temperature) till the reheating end and hence
\bea \label{momregrav}
\textit{p}_1(a_{re})=\frac{a_{in}}{a_{re}}\textit{p}_{in}=\left(\frac{T_{re}}{T_{rad}^{max}}\right)^{\frac{8}{3(1+\omega_\phi)}}m_\phi~~.
\eea
Upon substituting the above equation into Eqn.\ref{avgmom}, the average momentum of the dark matter particles at the end of the reheating is estimated as 
\bea \label{avgmomcom}
\langle\, \textit{p}\,\rangle_{re}=\xi\,\left(\,\frac{T_{re}}{T_{rad}^{max}}\,\right)^{\frac{8}{3\,(\,1+\,\omega_\phi\,)}}\,m_\phi+\,3\,(\,1-\xi\,)\,T_{re}~~,
\eea
where $\xi$ can be determined by solving Eqs. (\ref{finalgrav}), (\ref{finalgravf}) and (\ref{finalgravrad}). Since the momentum is redshifted by the expansion from the end of reheating till the present day, the value of the average momentum at present is 
\bea \label{avgmompresent}
\langle\, \textit{p}\,\rangle_{now}=\frac{a_{re}}{a_{now}}\,\langle\, \textit{p}\,\rangle_{re}=\left(\,\frac{43}{11\,g_s\,(T_{re}\,)}\,\right)^{1/3}\,\frac{T_{now}}{T_{re}}\,\langle\,\textit{p}\,\rangle_{re}~~.
\eea
Now that we have calculated the approximate expression for the average momentum of the dark matter particle at the present epoch, we can put constraints on the warm dark matter velocity depending upon the reheating equation state.  Using the warm dark matter bound, we further estimate the upper bound of the velocity of dark matter particles at present.  The detailed constraints on the upper limit of dark matter velocity for two different scenarios: production from inflaton and combined production from both inflaton and radiation bath, are depicted in Fig.\ref{maxv}.  For the case of production from inflaton decay maximum value of this upper bound turns out to be $\sim10^{-4}$ associated with $\omega_\phi\sim1$.  From Fig.\ref{maxv}, we can clearly see that for $\omega_\phi=0$, $v_{dm}^{max}$ for the combined case is dominated by the production from radiation bath and turns out to be independent of reheating temperature, $v_{dm}^{max}\sim 3.1\times 10^{-8}$.

\begin{figure}[t!]
\includegraphics[height=3.35cm,width=5.40cm]{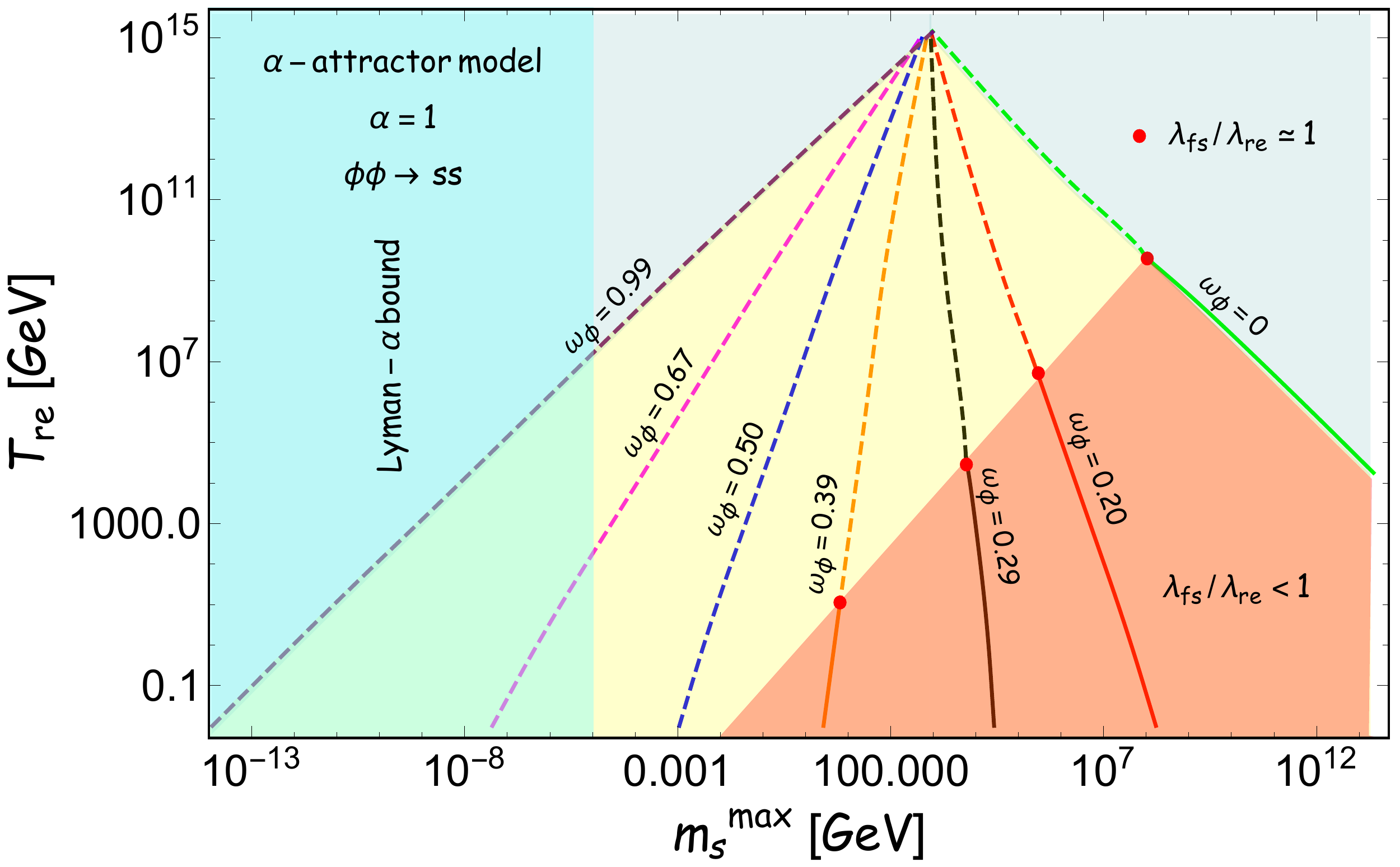}
\includegraphics[height=3.35cm,width=5.40cm]{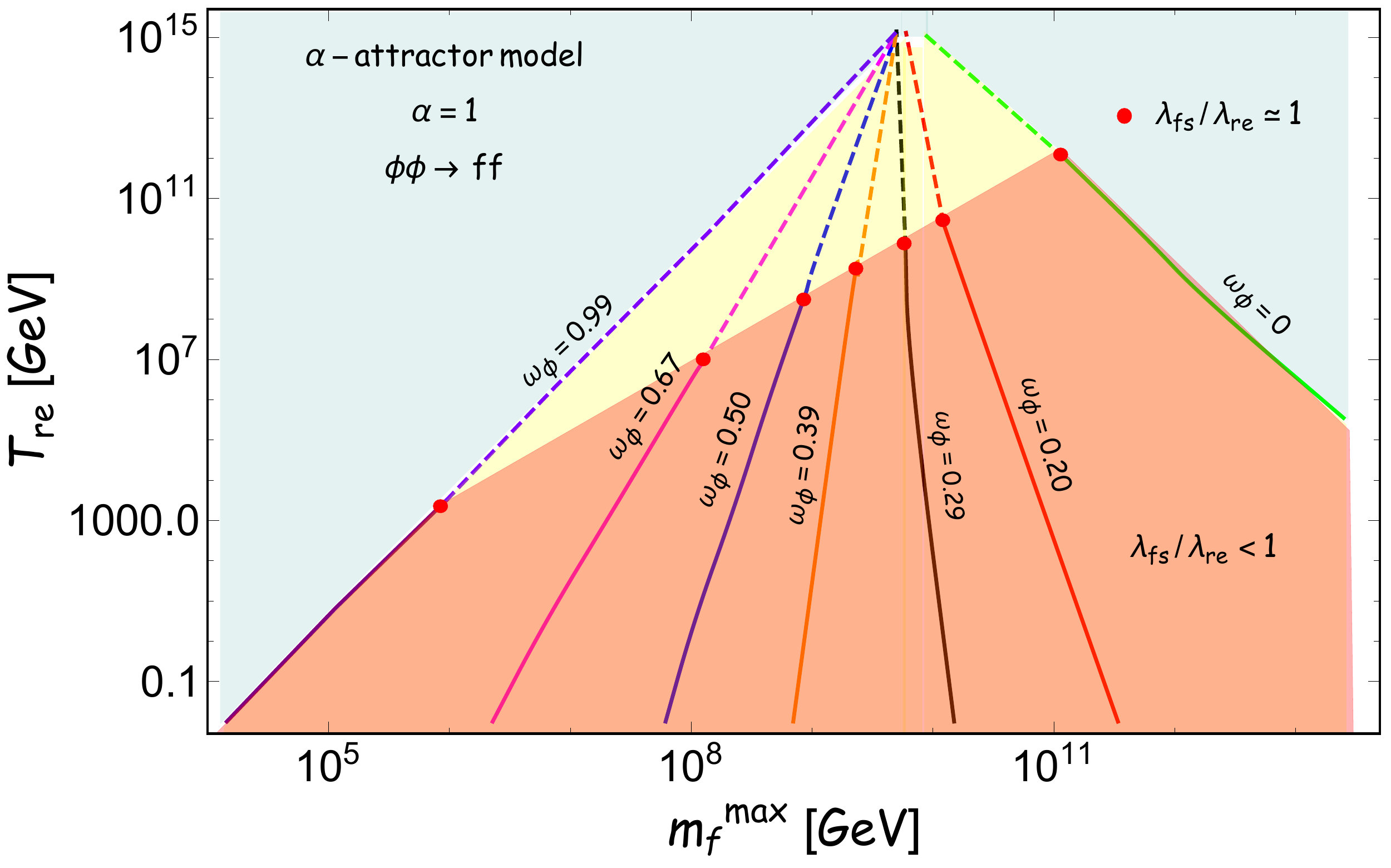}
\includegraphics[height=3.35cm,width=5.40cm]{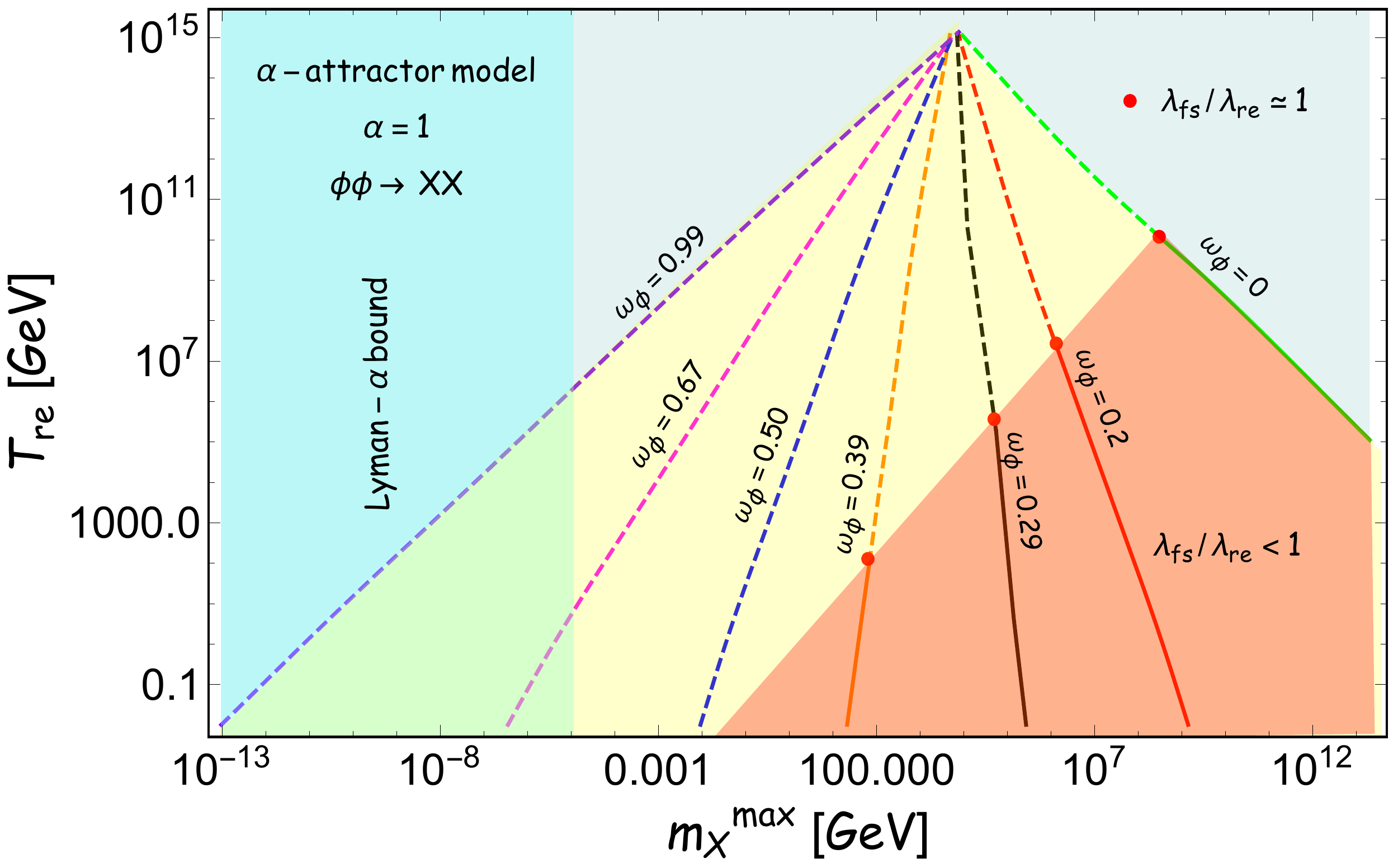}
\includegraphics[height=3.35cm,width=5.40cm]{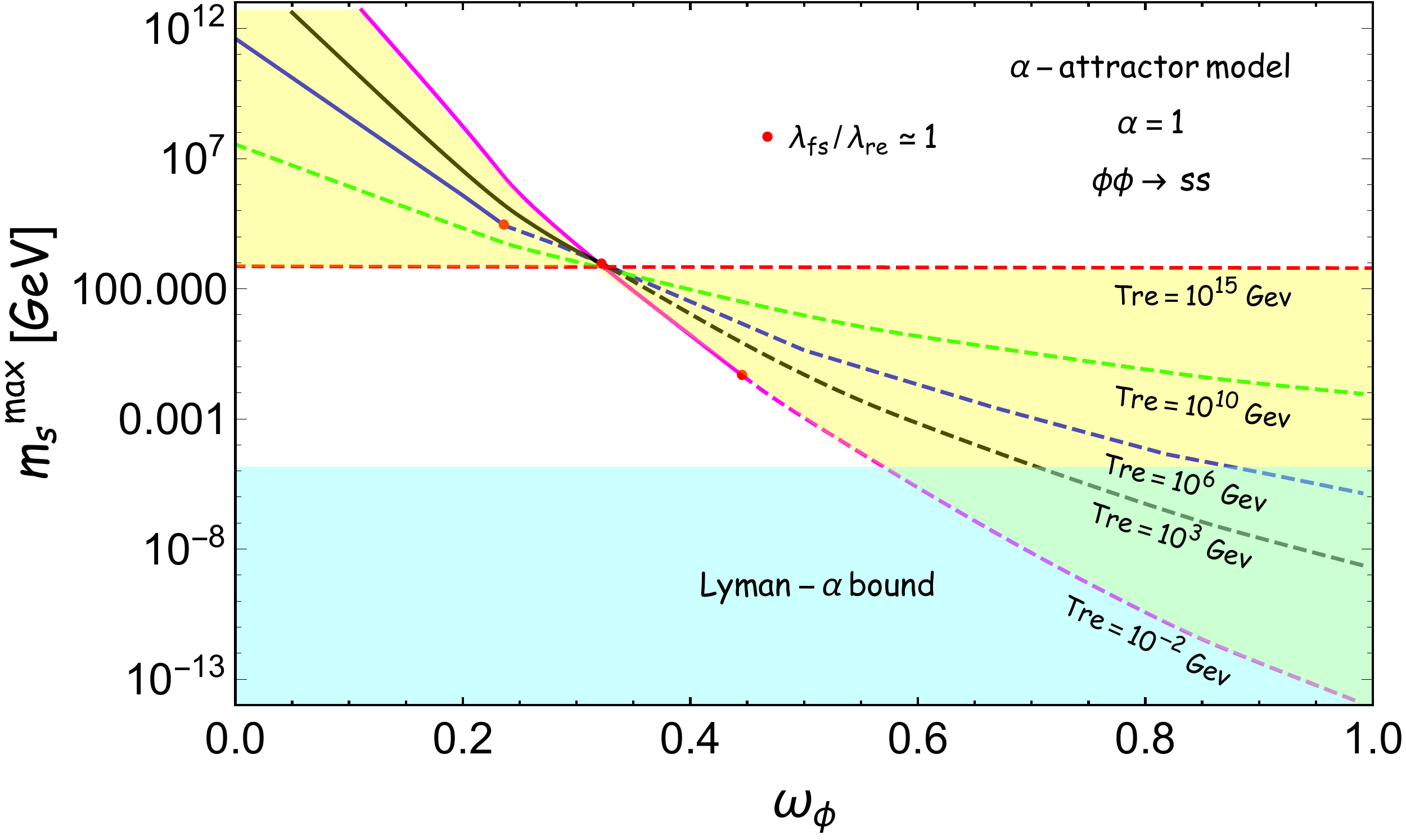}
\includegraphics[height=3.35cm,width=5.40cm]{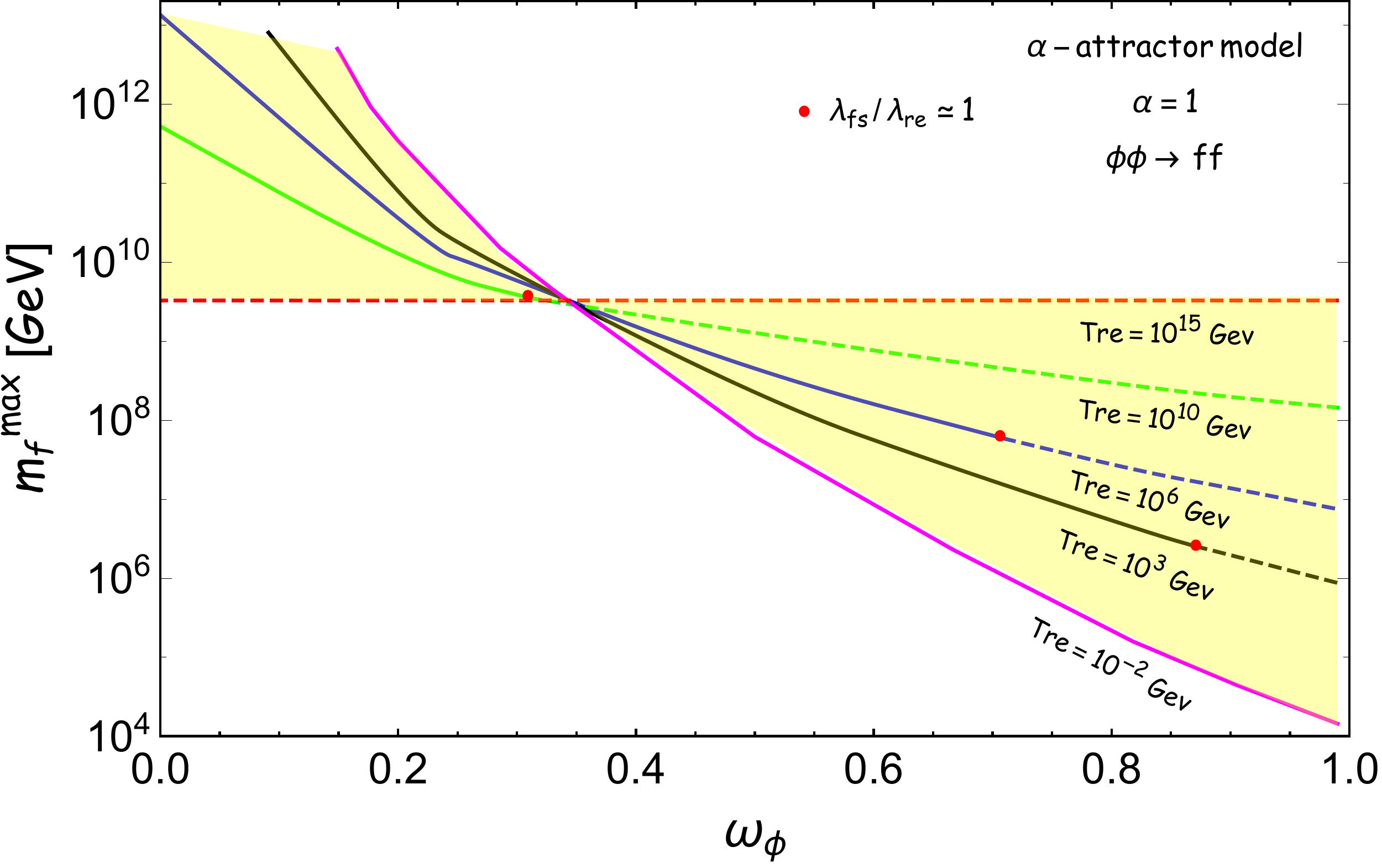}
\includegraphics[height=3.35cm,width=5.40cm]{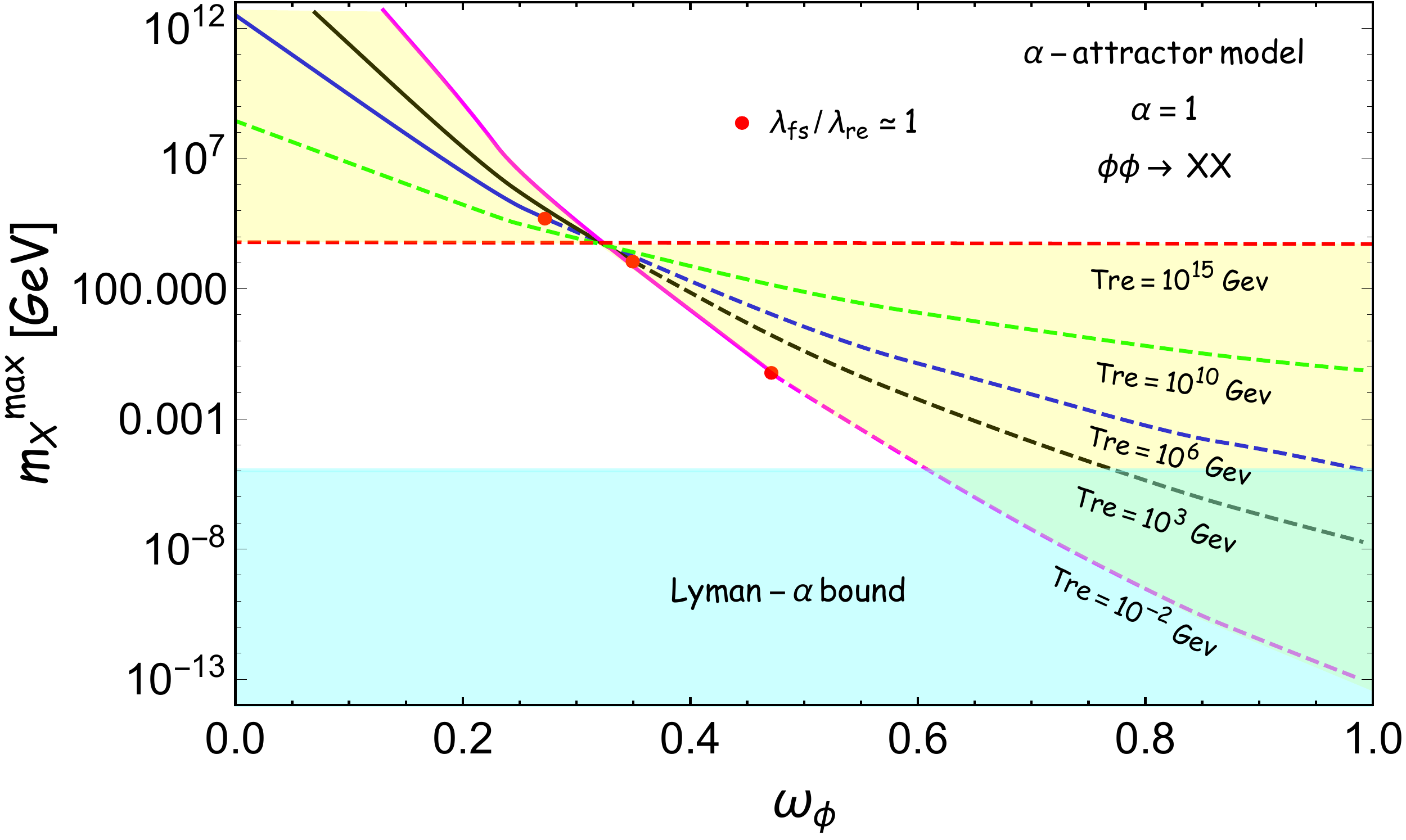}
\caption{{\bf Upper panel :} We have shown the region (indicated by orange color) in the $T_{re},\,m_Y^{max}$ plane, where the free-streaming effect does not hamper the small-scale structures formed during the reheating phase. The red circle corresponding to that point where $\lambda_{fs}/ \lambda_{re}\simeq1$. The other description of these figures is the same as fig.\ref{mmax}. {\bf Lower panel :} We have shown the parameters compatible with the condition $\lambda_{fs}/ \lambda_{re}<1$ through the solid line and dotted line for $\lambda_{fs}/ \lambda_{re}>1$ with different sets of reheating temperature. The additional description of these last three figures is given in fig.\ref{mfmaxomega}.} 
\label{free-streaming}
\end{figure}
\subsection{Free streaming of dark matter:} 
Understanding the free streaming behavior of the dark matter is important, as it plays crucial role in the process of structure formation. Larger the free steaming length, less probable will be to form the structure of around that length scale. If the dark matter particles have large initial momentum their free-streaming effect can erase the structure on scales smaller than the free-streaming horizon $\lambda_{fs}$. The free-streaming horizon strictly depends on the position where the dark matter particles decouple in the early universe. In this section we calculate the free-streaming horizon for different dark matter production scenarios during reheating. The free-streaming horizon is naturally related to the average momentum of the dark matter particles and can be approximately calculated by integrating from the time of decoupling $t_{kd}$ to the present $t_0$ as \cite{Bertschinger:2006nq,Green:2003un,Green:2005fa,Boyarsky:2008xj,Loeb:2005pm}
\bea \label{freestreaming}
\lambda_{fs}=k_{fs}^{-1}=\int_{t_{kd}}^{t_0}\,\frac{v}{a}\,dt=\int_{a_{kd}}^{a_0}\,\frac{\textit{p}}{E}\,\frac{da}{a^2\,H}=\int_{a_{kd}}^{a_0}\,\frac{\textit{p}}{\sqrt{\textit{p}^2+m_x^2}}\,\frac{da}{a^2\,H}~~,
\eea
where $a_{kd}$ represents scale factor associated with the decoupled time $t_{kd}$. Therefore, the Hubble parameter after reheating can be related with the current Hubble rate as,
\bea \label{Hubbleevolution}
H(\,a\,)=H_0\,\sqrt{\Omega_r\,a^{-4}+\,\Omega_m\,a^{-3}}=a^{-2}\,H_0\,\sqrt{\Omega_r}\,\sqrt{1+a/a_{eq}}~~.
\eea
Where the scale factor at the matter-radiation equality is identified as $a_{eq}=\Omega_r/\Omega_m$.  We ignore dark energy contribution to the expansion.

{\bf Standard Freeze-in from thermal bath:}
For comparison we consider this scenario first. Evaluation can be divided into two regime, the produced dark matter particles are relativistic after reheating ends $\textit{p}_{re}>>m_Y$, and as the universe expanses it becomes non-relativistic in nature $\textit{p}<<m_Y$. Therefore, the free-streaming length can be expressed as
\bea \label{free-streaming length radiation}
k_{fs}^{-1}=\int_{a_{re}}^{a_{nr}}\,\frac{da}{a^2\,H}+\,\int_{a_{nr}}^{a_0}\,\frac{\textit{p}}{m_Y}\,\frac{da}{a^2\,H}~~,
\eea
here $a_{nr}$ indicates the scale factor at the transition between two regimes where $\textit{p}_{nr}=m_Y$. In the regime where dark matter particles are relativistic, the contribution to the free-streaming length turns out as
\bea \label{relativistic radiation}
\int_{a_{re}}^{a_{nr}}\,\frac{da}{a^2\,H}=\frac{1}{H_0\,\sqrt{\Omega_r}}\,\int_{a_{re}}^{a_{nr}}\,\frac{da}{\sqrt{1+a/a_{eq}}}\simeq\frac{a_{re}}{H_0\,\sqrt{\Omega_r}}\,\left(\,\frac{a_{nr}}{a_{re}}-1\,\right)\simeq\frac{1}{k_{re}}\,\frac{\textit{p}_{re}}{m_Y}
\eea
To determine the above equation, we use the relation ${a_{nr}}/{a_{re}}={\textit{p}_{re}}/{\textit{p}_{nr}}={\textit{p}_{re}}/{m_Y}$, as after reheating the momentum associated with dark matter particles redshifts due to expansion. Further, considering $a_{re}\leq\,a\,\leq a_{nr}$, $1+a/a_{eq}\simeq1$ $(a<<a_{eq})$, the contribution to $k_{fs}^{-1}$ during the period when dark matter particles are non-relativistic (\,$\textit{p}<<m_Y$\,) becomes, 
\bea \label{non-relativistic radiation}
\int_{a_{nr}}^{a_0}\,\frac{\textit{p}}{m_Y}\,\frac{da}{a^2\,H}=\frac{\textit{p}_{re}\,a_{re}}{m_Y\,H_0\,\sqrt{\Omega_r}}\,\int_{a_{nr}}^{1}\,\frac{da}{a\,\sqrt{1+\,a/a_{eq}}}=\frac{2\,\textit{p}_{re}}{k_{re}\,m_Y}\,\left[sinh^{-1}\sqrt{\frac{a_{eq}}{a_{nr}}}-sinh^{-1}\sqrt{a_{eq}}\right]~.
\eea
In deriving the above equation, we used the relation $k_{re}\,a_{re}=H(\,a_{re}\,)\,a_{re}^2=H_0\,\sqrt{\Omega_r}\,$ (derived from Eqn.\ref{Hubbleevolution}). Upon substituting Eqns.(\ref{relativistic radiation}) and (\ref{non-relativistic radiation}) into Eqn.(\ref{free-streaming length radiation}), the expression for free-streaming length becomes,
\bea \label{free-streaming radfinal}
\lambda_{fs}=k_{fs}^{-1}\simeq\frac{1}{k_{re}}\,\frac{\textit{p}_{re}}{m_Y}\left[1+2\,\left\{\,sinh^{-1}\sqrt{\frac{a_{eq}}{a_{nr}}}-sinh^{-1}\sqrt{a_{eq}}\,\right\}\right]~~.
\eea
$k_{re} \sim 1/\lambda_{re}$ is associated with the typical length scale which will be entering during the end of reheating. Since, our starting assumption  is $\textit{p}_{re}>>m_Y$ , Eqn.(\ref{free-streaming radfinal}) indicates that ${\lambda_{fs}}/{\lambda_{re}} >1$ which implies that the free-streaming effect may erases the growth of the DM perturbations produced during the reheating phase  \cite{Erickcek:2011us,Erickcek:2015jza,Fan:2014zua}. 

Interestingly if the dark matter particles produced from the radiation bath is non-relativistic $\textit{p}_{re}<<m_Y$, 
\begin{equation} \label{free-streaming radfinal non-relativistic}
\lambda_{fs}\simeq\frac{2\,\textit{p}_{re}}{k_{re}\,m_Y}\left[\,sinh^{-1}\sqrt{\frac{a_{eq}}{a_{re}}}-sinh^{-1}\sqrt{a_{eq}}\,\right]\simeq\frac{2 \lambda_{re}  \textit{p}_{re}}{m_Y}\,sinh^{-1}\sqrt{\frac{T_{re}}{T_{eq}}}\simeq\frac{2\,\textit{p}_{re}}{k_{re}\,m_x}\,ln\left(\,2\,\sqrt{\frac{T_{re}}{T_{eq}}}\,\right)~.
\end{equation}
To derive the expression above, uses have been made of the relation ${a_{eq}}/{a_{re}}={T_{re}}/{T_{eq}}$, the approximation $sinh^{-1}x$  as $log_e(2\,x)$ in the limit of $x>>1$ and $sinh^{-1}\sqrt{{a_{eq}}/{a_{re}}} >sinh^{-1}\sqrt{a_{eq}}$. The condition for small scale structures of length scales $\lambda_{re}$ being formed if one satisfies 
\bea
\lambda_{re} > \lambda_{fs} \implies
\frac{T_{re}}{T_{eq}}\,<\,\frac{1}{4}\,e^{\frac{m_Y}{3\,T_{re}}}~~.
\eea 
Where, $T_{eq}\simeq 0.8$ eV at the radiation-matter equality. In addition above constraint can be converted into the constraint on the velocity of the dark matter particle during the end of reheating as, $v_{re}<\frac{1}{4\,ln\left(\,\frac{T_{re}}{T_{eq}}\,\right)}$. As an example, for reheating temperatures $T_{re}=(\,10^{-2},10^2,10^6\,)$ GeV, the upper bound on $v_{re}$ turns out as $v_{re}<(\,6\times10^{-2},10^{-2},7\times10^{-3}\,)$ accordingly.\\
\begin{figure}[t!]
\includegraphics[height=3.4cm,width=5.40cm]{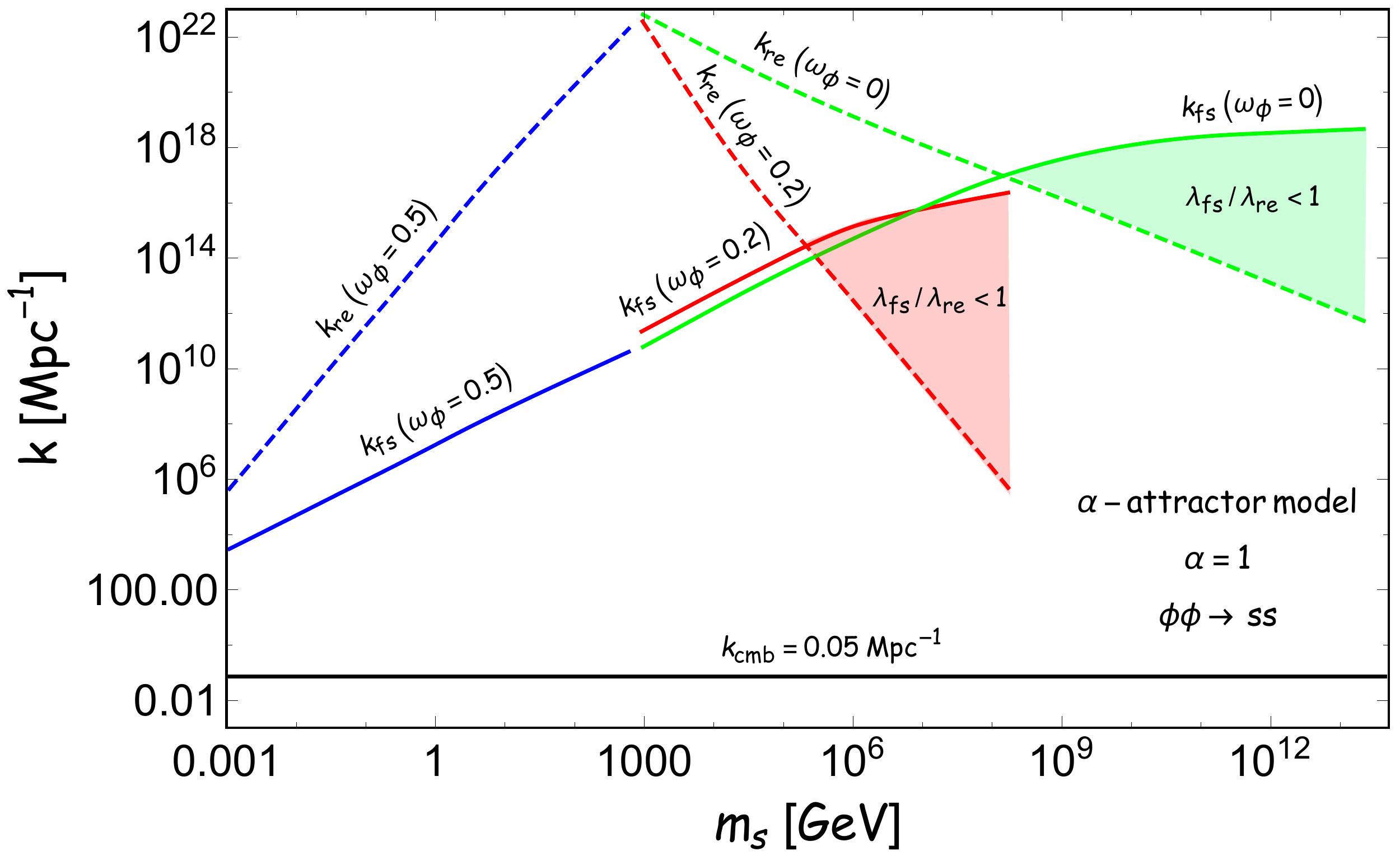}
\includegraphics[height=3.4cm,width=5.40cm]{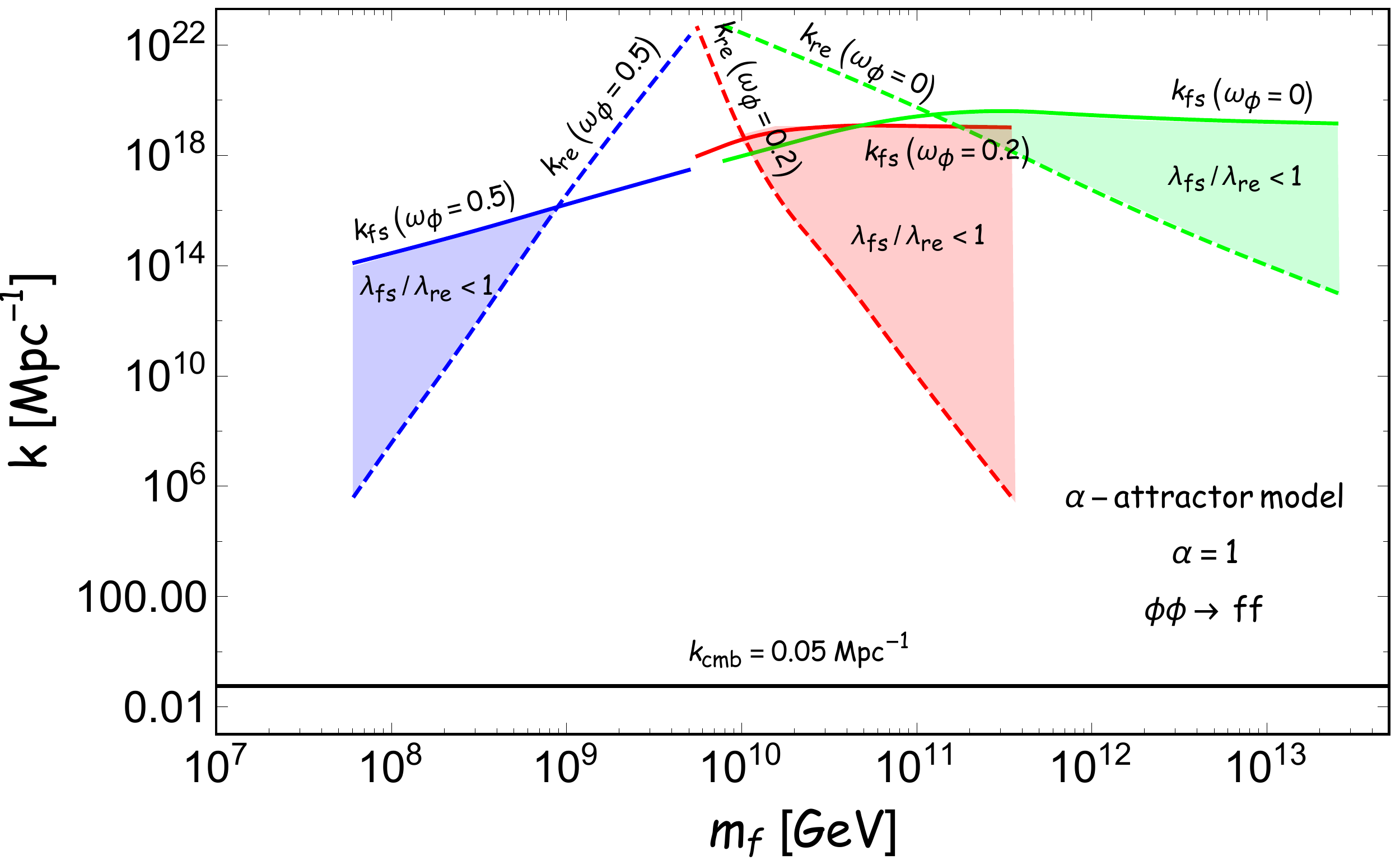}
\includegraphics[height=3.4cm,width=5.40cm]{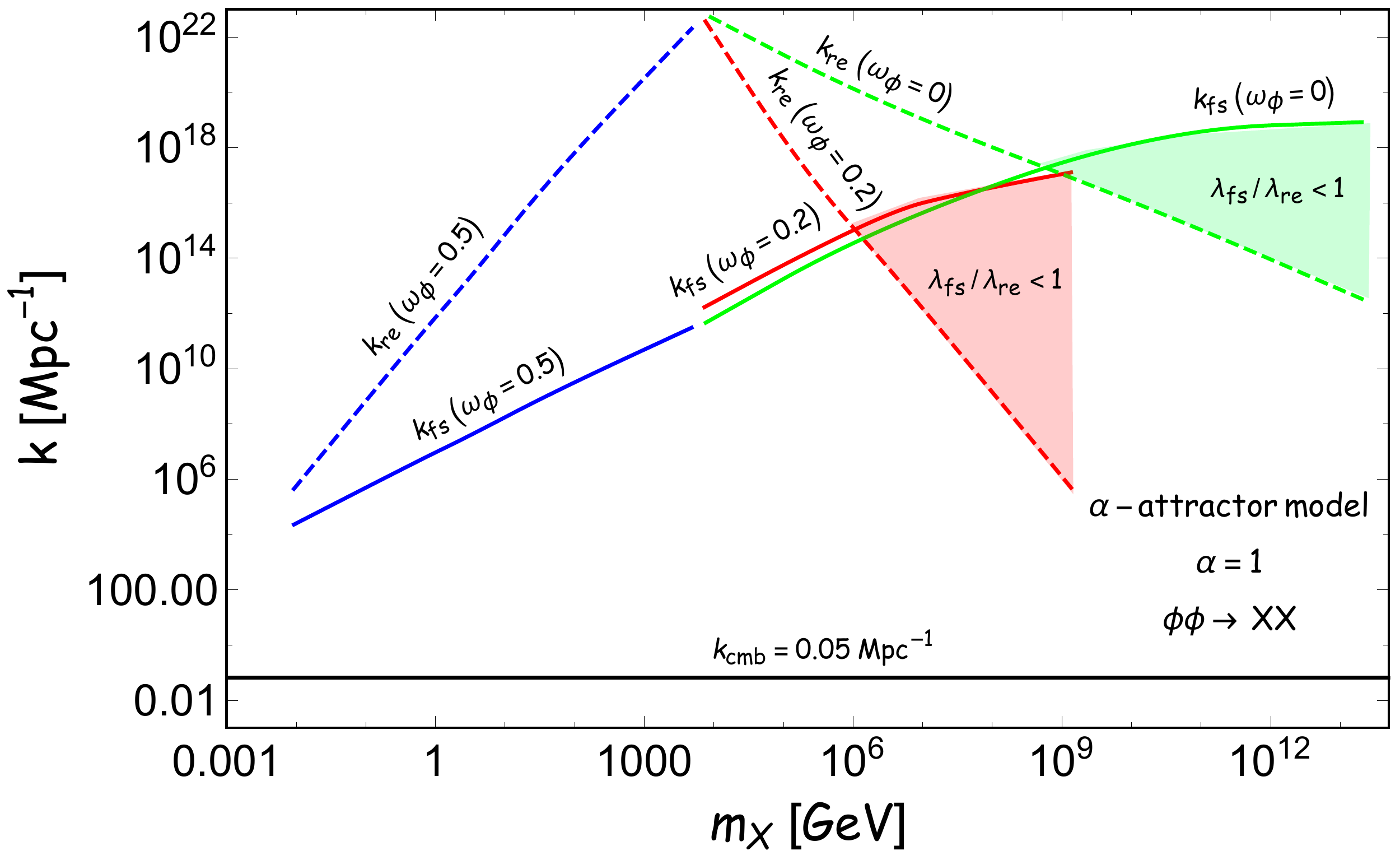}
\includegraphics[height=3.4cm,width=5.40cm]{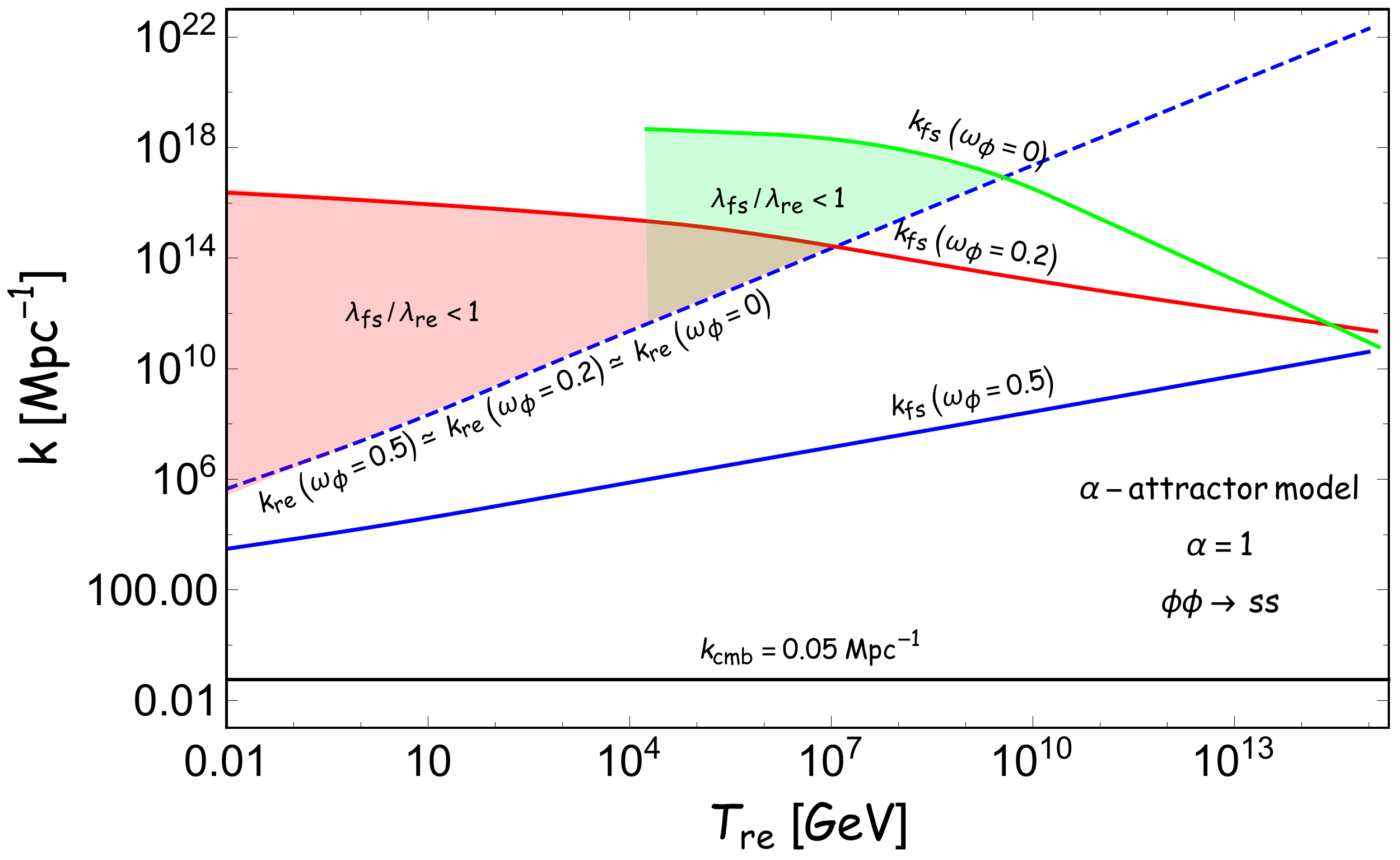}
\includegraphics[height=3.4cm,width=5.40cm]{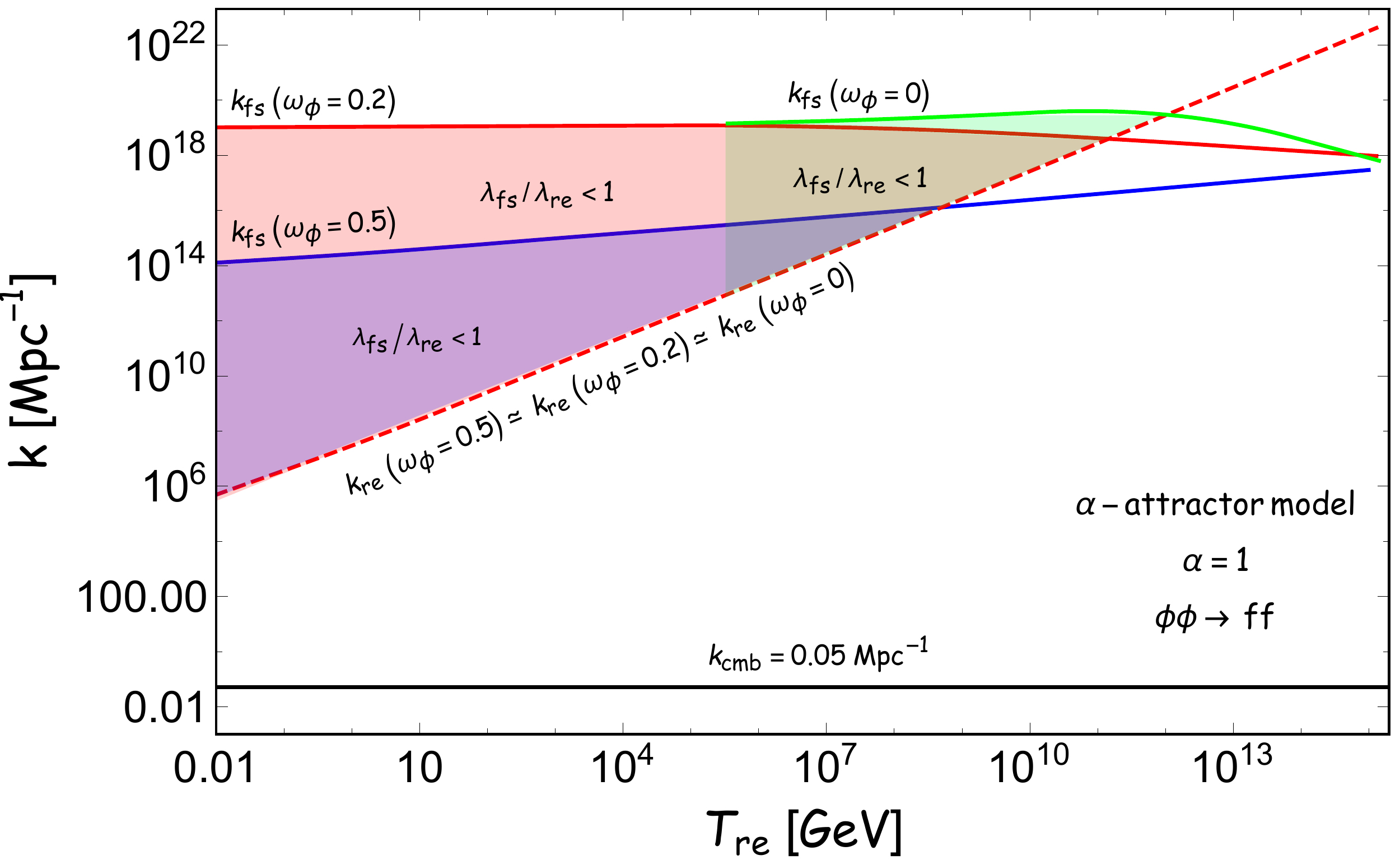}
\includegraphics[height=3.4cm,width=5.40cm]{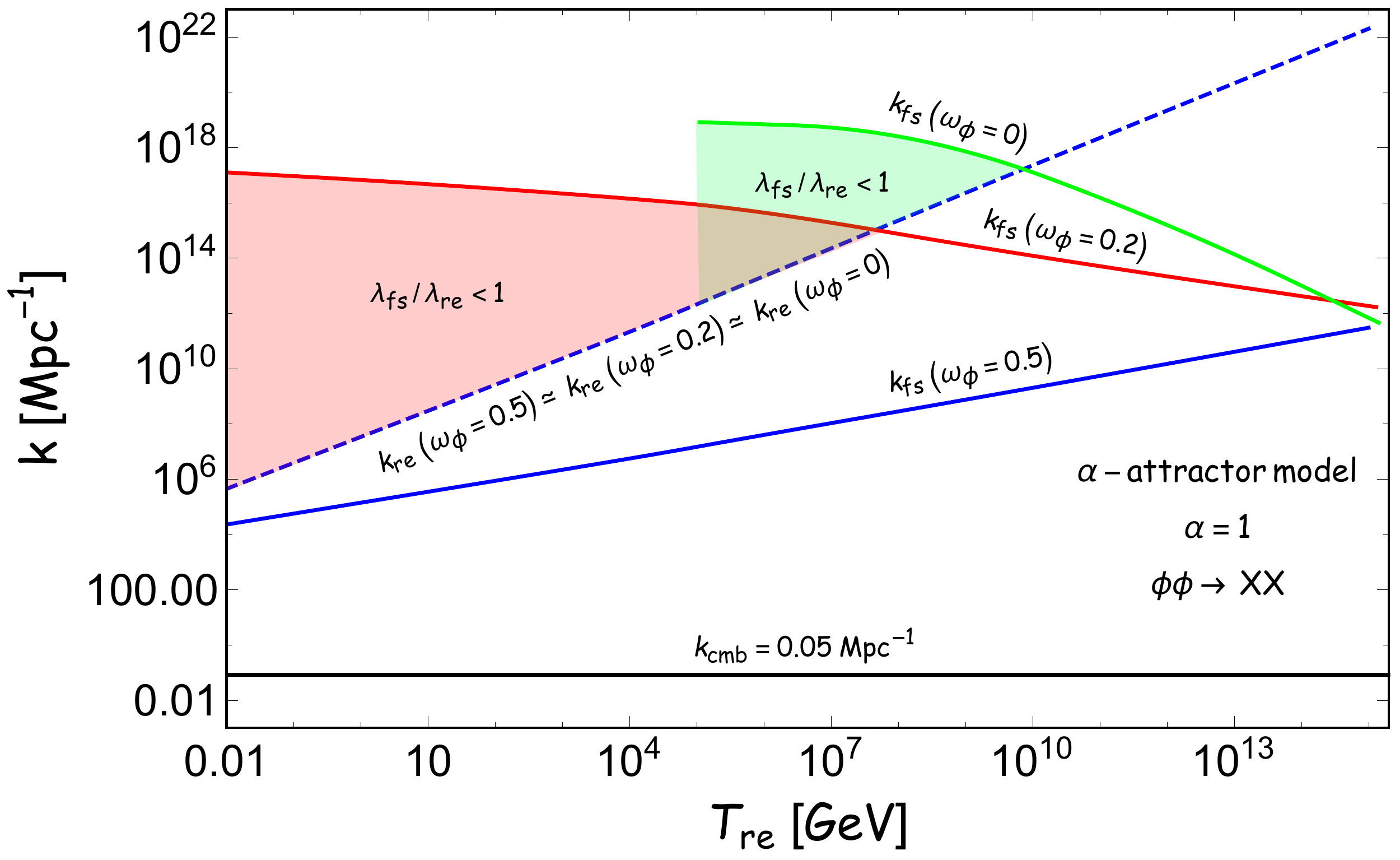}
\caption{We have plotted the variation of $k_{fs}$ and $k_{re}$ as a function of dark matter mass $m_Y$ and reheating temperature $T_{re}$ for three different gravitationally produced dark matter scenarios: $\phi\phi \to SS$ (scalar dark matter), $\phi \phi\to ff$ (fermionic dark matter) and $\phi \phi\to XX$ (vector dark matter) with three different inflaton equation of state $\omega_\phi=(0,\,0.2,\,0.5)$ (in green, red and blue). The shaded region indicates the parameter space in the $k-m_Y$ and $k-T_{re}$ plane, where the free-streaming length of DM particles does not erase structures on small scales formed during reheating era.}  
\label{kfree-streaming}
\end{figure}
{\bf Gravitational dark matter from inflaton:} As has been discussed earlier, the gravitationally produced dark matter from the inflaton mostly occurred at the beginning of the reheating when the temperature is approximately taken as maximum radiation temperature, $T_{in}=T_{rad}^{max}$. And the initial momentum of the DM particle would be the same as inflaton mass $\textit{p}_{in}=m_\phi$.  As the dark matter has no interaction with the radiation bath, the momentum of the dark matter particles decreases as $1/a$ after $a=a_{max}$.  Therefore, the free streaming will have non-trivial dependence on the reheating equation of state for this scenario. 
Considering dark matter particles are relativistic until the end end of the reheating, $\lambda_{fs}$ can be expressed as
\bea \label{free-streaming length gravitational}
\lambda_{fs}=k_{fs}^{-1}=\int_{a_{in}(a_{max})}^{a_{re}}\,\frac{da}{a^2\,H}+\,\int_{a_{re}}^{a_{nr}}\,\frac{da}{a^2\,H}+\,\int_{a_{nr}}^{a_0}\,\frac{\textit{p}}{m_x}\,\frac{da}{a^2\,H}~~.
\eea
In the reheating regime, the contribution to the free-streaming length is given by
\begin{equation} \label{free-streaming reheating}
\int_{a_{max}}^{a_{re}}\,\frac{da}{a^2\,H}=\frac{1}{H_{re}\,a_{re}^{\frac{3}{2}\,(\,1+\omega_\phi\,)}}\,\int_{a_{max}}^{a_{re}}\,a^{\frac{1}{2}\,(\,3\,\omega_\phi-1\,)}\,da=\frac{2}{(\,1+3\,\omega_\phi\,)\,k_{re}}\,\left[\,1-\left(\,\frac{a_{max}}{a_{re}}\,\right)^{\frac{1}{2}\,(\,3\,\omega_\phi+1\,)}\,\right]~~,
\end{equation}
where $H_{re}$ represents Hubble parameter at the end of the reheating. To determine the above equation, we assume the variation of the Hubble parameter during reheating phase as $H(\,a\,)=H_{re}\,\left(\, \frac{a_{re}}{a}\,\right)^{\frac{3}{2}\,(\,1+\omega_\phi\,)}$, under the assumption that the reheating phase is dominating by the inflaton equation of state $\omega_\phi$. In addition to that, for the perturbative reheating scenario, the scale factor at the point of maximum radiation temperature is calculated as 
\bea
a_{max}=a_{end}\,\left(\,\frac{8}{5-3\,\omega_\phi}\,\right)^{\frac{2}{5-3\,\omega_\phi}},\,a_{end}=\frac{e^{N_k}\,k}{H_k}~~,
\eea
where the scale factor at the end of the inflation symbolizes by $a_{end}$. Combining Eqns.(\ref{relativistic radiation}), (\ref{non-relativistic radiation}), (\ref{free-streaming length gravitational}), and (\ref{free-streaming reheating}), we obtain the expression of the free-streaming length for gravitationally produced dark matter as follows
\begin{equation} \label{free-streaming gravitational}
\lambda_{fs}\simeq\frac{1}{k_{re}}\left[\,\frac{2}{1+3\,\omega_\phi}\,\left\{\,1-\left(\,\frac{a_{max}}{a_{re}}\,\right)^{\frac{1}{2}\,(\,3\,\omega_\phi+1\,)}\,\right\}+\,\frac{\textit{p}_{re}}{m_x}\left\{1+2\left(\,sinh^{-1}\sqrt{\frac{a_{eq}}{a_{nr}}}-sinh^{-1}\sqrt{a_{eq}}\,\right)\right\}\,\right]\,.
\end{equation}
In this scenario, when gravitationally produced dark matter particles are relativistic at the time of production as well as at the end of the reheating, Eqn.\ref{free-streaming gravitational} indicates that $\lambda_{fs}/\lambda_{re} > 1$. On the other hand if the gravitationally produced dark matter particles are relativistic at the time of production but become non-relativistic at the time of reheating end  $(p_{re}=m_Y)$,  one obtains
\begin{eqnarray}
\frac{\lambda_{fs}}{\lambda_{re}}\simeq\frac{2}{1+3\,\omega_\phi}\,\left\{\,1-\left(\,\frac{a_{max}}{a_{re}}\,\right)^{\frac{1}{2}\,(\,3\,\omega_\phi+1\,)}\,\right\}+\,\frac{2\,\textit{p}_{re}}{m_x}\,\left(\,sinh^{-1}\sqrt{\frac{a_{eq}}{a_{re}}}-sinh^{-1}\sqrt{a_{eq}}\,\right)
\end{eqnarray}
Since ${a_{max}}/{a_{re}}<<1$, and with the help of equation \ref{free-streaming radfinal non-relativistic}, the ratio ${\lambda_{fs}}/{\lambda_{re}}$ can be approximately expressed as
\bea
\frac{\lambda_{fs}}{\lambda_{re}}\simeq\frac{2}{1+3\,\omega_\phi}+\,2\,ln\left(\,2\,\sqrt{\frac{T_{re}}{T_{eq}}}\,\right)~~.
\eea
For ${\lambda_{fs}}<{\lambda_{re}}$, constraints on the reheating temperature $T_{re}$ will be
\bea
T_{re}<\,\frac{1}{4}\,T_{eq}\,e^{\frac{3\,\omega_\phi-1}{3\,\omega_\phi+1}}~~.
\eea
From the above equation, we can clearly notice that the bound on reheating temperature turns out as $T_{re}<<10^{-2}$ GeV, which violates the BBN constraints. Therefore, we can conclude that if the DM particles are relativistic until the reheating end, the free steaming length will be large enough to suppress the small-scale structure naturally. 

Similar to the previous case, if gravitationally produced dark matter particles become non-relativistic any time during the reheating, 
\bea \label{grav reheat non-re}
\lambda_{fs}=\int_{a_{in}(a_{max})}^{a_{nr}}\,\frac{da}{a^2\,H}+\,\int_{a_{nr}}^{a_{re}}\,\frac{\textit{p}}{m_x}\,\frac{da}{a^2\,H}+\,\int_{a_{re}}^{a_0}\,\frac{\textit{p}}{m_x}\,\frac{da}{a^2\,H}~~.
\eea
The first term on the right-hand side of the above equation evaluated as
\bea \label{gravrelativistic}
\int_{a_{max}}^{a_{nr}}\,\frac{da}{a^2\,H}\simeq\frac{2}{(\,1+3\omega_\phi\,)\,k_{re}}\,\left[\,\left(\,\frac{a_{nr}}{a_{re}}\,\right)^{\frac{1}{2}\,(\,3\,\omega_\phi+1\,)}-\,\left(\,\frac{a_{max}}{a_{re}}\,\right)^{\frac{1}{2}\,(\,3\,\omega_\phi+1\,)}\,\right]~~,
\eea 
where
\bea \label{scalefactorratio}
\frac{a_{nr}}{a_{re}}=\frac{\textit{p}_{re}}{\textit{p}_{nr}}=\frac{\textit{p}_{re}}{m_x}~~,~~\frac{a_{max}}{a_{re}}=\frac{\textit{p}_{re}}{\textit{p}_{in}}=\frac{\textit{p}_{re}}{m_x}\,\frac{m_x}{m_\phi}~~.
\eea
Upon substitution of Eqn.(\ref{scalefactorratio}) into Eqn.(\ref{gravrelativistic}) and futher considering $m_\phi>>m_Y$, one finds
\bea\label{gravrelativistic final}
\int_{a_{max}}^{a_{nr}}\,\frac{da}{a^2\,H}\simeq\frac{2}{(\,1+3\omega_\phi\,)\,k_{re}}\,\left(\,\frac{\textit{p}_{re}}{m_x}\,\right)^{\frac{1}{2}\,(\,3\,\omega_\phi+1\,)}~~.
\eea
Accordingly, the second term on the right-hand side of Eqn.(\ref{grav reheat non-re}) estimated  as
\begin{equation}\label{grav non-relativistic}
\,\int_{a_{nr}}^{a_{re}}\,\frac{\textit{p}}{m_x}\,\frac{da}{a^2\,H}\simeq \frac{\textit{p}_{re}\,a_{re}}{m_x\,H_{re}\,a_{re}^{\frac{3}{2}\,(\,1+\omega_\phi\,)}}\int_{a_{nr}}^{a_{re}}\,a^{\frac{3}{2}\,(\,\omega_\phi-1\,)}\,da=\frac{2}{(\,3\,\omega_\phi-1\,)}\,\frac{\textit{p}_{re}}{k_{re}\,m_x}\,\left[\,1-\,\left(\frac{\textit{p}_{re}}{m_x}\right)^{\frac{1}{2}\,(\,3\,\omega_\phi-1\,)}\right]\,.
\end{equation}
Therefore, connecting Eqns.(\ref{grav reheat non-re}), (\ref{gravrelativistic final}), (\ref{grav non-relativistic})  and (\ref{free-streaming radfinal non-relativistic}), one can find the following expression of free-streaming length 
\bea \label{free grav final}
\lambda_{fs}\simeq\lambda_{re}\left[\frac{4}{1-9\,\omega_\phi^2}\,\left(\frac{\textit{p}_{re}}{m_x}\right)^{\frac{1}{2}\,(\,1+3\,\omega_\phi\,)}+\frac{\textit{p}_{re}}{m_x}\,\left\{\,\frac{2}{3\,\omega_\phi-1}+\,2\,ln\left(\,2\,\sqrt{\frac{T_{re}}{T_{eq}}}\,\right)\,\right\}\,\right]
\eea
For this case the condition ${\lambda_{fs}}/{\lambda_{re}}<1$, will lead to following constraint relation among the inflaton equation of state,  dark matter mass, and reheating temperature must follow the relation
\begin{equation} \label{temp restriction grav}
\left(\frac{T_{re}}{T_{rad}^{max}}\right)^{\frac{8}{3\,(\,1+\,\omega_\phi\,)}}\left[\frac{4}{1-9\,\omega_\phi^2}\left(\frac{T_{re}}{T_{rad}^{max}}\right)^{\frac{4\,(3\,\omega_\phi-1)}{3\,(\,1+\,\omega_\phi\,)}}\left(\,\frac{m_\phi}{m_x}\,\right)^{\frac{\,3\,\omega_\phi-1}{2}}+\frac{2}{3\,\omega_\phi-1}+2\,ln\left(2\,\sqrt{\frac{T_{re}}{T_{eq}}}\,\right)\right]<\frac{m_x}{m_\phi}\,.
\end{equation}
The above constraint can be further transformed into constraint on the velocity of dark matter particles as,
\bea \label{velocity grav}
\frac{4}{1-9\,\omega_\phi^2}\,v_{re}^{\frac{1+3\,\omega_\phi}{2}}+\,v_{re}\,\left\{\,\frac{2}{3\,\omega_\phi-1}+\,2\,ln\left(\,2\,\sqrt{\frac{T_{re}}{T_{eq}}}\,\right)\,\right\}<1~~.
\eea
\begin{table}[t!]
	\caption{Different inflaton equation of state and reheating temperature (Measured in GeV), associated bound on $v_{re}$, considering purely gravitational production of dark matter.\\}
  \begin{tabular}{|p{1.58cm}|p{1.3cm}|p{1.3cm}|p{1.3cm}|p{1.3cm}|p{1.3cm}|p{1.3cm}|p{1.3cm}|p{1.3cm}|p{1.3cm}}
\hline
Parameter &\multicolumn{3}{c|}{$\omega_\phi=0$} &\multicolumn{3}{c|}{$\omega_\phi=0.2$}&\multicolumn{3}{c|}{$\omega_\phi=0.5$}\\
\cline{2-10}
~& \multicolumn{1}{c|}{$T_{re}=10^{-2}$} & \multicolumn{1}{c|}{$T_{re}=10^{3}$} & \multicolumn{1}{c|}{$T_{re}=10^{6}$} &  \multicolumn{1}{c|}{ $T_{re}=10^{-2}$}&\multicolumn{1}{c|}{$T_{re}=10^{3}$}& \multicolumn{1}{c|}{$T_{re}=10^{6}$} & \multicolumn{1}{c|}{$T_{re}=10^{-2}$}&\multicolumn{1}{c|}{$T_{re}=10^{3}$}& \multicolumn{1}{c|}{$T_{re}=10^{6}$}\\
\hline

\quad $v_{re}^{max}$& \multicolumn{1}{c|}{$0.024$}& \multicolumn{1}{c|}{$0.017$}& \multicolumn{1}{c|}{$0.015$}& \multicolumn{1}{c|}{$0.040$}& \multicolumn{1}{c|}{$0.027$}& \multicolumn{1}{c|}{$0.022$}& \multicolumn{1}{c|}{$0.049$}& \multicolumn{1}{c|}{$0.031$}& \multicolumn{1}{c|}{$0.026$}\\
\hline
 \end{tabular}\\[.2cm]
  
  \label{maxvel}
\end{table}
\begin{table}[t!]
	\caption{Different reheating temperature (measured in units of GeV), associated limits on the inflaton equation of state and dark matter mass $m_{Y}$ (measured in units of GeV), emerging from the free-streaming effect.\\}
  \begin{tabular}{|p{1.75cm}|p{1.5cm}|p{1.5cm}|p{1.5cm}|p{1.5cm}|p{1.5cm}|p{1.5cm}|p{1.5cm}|p{1.5cm}|p{1.5cm}}
\hline
Parameters &\multicolumn{3}{c|}{$T_{re}=10^{-2}$} &\multicolumn{3}{c|}{$T_{re}=10^3$}&\multicolumn{3}{c|}{$T_{re}=10^6$}\\
\cline{2-10}
~& \multicolumn{1}{c|}{$\phi\phi\to SS$} & \multicolumn{1}{c|}{$\phi\phi\to ff$} & \multicolumn{1}{c|}{$\phi\phi\to XX$} &  \multicolumn{1}{c|}{ $\phi\phi\to SS$}&\multicolumn{1}{c|}{$\phi\phi\to ff$}& \multicolumn{1}{c|}{$\phi\phi\to XX$} & \multicolumn{1}{c|}{$\phi\phi\to SS$}&\multicolumn{1}{c|}{$\phi\phi\to ff$}& \multicolumn{1}{c|}{$\phi\phi\to XX$}\\
\hline
\quad $\omega_\phi^{max}$& \multicolumn{1}{c|}{$0.45$}& \multicolumn{1}{c|}{$1.00$}& \multicolumn{1}{c|}{$0.47$}& \multicolumn{1}{c|}{$0.33$}& \multicolumn{1}{c|}{$0.87$}& \multicolumn{1}{c|}{$0.35$}& \multicolumn{1}{c|}{$0.24$}& \multicolumn{1}{c|}{$0.71$}& \multicolumn{1}{c|}{$0.27$}\\
\hline
$m_{Y}^{max}$(min)& \multicolumn{1}{c|}{$10^{-2}$}& \multicolumn{1}{c|}{$1.4\times 10^4$}& \multicolumn{1}{c|}{$2.0\times10^{-2}$}& \multicolumn{1}{c|}{$650$}& \multicolumn{1}{c|}{$2.5\times10^6$}&  \multicolumn{1}{c|}{$10^{3}$}& \multicolumn{1}{c|}{$2.0\times10^{4}$}&  \multicolumn{1}{c|}{$6.0\times 10^{7}$}& \multicolumn{1}{c|}{$4.0\times10^{4}$}\\
\hline
  \end{tabular}\\[.2cm]
  
  \label{free-streaming constraints}
\end{table}
We now have all the necessary analytical along with the numerical results to understand the region in the parameter space of reheating temperature and dark matter mass and inflaton equation of state.  The condition $\lambda_{fs}/\lambda_{re}<1$ is expected to play important role in the formation of small-scale structures. As one would expect, the effect of free-streaming on the DM structures of length scale above the free-streaming horizon should be negligible.  The numerical value of scales around which free streaming may have an effect can be estimated from Fig.\ref{kfree-streaming} (shaded region) as a function of dark matter mass (upper panel) and reheating temperature (lower panel) for different kinds of dark matter particles with three distinct values of the inflaton equation of state $\omega_\phi=(0,\,0.2,\,0.5)$.  As an example, the permitted range of scales to sustain small scale structure lies in between $\{(5\times 10^{11},\,5\times 10^{18}),\,(5\times 10^{5},\,2\times 10^{16})\}$ $\mbox{Mpc}^{-1}$ for scalar dark matter, $\{(3\times 10^{12},\,8\times10^{18}),\,(5\times 10^{5},\,10^{17})\}$ $\mbox{Mpc}^{-1}$ for vector dark matter and  $\{( 10^{13},\,10^{19}),\,(5\times 10^{5},\,10^{19})\}$ $\mbox{Mpc}^{-1}$ for fermionic dark matter with EoS $\omega_\phi=(0,\,0.2)$ accordingly. Moreover, for $\omega_\phi=0.5$, there is no allowed range of scales above the free-streaming horizon for scalar and vector dark matter, whereas, for fermionic dark matter, the permitted range lies within $( 5\times10^{5},\,10^{14})$ $\mbox{Mpc}^{-1}$.  We should mention at this point that the detailed effects of free-streaming can be understood from the dynamics of the DM perturbation, which we will study in the future.  Anyway, free-streaming effects also impose constraints on the reheating and dark matter parameters $T_{re},\,\omega_\phi$, and $m_Y^{max}$, shown in fig.\ref{free-streaming}. In the upper three plots of Fig.\ref{free-streaming} the brown shaded region corresponds to $\lambda_{re} >\lambda_{fs}$. Therefore, any observation of small-scale DM halos will discard the yellow shaded regions and put constraints on the upper bound on reheating temperature.    
 For example, upper limit on the reheating temperature will be brought down from $10^{15}$ GeV $\to (3.7\times 10^9,\,10^{12},\,1.0\times10^{10})$ GeV for $\phi\phi\to SS/ff/XX$ for inflaton EoS $\omega_\phi=0$. Gravitational production has a one-to-one correspondence between reheating temperature and DM mass.  Hence, upper limit on reheating temperature lead to lower limit on the maximum possible DM mass $m_Y^{max}$ as $(9.0\times10^2,\,10^{10},\,7\times10^3)$ GeV $\to(10^8,\,10^{11},\,3.0\times 10^8)$ GeV for $\phi\phi\to SS/ff/XX$ respectively. The details constraints on the $T_{re}$ and $m_Y^{max}$ for different sets of the inflaton equation of state can be read off from the fig.\ref{free-streaming} (first three plot).  In the last three plots of fig.\ref{free-streaming}, we observe the possible constraints on the inflaton equation of state $\omega_\phi$ and $m_Y^{max}$ due to free-streaming effect for different sets of $T_{re}$. The numerical values of the possible limitation on $\omega_\phi,\, m_{Y}^{max}$ for three distinct reheating temperatures $T_{re}=(10^{-2},\,10^3,\,10^6)$ GeV are provided in Table -\ref{free-streaming constraints}. In addition to that, in Table-\ref{maxvel}, we have shown the possible constraints on the maximum DM velocity at the end of reheating for different sets of $(T_{re},\,\omega_\phi)$ .\\ At the end, we would like to point out that during the reheating phase, there is a growth in DM density perturbation due to gravitational instability.  The early DM microholes can be formed from that enhanced perturbation if the free-steaming length is smaller than the horizon.  This growth in perturbation modified the dark matter annihilation rate by several orders \cite{Erickcek:2015jza} and strictly depended on the microhalos' formation time.  Our eventual plan in the future is to study the growth of the dark matter perturbation in the present context.

\section{conclusions}
In this paper, our focus is on the two main topics of DM phenomenology.  In the first half, we studied the production of DM matter from the decay of inflaton mediated by gravitational interaction.  For completeness, we also include the production from radiation bath.  This is the reason in the $(\langle \sigma v\rangle~\mbox{Vs} ~m_Y)$ parameter space the gravitationally produced dark matter appeared to have unique mass value $m_Y^{max}$ (see Figs.\ref{scalarcrosssection3}, \ref{alpha1}) for which the present dark matter abundance is satisfied. The value of $m_Y^{max}$ is uniquely determined by the inflationary energy scale $H_{end}$, and inflaton effective equation of state during reheating $\omega_{\phi}$ (see Fig.\ref{mmax}, \ref{mfmaxomega}), which are expressed in Eq.\ref{mymax}. We studied the constraint on the DM mass considering vector, scalar, and fermion type dark matter considering both CMB power spectrum and the dark matter abundance.  For bosonic dark matter the observationally viable mass range turned out to be within $(10^{13} - 10^{-8})$ GeV. Therefore, gravitationally produced dark matter of mass in the eV range can be identified as axion field.  However, in order to obtain such a low bosonic dark matter mass through gravitational production, we found that reheating equation of state needs to be closed to unity which is equivalent to kination domination.  We will study this fact in detail in the future.  For fermionic dark matter mass range turned out to be $m_f^{max} = (10^{13} - 10^{4})$ GeV. Importantly, it is observed that allowed DM mass range shrinks to a point as $\omega_{\phi}$ approaches towards 1/3, which are clearly observed in Fig.\ref{mfmaxomega}. We have discussed single component and two-component dark matter scenarios and discussed the constraints on the dark matter parameters consistent with both CMB and dark matter abundance.

In the second half of the paper, we discussed the phase space distribution and the free streaming properties.  These are the properties that are believed to capture the microscopic properties of DM.  The formation of structure at all scales is crucially dependent on these intrinsic properties of the DM, which has gained interest in the recent past.  The phase-space distribution has been shown to be crucially dependent on the production mechanism and the background dynamics (see Fig.\ref{momentum distribution zero}).  The bosonic DM phase-space distribution function contains the equation of state-dependent peak at the initial moment of dark matter production, and the associated momentum of the particle is of the order of inflaton mass with which DM particles will subsequently start to free stream.  Interestingly the fermionic phase-space distribution function contains an additional peak in the later time, which arises due to fermionic decay width $\Gamma_{\phi\phi\rightarrow ff}$ non-trivially depending upon the decaying inflaton mass.  This secondary peak height is naturally dependent upon the reheating temperature; as the reheating temperature reduces, the peak height increases, which can be observed in Fig.\ref{momentum distribution zero}. 
Considering free streaming horizon, we have divided the allowed range of dark matter mass in terms of $T_{re}$ and $\omega_{\phi}$ (see Fig.\ref{free-streaming}) into two sub-ranges for $\lambda_{re} >\lambda_{fs}$ depicted by the brown shaded region in the upper panel and solid lines in the lower panel and for $\lambda_{re} <\lambda_{fs}$ depicted by the yellow shaded region in the upper panel and dotted lines in the lower panel. Finally, in Fig.\ref{kfree-streaming}, we plotted allowed ranges of scales associated with the free-streaming horizon around which small DM halos can be formed.  Shaded regions correspond to $\lambda_{fs} <\lambda_{re}$ which indicates that due to gravitational pull, small scale DM halos can be formed associated with those scales during reheating. If those small-scale structures are detected, DM matter mass parameter space, inflaton equation of state, and reheating temperature will be significantly constrained. 
\vskip1cm

{\bf \large Appendix}
\section{Analytic expression of maximum dark matter mass $m_Y^{max}$}\label{maxdark matter}
The expression for the relic abundance Eq.\ref{darkmatter relic} indicates that the dark matter abundance increases with increasing the dark matter mass. Consequently, there should exist a maximum allowed dark matter mass $m_{Y}^{max}$ associated with each viable value of the spectral index or reheating temperature. The evolution of the gravitationally produce dark matter number density follows form the equation
\bea \label{Boltzgrav}
d\,(\,n_{Y}\,a^3\,)=\frac{\Gamma_{\phi\phi\to YY}}{m_\phi}\,\frac{\rho_\phi\,(\,1+\,\omega_\phi\,)}{H}\,a^2\,da~~.
\eea
{\bf Comoving number density of scalar dark matter:} The comoving number density at the end of the reheating era is followed by the equations (\ref{decaywidthscalar}), (\ref{Boltzgrav}) and found  to be 
\begin{equation}\label{gravnumber}
n_{s}^{re}\,A_{re}^3=\int_{1}^{A_{re}}\frac{\rho_\phi^2\,(\,1+\,\omega_\phi\,)}{1024\,\pi\,M_p^4}\,\left(\,1+\,\frac{m_s^2}{2\,m_\phi^2}\,\right)\,\sqrt{1-\,\frac{m_s^2}{m_\phi^2}}\,\frac{A^2\,dA}{H}\,\approx \int_{1}^{A_{re}}\frac{\rho_\phi^2\,(\,1+\,\omega_\phi\,)}{1024\,\pi\,M_p^4}\,\frac{A^2\,dA}{H}~~.
\end{equation}
Ignoring the sub dominated effect of the dark matter production into the evolution of the inflaton energy density, the inflaton energy density shall follow the following equation
\bea \label{inflatonevolution}
\rho_\phi=\rho_\phi^{end}\,A^{-3\,(\,1+\,\omega_\phi\,)}\,e^{-\Gamma_\phi\,(\,1+\,\omega_\phi\,)\,(\,t-\,t_{end}\,)}\,\approx\,\rho_\phi^{end}\,A^{-3\,(\,1+\,\omega_\phi\,)}~~,
\eea
where $\rho_\phi^{end}$ is the inflaton energy density at the end of the inflation. As the initial stage of the perturbative reheating is dominated by the inflaton energy density, the main contribution in the gravitationally produced dark matter sector is coming at the initial stage. Therefore, we can ignore the effect of the decay constant $\Gamma_\phi$ in determining the gravitationally produced dark matter number density. The Hubble parameter during perturbative reheating can be approximated as
\bea \label{Hubble}
H=H_{end}\,A^{-\frac{3}{2}\,(\,1+\,\omega_\phi\,)}~~,
\eea
where $H_{end}=\sqrt{\rho_\phi^{end}/3\,M_p^2}$ is the Hubble parameter at the end of the inflation. Upon substituting the equations (\ref{inflatonevolution}) and (\ref{Hubble}) in the expression of the comoving gravitationally produced dark matter number density (\,Eqns.\ref{gravnumber}\,), we obtain
\begin{equation}\label{finalgrav}
n_{s}^{re}\,A_{re}^3\approx\frac{(\,\rho_\phi^{end}\,)^2\,(\,1+\,\omega_\phi\,)}{1024\,\pi\,M_p^4\,H_{end}}\,\int_{1}^{Are}A^{-\frac{1}{2}(\,5+\,3\,\omega_\phi\,)}\,dA\,=\frac{3}{512\,\pi}\,\frac{(\,1+\,\omega_\phi\,)}{(\,1+\,3\,\omega_\phi\,)}\,H_{end}^3\,\left[1-\,A_{re}^{-\frac{3}{2}\,(\,1+\,3\,\omega_\phi\,)}\,\right].
\end{equation}\\
{\bf Comoving number density of fermionic dark matter:} The relic abundance of the dark matter is obtained from the comoving dark matter number density, calculated at the end of the reheating. Inserting the expression for the decay width Eq.\ref{fermion}) into the Eq.\ref{Boltzgrav}, corresponding number density of the dark matter for this present scenario turns out to be 
\begin{equation}\label{gravnumberf}
n_{f}^{re}\,A_{re}^3=\int_{1}^{A_{re}}\frac{\rho_\phi^2\,m_f^2\,(\,1+\,\omega_\phi\,)}{4096\,\pi\,M_p^4\,m_\phi^2}\,\left(\,1-\,\frac{m_f^2}{m_\phi^2}\,\right)\,\frac{A^2\,dA}{H}\,\approx \int_{1}^{A_{re}}\frac{\rho_\phi^2\,m_f^2\,(\,1+\,\omega_\phi\,)}{4096\,\pi\,M_p^4\,m_\phi^2}\,\frac{A^2\,dA}{H}~~.
\end{equation}
The inflaton mass $m_\phi^{2}$ can be calculated from the second derivative of the inflaton potential. Since reheating happens near the minimum of the potential we first expand the inflaton potential in the limit of $\phi<<M_p$ as
\be \label{alphasimplified}
V(\phi)\simeq \lambda\,\phi^{2n}\,,
\ee
where $\lambda=\Lambda^4\,\left(\sqrt{\frac{2}{3\,\alpha}}\,\frac{1}{M_p}\right)^{2n}$. Therefore, 
\be \label{massapprox}
m_\phi^2=V''(\phi_0(t))\simeq 2n\,(\,2n-1\,)\,\lambda^{\frac{1}{n}}\,
\rho_\phi^{\frac{n-1}{n}}
\ee
 Upon substituting the equations (\ref{massapprox}), (\ref{Hubble}) and (\ref{inflatonevolution}) into the expression (\ref{gravnumberf}), one can find the gravitationally produced comoving fermionic dark matter number density at the end of reheating as
\begin{equation} \label{finalgravf}
n_f^{re}A_{re}^3=\frac{H_{end}^3\,m_f^2\,\lambda^{\frac{\omega_\phi-1}{\omega_\phi+1}}\,\nu(\omega_\phi)}{4096\,\pi\left(1+3\,\omega_\phi\right)\left(H_{end}^2M_p^2\right)^{\frac{2\,\omega_\phi}{1+\omega_\phi}}}\left[\,1-A_{re}^{-\frac{3}{2}\left(1-\omega_\phi\right)}\,\right]\simeq\frac{3}{2048\,\pi}\,\frac{1+\omega_\phi}{1-\omega_\phi}\,H_{end}^3\,\left(\frac{m_f}{m_\phi^{end}}\right)^2,
\end{equation}
where $\nu(\omega_\phi)=3^{\frac{1-\omega_\phi}{1+\omega_\phi}}\,(1-\omega_\phi)$ and $m_\phi^{end}$ indicates effective mass calculated at the end of the inflation. We use the relation $\omega_\phi=(n-1)/(n+1)$, to find the above relation of comoving dark matter number density in terms of $\omega_\phi$.\\
{\bf Comoving number density of vector dark matter:} For vector dark matter, the comoving number density can be written as (Combining Eqns.\ref{gauge} and \ref{Boltzgrav}) 
\begin{equation}\label{gravnumbervector}
n_{X}^{re}\,A_{re}^3=\int_{1}^{A_{re}}\frac{\rho_\phi^2\,(\,1+\,\omega_\phi\,)}{32768\,\pi\,M_p^4}\,\sqrt{1-\frac{m_X^2}{ m_\phi^2}}\,\left(4+\,4\,\frac{m_X^2}{ m_\phi^2}+\,19\,\frac{m_X^4}{ m_\phi^4}\right)\frac{A^2\,dA}{H}\,\approx \int_{1}^{A_{re}}\frac{\rho_\phi^2\,(\,1+\,\omega_\phi\,)}{8192\,\pi\,M_p^4}\,\frac{A^2\,dA}{H}~~.
\end{equation}
We can see that in the limit of $m_X<<m_\phi$, the above expression can be related with the comoving number density for the scalar dark matter (Eqn.\ref{gravnumber}) through a 1/8 factor. Therefore, 
\bea \label{numberdensityvector}
n_{X}^{re}\,A_{re}^3=\frac{1}{8}\,n_{s}^{re}\,A_{re}^3=\,-\frac{3}{4096\,\pi}\,\frac{(\,1+\,\omega_\phi\,)}{(\,1+\,3\,\omega_\phi\,)}\,H_{end}^3\,\left[\,A_{re}^{-\frac{3}{2}\,(\,1+\,3\,\omega_\phi\,)}-1\,\right]~~.
\eea\\
{\bf Expression for $m_{Y}^{max}$:} As we mentioned earlier the dark matter relic $\Omega_{Y}h^2$ could be expressed in terms of present radiation abundance $\Omega_R\,h^2$ as
\bea
\Omega_{Y}\,h^2\,=\frac{\rho_{Y}\,(\,A_{re}\,)}{\rho_R\,(\,A_{re}\,)}\,\frac{T_{re}}{T_{now}}\,\Omega_r\,h^2\,=\,\frac{m_{Y}\,A_{re}^{-3}\,(\,n_{Y}^{re}\,A_{re}^3\,)}{\beta\,T_{re}^3\,T_{now}}\,\Omega_r\,h^2~~,
\eea
where $\beta={\pi^2 g_{re}}/{30}$. In the context of the perturbative reheating dynamics, one can obtain the approximate analytical expression for the reheating temperature $T_{re}$ and the normalized scale factor $A_{re}$ at the end of the reheating to be (in this context, see Ref. \cite{Haque:2020zco})
\begin{equation}
T_{re}=\mathcal{G}\,Are^{-1},\,\mathcal{G}=\left(\frac{43}{11\, g_{s, re}}\right)^{\frac{1}{3}}\left(\,\frac{a_0\,T_0}{k}\,\right) H_k\,e^{-N_k},\,A_{re}=\left(\,\frac{12\,M_p^2\,H_{end}^2\,(\,1+\,\omega_\phi\,)^2}{\mathcal{G}^4\,\beta\,(\,5-\,3\,\omega_\phi\,)^2}\,\right)^{\frac{-1}{(1-\,3\,\omega_\phi)}}.
\end{equation}
Inserting expression of the reheating temperature into the expression of the present-day dark matter relic (admitting only gravitationally produced dark matter), the maximum allowed dark matter mass can be written as
\bea
m_{Y}^{max}=\frac{\mathcal{G}\,\beta\,T_{now}}{n_{Y}^{re}\,A_{re}^3}\,\frac{\Omega_{Y}\,h^2}{\Omega_r\,h^2}\,.
\eea
By utilizing the above equations with the expression of the comoving number density for gravitationally produced dark matter (Eqns. \ref{finalgrav}, \ref{finalgravf} and \ref{numberdensityvector}), we can easily fix $m_{Y}^{max}$.

\section{An analytical expression for the dark matter number density: produced from radiation bath}
The relevant Boltzmann equation for the dark matter particles produced from the radiation bath during perturbative reheating can be expressed as
\bea \label{numberdensityrad}
d(\,n_x\,a^3\,)=-a^3\, \langle\,\sigma\,v\,\rangle \,\left[\,n_x^2-(\,n_x^{eq}\,)^2\,\right]\,dt=-a^2\, \langle\,\sigma\,v\,\rangle \,\left[\,n_x^2-(\,n_x^{eq}\,)^2\,\right]\,\frac{dt}{H}~~.
\eea
Let us assume that the dark matter particles are relativistic (\,$m_x<<T$\,) and never reach chemical equilibrium (\,$n_x<<n_x^{eq}$\,) during reheating. Therefore Eqn.\ref{numberdensityrad} can be approximated as
\bea \label{numberradfinal}
d(\,n_x\,a^3\,)=\frac{a^3\, \langle\,\sigma\,v\,\rangle \,(\,n_x^{eq}\,)^2}{a\,H}\,da\simeq \frac{g^2}{\pi^4}\,\frac{a^2\,\langle\sigma\,v\,\rangle\,T^6}{H}\,da~~,
\eea
where we use equilibrium distribution of the dark matter in the relativistic limit
\bea
n_x^{eq}=\frac{g\,T^3}{\pi^2}~~.
\eea
Here $g$ counts the number of degrees of freedom associated with the dark matter particles. In the perturbative reheating scenario, the analytical expression for the radiation temperature during reheating can be obtained as
\bea \label{radtemp}
T=\gamma_3^{1/4}\,A^{-\frac{3}{2}\,(1+\omega_\phi)},\,\gamma_3=\frac{6}{5-3\,\omega_\phi}\,\frac{M_p^2\,H_{end}}{\beta}\,\Gamma_\phi\,(\,1+\omega_\phi\,)~~.
\eea
Connecting Eqns (\ref{numberradfinal}), (\ref{radtemp}) and (\ref{Hubble}) , the comoving dark matter number density is found to be 
\bea \label{finalgravrad}
\,n_x\,A_{re}^3= \frac{g^2}{\pi^4}\,\frac{\gamma_3^{3/2}\,\langle\,\sigma\,v\,\rangle}{H_{end}}\,\int_1^{A_{re}}A^{\frac{1}{4}\,(\,5-3\,\omega_\phi)}\,dA=\gamma_4\,\langle\,\sigma\,v\,\rangle\,\left(\,A^{\frac{3}{4}\,(\,3-\,\omega_\phi)}-1\,\right)~~,
\eea
where $\gamma_4=\frac{4}{3\,(\,3-\omega_\phi\,)}\,\frac{g^2}{\pi^4}\,\frac{\gamma_3^{3/2}}{H_{end}}$.

\section{Comoving number density of the gravitationally produced dark matter from SM scattering:}\label{smcal}
The evolution of the gravitational produced dark matter number density from radiation bath is followed by the Eqn.\ref{grav2} as
\bea\label{gravsm}
d(n_{Y(R)}\,A^3)=\,\gamma\,\frac{T^8}{M_p^4}\,\frac{A^2\,dA}{H}\,.
\eea
In the perturbative reheating scenario, the analytical expression for the radiation temperature during reheating can be obtained as
\bea \label{radtemp1}
T=\gamma_3^{1/4}\,A^{-\frac{3}{8}\,(1+\omega_\phi)},\,\gamma_3=\frac{6}{5-3\,\omega_\phi}\,\frac{M_p^2\,H_{end}}{\beta}\,\Gamma_\phi\,(\,1+\omega_\phi\,)~~.
\eea
Upon substitution of the Eqn.\ref{radtemp1} along with Eqn.\ref{Hubble} in equation \ref{gravsm}, the comoving number density turns out to be
\bea \label{smgrav}
n_{Y(R)}^{re}\,A_{re}^3=\frac{\gamma\,\gamma_3^2}{M_p^4\,H_{end}}\,\int_1^{A_{re}}\,A^{\frac{1}{2}\,(\,1-3\omega_\phi\,)}\,dA=\frac{2}{3\,(1-\omega_\phi)}\frac{\gamma\,\gamma_3^2}{M_p^4\,H_{end}}\,\left[A_{re}^{\frac{3}{2}\,(1-\omega_\phi)}-1\right]
\eea
As the normalized scale factor at the end of the reheating $A_{re}>>1$ (except for the temperature associated with the instantaneous reheating), the above equation simplified as
\bea\label{finalsmgrav}
n_{Y(R)}^{re}\,A_{re}^3=\frac{2}{3\,(1-\omega_\phi)}\frac{\gamma\,\gamma_3^2}{M_p^4\,H_{end}}A_{re}^{\frac{3}{2}\,(1-\omega_\phi)}\simeq \frac{2\gamma}{3\,(\,1-\omega_\phi\,)}\frac{e^{\frac{3}{2}\,N_{re}\,(\,3+\,\omega_\phi\,)}\,T_{re}^8}{M_p^4\,H_{end}}\,.
\eea
To find the above-simplified form, we use the approximate analytic expression for reheating temperature $T_{re}=\gamma_3^{1/4}\,A_{re}^{-\frac{3}{8}\,(1+\omega_\phi)}$ \cite{Haque:2020zco,Maity:2018exj}.\\
 \hspace{0.5cm}

\end{document}